\documentclass[landscape,twocolumn,twocolappendix]{aastex63}
\usepackage{multirow}
\usepackage{rotating}
\usepackage{xcolor}

\newcommand{\halpha}{H$\alpha$}
\newcommand{\pabeta}{Pa$\beta$}

\acceptjournal{ApJS}

\shorttitle{HST Imaging Survey of Local SF Galaxies}
\shortauthors{Giménez-Arteaga et al.}

\begin{document}

\title{High Resolution HST Imaging Survey of Local Star-Forming Galaxies I: Spatially-Resolved Obscured Star Formation with H$\alpha$ and Paschen-$\beta$ Recombination Lines}

\shorttitle{HST Imaging Survey of Local SF Galaxies - Obscured Star Formation}

\correspondingauthor{Clara Giménez-Arteaga}
\email{clara.arteaga@nbi.ku.dk}

\author[0000-0001-9419-9505]{Clara Giménez-Arteaga}
\affiliation{Cosmic Dawn Center (DAWN), Jagtvej 128, DK2200 Copenhagen N, Denmark}
\affiliation{Niels Bohr Institute, University of Copenhagen, Blegdamsvej 17, DK2100 Copenhagen \O, Denmark}

\author[0000-0003-2680-005X]{Gabriel B. Brammer}
\affiliation{Cosmic Dawn Center (DAWN), Jagtvej 128, DK2200 Copenhagen N, Denmark}
\affiliation{Niels Bohr Institute, University of Copenhagen, Blegdamsvej 17, DK2100 Copenhagen \O, Denmark}

\author[0000-0001-9002-3502]{Danilo Marchesini}
\affiliation{Department of Physics $\&$ Astronomy, Tufts University, 574 Boston Avenue Suites 304, Medford, MA 02155, USA}

\author[0000-0002-9090-4227]{Luis Colina}
\affiliation{Centro de Astrobiolog\'ia, (CAB), CSIC--INTA, Departamento de Astrof\'\i sica, Cra. de Ajalvir Km.~4, 28850 -- Torrej\'on de Ardoz, Madrid, Spain}
\affiliation{Cosmic Dawn Center (DAWN), Jagtvej 128, DK2200 Copenhagen N, Denmark}

\author{Varun Bajaj}
\affiliation{Space Telescope Science Institute, 3700 San Martin Drive, Baltimore, MD 21218, USA}

\author[0000-0002-0245-6365]{Malte Brinch}
\affiliation{Cosmic Dawn Center (DAWN), Jagtvej 128, DK2200 Copenhagen N, Denmark}
\affiliation{DTU-Space, National Space Institute, Technical University of Denmark, Elektrovej 327, 2800 Kgs. Lyngby, Denmark}

\author[0000-0002-5189-8004]{Daniela Calzetti}
\affiliation{Department of Astronomy, University of Massachusetts, 710 N. Pleasant Street, LGRT 619J, Amherst, MA 01002, USA}

\author{Daniel Lange-Vagle}
\affiliation{Department of Physics $\&$ Astronomy, Tufts University, 574 Boston Avenue Suites 304, Medford, MA 02155, USA}

\author[0000-0001-7089-7325]{Eric J. Murphy}
\affiliation{National Radio Astronomy Observatory, 520 Edgemont Road, Charlottesville, VA 22903, USA}

\author[0000-0002-0362-5941]{Michele Perna}
\affiliation{Centro de Astrobiolog\'ia, (CAB), CSIC--INTA, Departamento de Astrof\'\i sica, Cra. de Ajalvir Km.~4, 28850 -- Torrej\'on de Ardoz, Madrid, Spain}

\author[0000-0003-1580-1188]{Javier Piqueras-López}
\affiliation{Centro de Astrobiolog\'ia, (CAB), CSIC--INTA, Departamento de Astrof\'\i sica, Cra. de Ajalvir Km.~4, 28850 -- Torrej\'on de Ardoz, Madrid, Spain}

\author{Gregory F. Snyder}
\affiliation{Space Telescope Science Institute, 3700 San Martin Drive, Baltimore, MD 21218, USA}

\begin{abstract}
We present a sample of 24 local star-forming galaxies observed with broad- and narrow-band photometry from the \textit{Hubble Space Telescope}, that are part of the GOALS survey of local luminous and ultra-luminous infrared galaxies. With narrow-band filters around the emission lines H$\alpha$ (and [\ion{N}{2}]) and Pa$\beta$, we obtain robust estimates of the dust attenuation affecting the gas in each galaxy, probing higher attenuation than can be traced by the optical Balmer decrement H$\alpha$/H$\beta$ alone by a factor of $>1$ mag. We also infer the dust attenuation towards the stars via a spatially-resolved SED-fitting procedure that uses all available \textit{HST} imaging filters. We use various indicators to obtain the star formation rate (SFR) per spatial bin, and find that \pabeta\ traces star-forming regions where the H$\alpha$ and the optical stellar continuum are heavily obscured. The dust-corrected Pa$\beta$ SFR recovers the 24$\mu$m-inferred SFR with a ratio $-0.14\pm0.32$ dex and the SFR inferred from the $8\mathrm{-}1000\,\mu\mathrm{m}$ infrared luminosity at $-0.04\pm0.23$ dex. Both in a spatially-resolved and integrated sense, rest-frame near infrared recombination lines can paint a more comprehensive picture of star formation across cosmic time, particularly with upcoming JWST observations of Paschen-series line emission in galaxies as early as the epoch of reionization.
\end{abstract}


\section{Introduction} \label{sec:intro}

Past and recent multi-wavelength extra-galactic surveys both on the ground and in space (e.g., COSMOS, \citealt{2016ApJS..224...24L}; \citealt{2021arXiv211013923W}; CANDELS/3D-HST, \citealt{2011ApJS..197...35G}; \citealt{2011ApJS..197...36K}; \citealt{2014ApJS..214...24S}; UltraVISTA, \citealt{2012A&A...544A.156M}; \citealt{2013ApJ...777...18M}; UKIDSS, \citealt{2007MNRAS.379.1599L}; HFF, e.g., \citealt{2018ApJS..235...14S}), have allowed us to closer examine the evolutionary processes in many thousands of stellar-mass-selected galaxies back to when the Universe was only 1.5 Gyr old. Well sampled photometry of these objects enables us to robustly model their spectral energy distributions (SEDs) and thus estimate their redshifts and study the properties of their stellar populations (e.g. stellar mass, star formation rate/history, dust attenuation) over a wide range of redshifts. We can then study how these galaxies form, evolve and interact across cosmic time. However, the SED-fitting approach relies on a number of assumptions (e.g., the initial mass function, the dust geometry and extinction law, star formation history; see e.g. \citealt{2013ARA&A..51..393C} for a review), which are often overly simplified and surprisingly poorly constrained, even in the nearby Universe. 

To derive the intrinsic properties of a galaxy, one must have an in-depth understanding of the dust content of the sources, since it modifies the galaxies’ inherent SEDs in a wavelength-dependent way, in terms of both extinction and reddening. The dust in a galaxy absorbs the stellar emission in the UV and re-emits it at longer wavelengths. A commonly used dust obscuration prescription is the local starburst attenuation curve (e.g., \citealt{2000ApJ...533..682C}). For normal star-forming galaxies, a Milky Way-like or Magellanic-Clouds attenuation curve may be more appropriate \citep[e.g.,][]{2013ApJ...775L..16K}. Therefore, even in the local Universe the attenuation curve is not universally \cite{2000ApJ...533..682C}. As we look back in time, the conditions we observe and measure in galaxies change (star formation rate, metallicity, dust content, etc.), and it remains unknown whether these attenuation curves are valid for high-redshift galaxies or the variations in these properties make them also dependent on redshift \citep[see, e.g.,][]{2019ApJ...881...18R, 2020ARA&A..58..529S}.
  
Hydrogen recombination lines can be used to derive both the dust extinction and geometry. The most common pair of recombination lines used for this purpose is the Balmer decrement, H$\alpha$/H$\beta$. However, recent works find that the classic Balmer decrement underestimates the attenuation in galaxies when compared to the Balmer-to-Paschen decrement (H$\alpha$/Pa$\beta$). For instance, \citealt{2013ApJ...778L..41L} (hereafter L13) show that the ratios between redder lines (e.g. H$\alpha$/Pa$\beta$) can probe larger optical depths than bluer lines such as the Balmer decrement by $>$1 mag. \cite{1996ApJ...458..132C} find equivalent results with the \pabeta/Br$\gamma$ ratio on a sample of local starburst galaxies. The Balmer-to-Paschen decrement (H$\alpha$/Pa$\beta$) technique has been demonstrated for local-group HII regions (\citealt{2011AJ....142..132P}; \citealt{2014MNRAS.445...93D}), for a portion of the nearby starburst M83 (L13), and more recently employed at somewhat higher redshifts \citep{2020arXiv200900617C}. However, to date this technique has not been used extensively for systematic study of the extinction properties of local galaxies, nor with the quality and high spatial resolution of the \textit{Hubble Space Telescope} (HST). Local Seyfert, luminous and ultra-luminous infrared galaxies often have significant attenuation with $A_V > 2$ mag over a large fraction of the visible extent of the galaxies \citep[e.g.,][]{ 2013ApJ...771...62K, 2019A&A...622A.146M, 2019A&A...623A.171P, 2020A&A...643A.139P}; all above mentioned arguments suggest that even larger attenuations might be present in these classes of sources.

The strength of using hydrogen recombination lines to infer dust obscuration lies in the fact that the intrinsic line ratios show relatively little variation across a broad range of physical conditions of the ionised gas. As shown in L13, assuming Case B recombination, with changes in the temperature between 5$\times10^3$ and $10^4$ K, and in the electron density of an HII region between 10$^2$ and 10$^4$ cm$^{-3}$, the H$\alpha$/Pa$\beta$ line ratio only changes between 16.5 and 17.6 ($\sim7\%$, \citealt{1989agna.book.....O}). For the analysis below, we adopt an electron density $n_e=10^3$ cm$^{-3}$ and temperature $T_e$=7500 K (as in e.g. L13 for M83), resulting in an intrinsic ratio (H$\alpha$/Pa$\beta)_{int}$=17.56 \citep{1989agna.book.....O}. Under the assumption of constant electron density and temperature, deviations measured from this intrinsic value can thus be directly associated to dust attenuation.

Near infrared Paschen-series lines (e.g., Pa$\alpha$ and Pa$\beta$ at wavelengths 1.876 and 1.282~$\mu\mathrm{m}$, respectively) have been used to reveal star formation activity that is otherwise obscured for visible hydrogen recombination lines such as H$\alpha$ and H$\beta$, as well as for the optical emission from the stellar continuum \citep[e.g.,][]{ 2015ApJS..217....1T, 2016A&A...590A..67P, 2020arXiv200900617C}. They have also been compared to IR-based SFR indicators \citep[e.g., $L(8\mathrm{-}1000\,\mu\mathrm{m})$;][]{1998ARA&A..36..189K}, to check whether one can recover the star formation activity after applying dust corrections. Other studies directly suggest that one cannot use rest-optical lines to estimate physical properties of entire starburst systems, and one should instead aim to obtain rest-frame near-IR observations (\citealt{2017ApJ...838L..18P}, \citealt{2018ApJ...862L..22C}). Other recent works have used radio emission (specifically free-free emission) to measure star formation, in addition to IR and recombination lines \citep[e.g.,][]{2018ApJS..234...24M, 2019ApJ...881...70L, 2021ApJ...923..278L, 2021ApJ...916...73S}. Free-free emission does not have the caveat of being affected by extinction as is the case for recombination lines. However, no study has had both the high quality and spatial resolution that HST can offer. This work opens an exciting avenue in spatially-resolved studies that will become increasingly available with upcoming JWST observations.

Recent works show larger amount of dust obscuration with increasing redshift among the most massive galaxies (log($M/M_{\odot})>$ 10.5; \citealt{2009ApJ...706L.173B}; \citealt{2014ApJ...794...65M}; \citealt{2014ApJS..214...24S}; \citealt{2022ApJ...924...25M}). The detection of many of these objects in the \textit{Spitzer}/MIPS 24$\mu$m band implies that they have $L_{IR} > 10^{11}L_{\odot}$, typical of luminous and ultra-luminous infrared galaxies (LIRGs and ULIRGs; \citealt{1996ARA&A..34..749S}). At $z<1$, this population of galaxies seems to have generally diminished (\citealt{2014ApJ...794...65M}, \citealt{2017ApJ...837..147H}). The major complication to comprehend these systems is that at high-redshift they might be multiple unresolved objects in the process of merging \citep[][]{2017Natur.545..457D, 2018ApJ...868...46S, 2019ApJ...871..201M, 2020MNRAS.491L..18J, 2021A&A...645A..33B}, as well as how little we know about the dust distribution in them, leading to oversimplified assumptions of their dust modelling.

Understanding the dust distribution in local LIRGs and ULIRGs is imperative to further our comprehension of the population of massive galaxies at the peak of star formation ($z\sim2$, \citealt{2014ARA&A..52..415M}). Beyond $z>1$, U/LIRGs begin to play an important part in the evolution of the star formation history of the Universe, increasing its contribution with redshift, to the point of even dominating the SFR activity at $z\sim2$ \citep[e.g.,][]{2005ApJ...630...82P,2013A&A...553A.132M,2021ApJ...909..165Z}. These LIRGs studies show that obscured star formation contributes the most to the SFR density. Instruments such as WFC3 onboard of the HST allow us to closely examine this by obtaining H$\alpha$ and Pa$\beta$ extinction maps and star formation estimates. Although separated by billions of years of evolution, local highly luminous objects appear similar to very distant massive dusty galaxies, in terms of for example their high infrared luminosity, large amounts of dust obscuration, \halpha\ size and surface brightness \citep[see e.g.,][]{2012A&A...541A..20A}. Therefore, by exploiting the high spatial resolution and high signal-to-noise ratio (S/N) of these local systems, we can at the same time aid our closer examination of the distant Universe. On the other hand, these local objects appear to be different to higher redshift systems in terms of the distribution of the star formation and the dust temperature \citep{2010ApJ...725..742M,2022arXiv220402055B}. Furthermore, high redshift U/LIRGs are more extended and display a cooler IR SED \citep{2011A&A...533A.119E}. 

In this paper we present the observations and first results of a multi-wavelength study of 24 nearby galaxies ($z < 0.035$) observed with HST, to study the spatially-resolved properties of their dust-obscured stellar populations and star formation activity.  Using data from our own recent program along with archival observations, we obtain, at a minimum, narrow-band images centered on the redshifted \halpha\ and \pabeta\ recombination lines and corresponding optical and near-infrared broad-band continuum images in filters similar to rest-frame $I$- and $J$-bands.  Archival observations of a subset of the full sample provide additional broad-band images extending the SED sampling to the near-ultraviolet wavelengths.  We develop a spatial binning procedure with Voronoi tessellation that probes spatial scales as small as 40 pc at the median redshift of our sample ($z=0.02$), and a SED-fitting technique that robustly infers emission line fluxes from the narrow-band images.

This paper is structured as follows: In Section \ref{sec2}, we introduce the HST observations and data processing procedure. Section \ref{sec3} describes the methodology we develop for the spatially-resolved image analysis, including spatial binning and multi-band SED-fitting software. In Section \ref{sec4}, we present the main results and properties of our sample, as well as discussing the implications of our findings. Finally, Section \ref{sec6} presents a summary of our work and corresponding conclusions. Throughout this paper, we assume a \cite{2003PASP..115..763C} initial mass function (IMF) and a simplified $\Lambda$CDM cosmology with $H_0 = 70$ km/s/Mpc, $\Omega_m$= 0.3 and $\Omega_{\Lambda} = 0.7$. We use the biweight location and scale statistics defined by \citet{coldbeers}, when referring to derived mean values and their uncertainties.

\section{Data and observations} \label{sec2}

We conducted an HST snapshot survey program in Cycle 23 (HST-14095, PI: Gabriel Brammer; \citealt{2015hst..prop14095B}) to obtain narrow-band images of nearby galaxies centred on the (redshifted) \halpha\ and \pabeta\ hydrogen recombination lines. The full target list of the survey was defined as essentially any nearby galaxy ($z<$ 0.05) that had existing wide-field HST archival imaging in either narrow-band \halpha\  from ACS/WFC or WFC3/UVIS, or \pabeta\ from WFC3/IR, to which we added the missing narrow-band and an associated broad-band continuum filter. The combination of the parent selection and the fact that a random subset of them were observed as snapshots, yields a heterogeneous sample in terms of star formation rate, stellar mass, morphology, etc. ($0.01<$ SFR [$M_\odot$/yr] $<200$; $10^{7.5}< M_* [M_\odot] < 10^{15.5}$; and in terms of morphology, from spirals to irregulars and mergers). The observations are available for a sample of 53 galaxies, 24 of which are in the GOALS survey of local luminous and ultraluminous infrared galaxies \citep[$z < 0.088$, $L_\mathrm{IR} > 10^{11}~L_\odot$]{2009PASP..121..559A}.

\begin{table*}[!th]
\centering
\begin{tabular}{l c c c c c c c}
\hline
\hline
\multirow{3}{*}{Target} & R.A. & Dec.& \multirow{3}{*}{$z$} & \multirow{3}{*}{log$(\frac{L_{IR}}{L_{\odot}})$} & UV & Optical & IR \\ 
 & [deg] & [deg] & & & WFC3/ & WFC3/UVIS, ACS/WFC & WFC3/IR \\
 & & & & & UVIS & & \\
\hline
Arp~220\textsuperscript{**} & 233.7375 & 23.5031 & 0.01813 & 12.27 & F336W & F435W, FQ508N, F621M, & F110W, F130N, \\ 
 & & & & & & F665N, F680N, F814W & F160W \\
ESO550-IG025\textsuperscript{**} & 65.33333 & -18.81333 & 0.03209 & 11.50 & F225W & F435W, f673n, F814W & f110w, f132n \\
IRAS03582+6012 & 60.63374 & 60.34383 & 0.03001 & 11.42 & & f625w, f673n & F110W, F130N, \\
 & & & & & & & F132N, F160W \\
IRAS08355-4944\textsuperscript{**} & 129.25761 & -49.90841 & 0.02590 & 11.61 & F225W & F435W, f673n, F814W & F110W, F130N, F132N \\ 
IRAS12116-5615\textsuperscript{**} & 129.25761 & -49.90841 & 0.02710 & 11.64 & & F435W, f673n, F814W & f110w, f132n \\ 
IRAS13120-5453\textsuperscript{*} & 198.77660 & -55.15628 & 0.03076 & 12.31 & F225W & F435W, f673n, F814W & F110W, F130N, F132N \\ 
IRAS18090+0130\textsuperscript{**} & 272.91010 & 1.52771 & 0.02889 & 11.64 & & F435W, f673n, F814W & f110w, f132n \\ 
IRAS23436+5257 & 356.52380 & 53.23235 & 0.03413 & 11.56 & F225W & F435W, f673n, F814W & F110W, F130N, F132N  \\ 
IRASF10038-3338\textsuperscript{**} & 151.51938 & -33.88503 & 0.03410 & 11.77 & & F435W, FR656N, f673n, & F110W, F130N, F132N\\
 & & & & & & F814W, FR914M &  \\
IRASF16164-0746\textsuperscript{*} & 244.799 & -7.90081 & 0.02715 & 11.61 & & F435W, f673n, F814W & F110W, F130N, F132N \\
IRASF16399-0937\textsuperscript{**} & 250.66709 & -9.7205 & 0.02701 & 11.62 & & F435W, FR656N, f673n, & F110W, F130N, F132N \\ 
 & & & & & & F814W, FR914M &  \\
MCG-02-01-051 & 4.71208 & -10.376803 & 0.02722 & 11.47 & & F435W, f673n, F814W & F110W, F130N, F132N \\ 
MCG+12-02-001\textsuperscript{**} & 13.51643 & 73.084786 & 0.01570 & 11.49 & F225W & F435W, f665n, F814W & f110w, f130n \\
NGC1614\textsuperscript{**} & 68.50011 & -8.57905 & 0.01594 & 11.64 & F225W & F435W, f665n, F814W & f110w, f130n  \\ 
NGC2146\textsuperscript{**} & 94.65713 & 78.35702 & 0.00298 & 11.11 & F225W, & F658N, F814W & F110W, F128N,  \\
 & & & & & F336W & & F160W, F164N \\
NGC2623\textsuperscript{*} & 129.60039 & 25.75464 & 0.01851 & 11.59 & & F435W, F555W, FR656N,& f110w, f130n \\ 
 & & & & & & f665n, F814W & \\
NGC5256\textsuperscript{*} & 204.57417 & 48.27806 & 0.02782 & 11.55 & & F435W, f673n, F814W & F110W, F130N, F132N \\ 
NGC5331\textsuperscript{**} & 208.06729 & 2.10092 & 0.03304 & 11.65 & & F435W, f673n, F814W & F110W, F130N, F132N \\ 
NGC6090\textsuperscript{**} & 242.91792 & 52.45583 & 0.02984 & 11.57 & F336W & F435W, F502N, F550M, & F110W, F130N, \\
 & & & & & & FR656N, f673n, F814W & F132N, F160W \\
NGC6240\textsuperscript{*} & 253.24525 & 2.40099 & 0.02448 & 11.92 & FQ387N & F435W, F467M, FQ508N, & F110W, F130N, F132N \\
 & & & & & & F621M, F645N, F673N, &  \\
 & & & & & & F680N, F814W &  \\
NGC6670\textsuperscript{**} & 278.3975 & 59.88881 & 0.02860 & 11.64 & F225W & F435W, f673n, F814W & f110w, f132n \\ 
NGC6786 & 287.7247 & 73.41006 & 0.02511 & 11.48 & & F435W, f673n, F814W & F110W, F130N, \\
 & & & & & & & F132N, F160W \\
NGC7592\textsuperscript{*} & 349.59167 & -4.41583 & 0.02444 & 11.39 & & f625w, f673n & f110w, f132n \\ 
VV340A\textsuperscript{*} & 224.25127 & 24.606853 & 0.03367 & 11.73 & & F435W, f673n, F814W & F110W, F130N, F132N \\  
\hline
\hline
\multicolumn{8}{l}{\textsuperscript{*}\footnotesize{Targets that host an AGN, X-ray selected based on \cite{2011iwasawa}, \cite{2018torres} or \cite{2021MNRAS.506.5935R}.}}\\
\multicolumn{8}{l}{\textsuperscript{**}\footnotesize{Targets that do not host an AGN, based on \cite{2011iwasawa}, \cite{2018torres} or \cite{2021MNRAS.506.5935R}.}}\\
\end{tabular}
\caption{Sources in the sample of nearby star-forming galaxies presented in this work that are also part of the GOALS Survey \citep{2009PASP..121..559A}. The infrared luminosity is from GOALS (normalised to our cosmological parameters using \cite{2006PASP..118.1711W}). The redshift is from \cite{2017ApJS..229...25C}. The filters added by the Cycle 23 SNAP program (HST-14095, \citealt{2015hst..prop14095B}) are indicated with small caps, whereas the rest are archival data.}
\label{tab_sample}
\end{table*}

The number of available imaging filters from the ultraviolet through the near infrared varies significantly across the full sample. At the very least, each target has narrow-band images for \halpha, \pabeta, and an associated continuum filter for each line. The filters used in the continuum also vary due to the availability of archival observations -- the optical continuum may be F606W, F621M, F555W, F625W or F814W and the near-infrared continuum is either the broad F110W filter and/or off-band narrow-band images (the continuum filters were WFC3/UVIS F814W for \halpha\ and WFC3/IR F110W for \pabeta\ for the observations from our program). The narrow-band filters were chosen as appropriate for the redshift of a given source. The full sample and filter coverage are summarized in Tables~\ref{tab_sample} and \ref{tab_sample2}.

Each image was processed/created using the \texttt{DrizzlePac} \citep{2012drzp.book.....G}  modules \texttt{TweakReg} for relative alignment and \texttt{AstroDrizzle} for image combination and stacking. The original pixel size of the images obtained with the WFC3/IR channel is 0\farcs128 pixel$^{-1}$. The images are drizzled to combined mosaics with 0\farcs1/pixel for the IR filters and 0\farcs05/pixel for the optical ACS/WFC and WFC3/UVIS filters\footnote{All of the aligned image mosaics are available at \\ \url{http://cosmos.phy.tufts.edu/dustycosmos/}}. Our sample spans redshifts from 0.0001 to 0.035, with a median $z\sim0.02$. This corresponds to a physical size of 0.3 to 70~pc/pixel, $\sim$40~pc/pixel at the median redshift (for the 0\farcs1 WFC3/IR pixels).

\begin{figure}[t]
\centering
\includegraphics[width=\columnwidth]{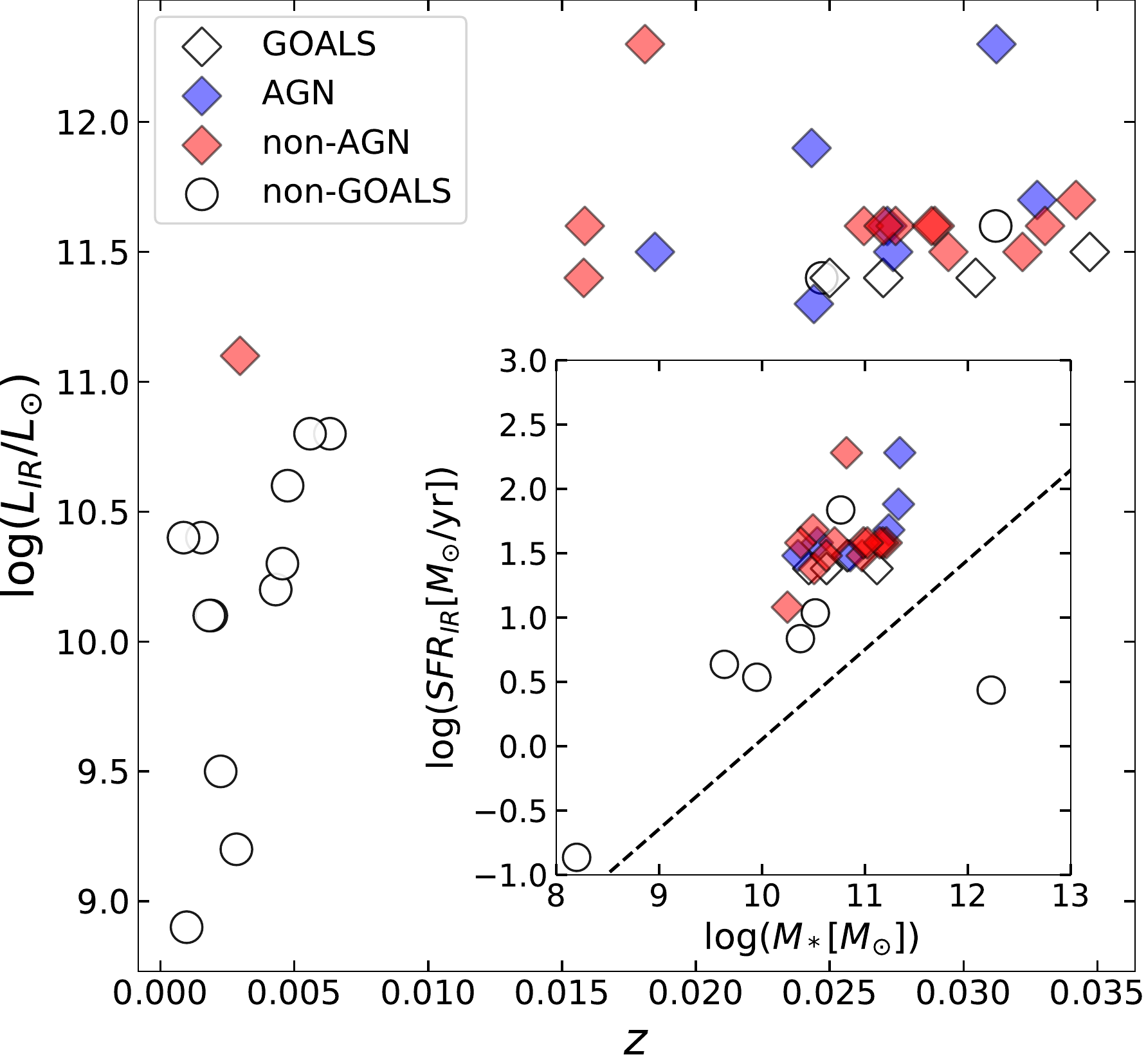}
\caption{Infrared luminosity as a function of redshift for the galaxies where $L_{IR}$ is available, as tabulated in Tables \ref{tab_sample} and \ref{tab_sample2}. The targets from the GOALS sample are indicated with the diamond markers (as opposed to white circles). The X-ray selected AGNs are blue-coloured, whereas the non-AGN are red. The inset plot shows the SFR$_{IR}$ (derived from $L_{IR}$ using Eq.\ref{eq_ir}) versus the stellar mass (inferred with our SED-fitting code), with the $z=0.02$ main sequence from \cite{2012ApJ...754L..29W} indicated as the dashed line. \label{fig_sample}}
\end{figure}

Our galaxies span a wide range in infrared luminosity (8.9 $<$ log($L_{IR}$/$L_{\odot}$) $<$ 12.3), as can be seen in Figure \ref{fig_sample}, which yields a very heterogeneous sample, also in terms of e.g., stellar mass and morphology. A large number of our galaxies are undergoing starbursts (as we would expect for the GOALS sample selection criteria, which are all above the main sequence in Figure \ref{fig_sample}), most likely driven by mergers in some cases, since we can see paired galaxies in our HST images. In the work presented in this first paper, where we focus on the study of the star formation activity inferred with different tracers, we only use the targets from the GOALS sample in our analysis (Table \ref{tab_sample}), due to the availability of measurements for their infrared luminosities and 24$\mu$m fluxes. We present additional galaxies in the Appendix (Table \ref{tab_sample2}), for which we have processed the HST data as explained above, but that are not part of the analysis in this work. These targets often have spatial extent larger than the Field of View (FoV) of the instrument.

\section{Methodology} \label{sec3}

In this work we develop two main methodologies to first derive spatial bins that homogenize the measurement signal-to-noise across the extent of each target galaxy, and second to analyse the (binned) multi-wavelength SEDs, that include measurements from as many as 12 broad- and narrow-band imaging filters from NUV to NIR wavelengths.

\subsection{Voronoi Binning} \label{sec:voronoi}
Modern instruments allow us to spatially resolve nearby extended sources, although these observations can display a significant anisotropy in the signal-to-noise-ratio (S/N) across the target. Sometimes these variations differ by orders-of-magnitude, and some pixels often have poor S/N. To resolve this, the spatial elements can be grouped locally (binned) to obtain a more uniform S/N across the image, however this results in a loss of spatial resolution. 

Disk galaxies typically have ``exponential'' surface brightness profiles \citep{1981ApJS...46..177B}, thus the outskirts are much fainter than the centers. With this in mind, we implement an adaptive binning scheme: for lower S/N, larger bins are created, whereas for high S/N regions, the bins are smaller and a high resolution is maintained. For this, we adapt the Voronoi binning procedure from \cite{2003MNRAS.342..345C}. This method implements adaptive spatial binning of integral-field spectroscopic (IFS) data to achieve a specified constant signal-to-noise ratio per spatial bin.

The \cite{2003MNRAS.342..345C} Voronoi algorithm\footnote{\url{https://www-astro.physics.ox.ac.uk/~mxc/software/}} is inefficient for handling datasets with millions of data points as is the case for the large image mosaics used here. Furthermore, the algorithm is not robust in the regime of low S/N per original data point \citep{2003MNRAS.342..345C}, as in the outskirts of the galaxies in our HST images. Therefore, we adopt a hybrid approach of block-averaging the galaxy mosaics in progressively-shrinking box sizes and applying the Voronoi algorithm on the blocked image. A detailed explanation of our binning procedure\footnote{\url{https://github.com/claragimenez/voronoi}} is provided in Appendix \ref{sec:app_voronoi}.

\begin{figure*}[t]
\centering
\includegraphics[width=\textwidth]{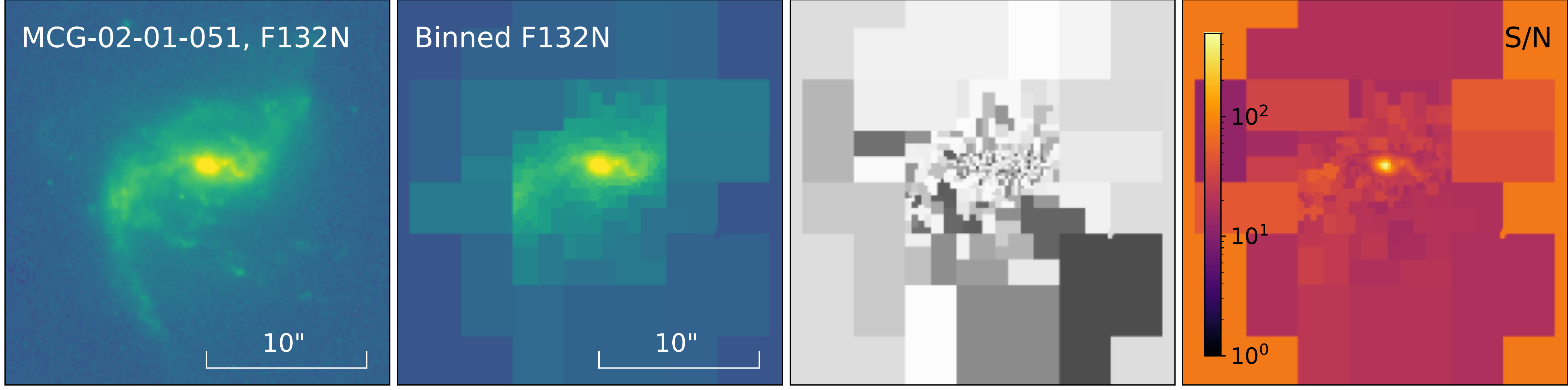} 
\includegraphics[width=\textwidth]{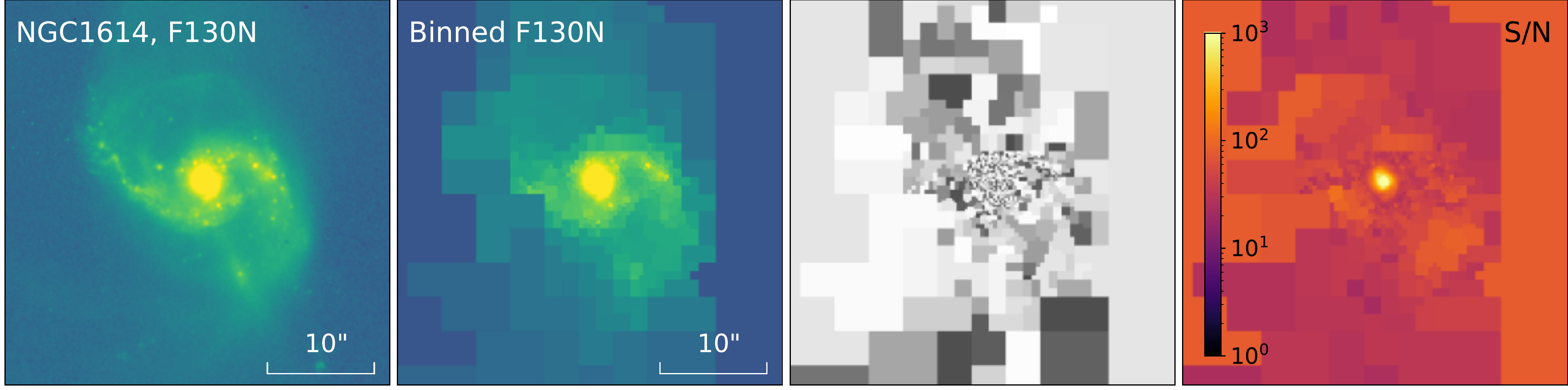}
\caption{Voronoi binning for the targets MCG-02-01-051 (top panels) and NGC1614 (bottom panels). From left to right: original HST narrow-band image that targets Pa$\beta$; binning on the narrow-band image where we impose a S/N threshold of 20; Voronoi binning pattern, where the colourmap separates each individual bin from its neighbours; resulting signal-to-noise maps on the narrow-band image after binning. \label{fig_binning}}
\end{figure*}

For the spatial binning, we set a target signal-to-noise (S/N) ratio of at least 20 per bin in the narrow-band image that includes \pabeta, to ensure a minimum constant S/N across the resulting \pabeta\ emission line image once the continuum is subtracted, where the per-pixel uncertainties are provided by the \textit{HST} imaging calibration pipelines \citep[e.g., \S 3.3.5 of the WFC3 Data Handbook;][]{wfc3dhb}. The spatial bins derived from the \pabeta\ narrow band image are then applied to all other available filters for a given target, so that we can measure photometry for each spatial bin across the filters. The threshold S/N imposed on the \pabeta\ narrow band image is not necessarily achieved in all the other filters (e.g. UV filters, where S/N can be low), whereas it can also be higher in broad band filters such as F110W. Rather than matching the wavelength-dependent PSFs of the different filters, which involves a convolution with an imperfect matching kernel and modifies the pixel variances in a nontrivial way, we adopt a minimum bin size of $2\times2$ 0\farcs1 pixels so that the bins are larger than the PSF FWHM of any of the individual images. Figure \ref{fig_binning} shows a demonstration of the binning scheme for the targets MCG-02-01-051 (top panels) and NGC1614 (bottom panels).

\subsection{Continuum Subtraction}

The raw flux density measured in a narrow-band filter centered on an emission line includes contributions from both the line itself and any underlying continuum, where the fractional contribution of the continuum increases with the width of the filter bandpass. To produce pure hydrogen emission line images (\halpha\ and \pabeta\ line fluxes in our case), we first need to remove the contribution from the stellar continuum. A common approach to deal with this issue is to perform a ``simple'' background subtraction, in which a nearby broad-band filter (or an interpolation of multiple broad-band filters) is scaled and subtracted from the respective narrow-band filter containing the emission line of interest using some estimate of the continuum shape across the various bandpasses. A linear continuum shape is generally either assumed explicitly or implicitly (e.g., a flat spectrum that defines the photometric calibration of the various filters) or estimated empirically from the colour of a pair of continuum filters bracketing the narrow line bandpass \citep[as is done in e.g.,][]{2013ApJ...778L..41L,2021ApJ...913...37C}. The adopted broad- or medium-band continuum filters may or may not contain the line of interest. A narrow band filter adjacent to the narrow band line filter can provide a relatively robust continuum estimate, though it is relatively more expensive to reach a given continuum sensitivity as the bandpass shrinks.  In all of the cases above, the continuum shape is likely the dominant source of systematic uncertainty on deriving the emission line flux from the narrow-band image.

In this work we develop a new approach to derive the continuum and emission line contributions that uses all of the available information by fitting population synthesis templates to the spatially resolved photometry in all available filters for a given target. It is inspired by the photometric redshift fitting code EAZY \citep{2008ApJ...686.1503B}, with the motivation to precisely fit the contribution from the stellar continuum and the lines separately, in order to produce robust line fluxes without stellar continuum contamination. Our code fits a set of continuum and line emission templates on a bin-to-bin basis, in order to infer spatially-resolved physical properties of our nearby galaxies. In the next section, as well as in Appendix \ref{sec:app_sedfit}, we explain our SED-fitting code in further detail.

\begin{figure}[!t]
\includegraphics[width=\columnwidth]{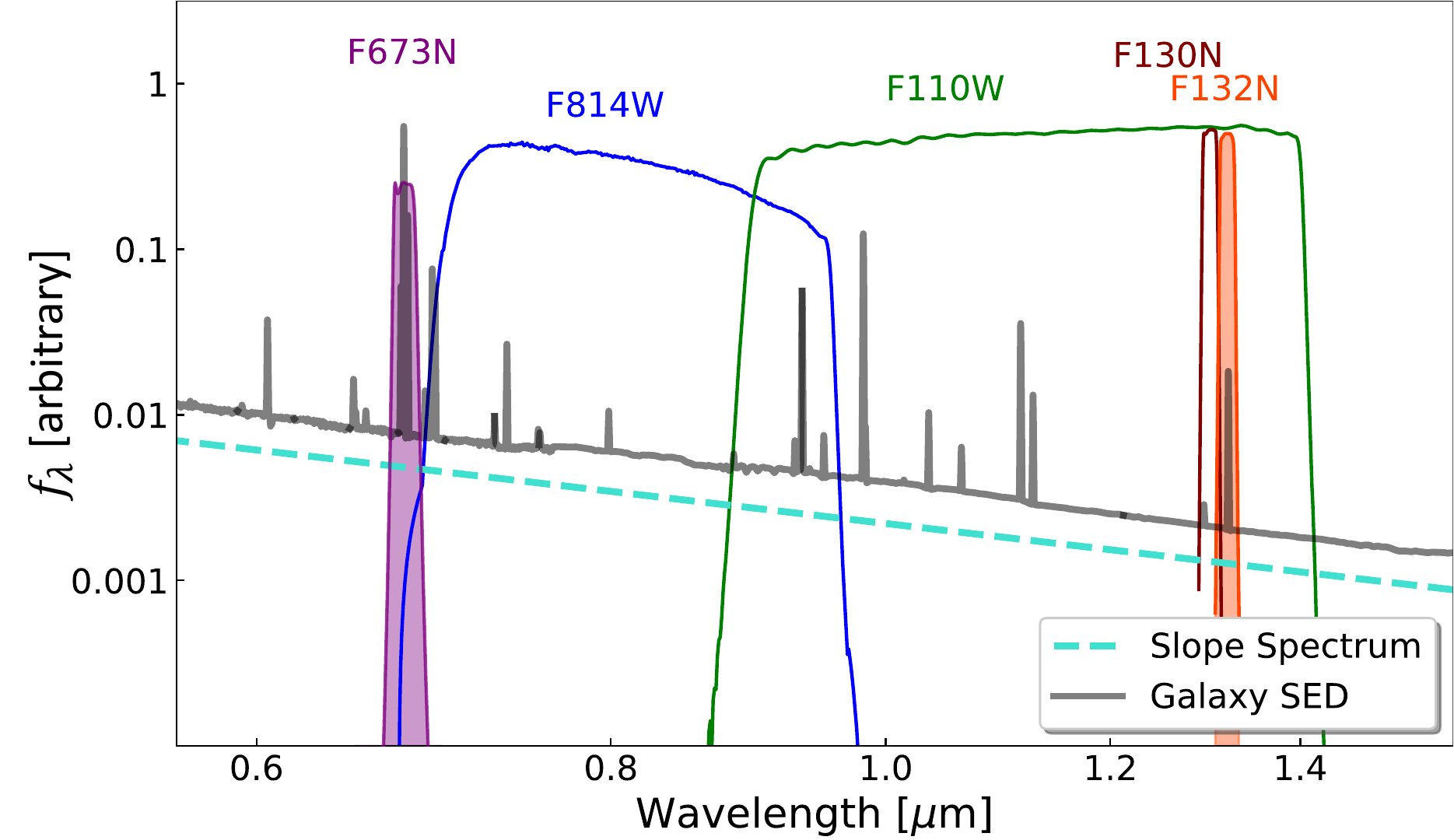}
\caption{Transmission of some of the HST filters that cover our sample, indicated by the coloured solid lines and the name for each filter, taken from the SVO Filter Profile Service \citep{2012ivoa.rept.1015R}. The solid grey line corresponds to a galaxy SED redshifted to $z=0.03$. The dashed turquoise line corresponds to a slope spectrum model used to fit the continuum in the 'simple' approach. The position of the H$\alpha$ and Pa$\beta$ emission lines are indicated by the coloured narrow bandpasses.  \label{fig:filter_curves}}
\end{figure}

Figure \ref{fig:filter_curves} shows an example of the HST coverage that we can have for one of our targets. We have two narrow-band filters targeting H$\alpha$ (purple colouring) and Pa$\beta$ (orange colouring), as well as broad-band neighbouring filters, F814W and F110W in this case. A flat sloped spectrum (turquoise dashed line) and a redshifted galaxy SED (black line) are plotted, and could be used in ``simple'' continuum subtraction methods to remove the stellar continuum emission from the narrow-band filters. As we can see, the F110W broad-band filter contains the emission line, and would lead to over-subtraction if used for this purpose. On the other hand, we could have off-band narrow-band filter coverage, such as F130N in Figure \ref{fig:filter_curves}, which would allow an analytical solution to subtract the continuum (in this case, iterative subtraction methods are used), but this is not always available, therefore justifying our need to develop a novel methodology. Sections \S \ref{section_ha_nii} and \S \ref{section_pab} display comparisons between different subtraction methods in further detail.

\subsection{SED-Fitting} \label{sec:sedfit}
As introduced in the previous section, we develop an SED-fitting code to accurately fit the continuum and line emissions in order to obtain the H$\alpha$ and Pa$\beta$ line fluxes. Furthermore, our method allows us to derive spatially-resolved physical properties, fitting each galaxy on a bin-to-bin basis, obtaining local estimates such as the star-formation rate (SFR) and stellar mass.

We develop an adaptation of the python-based version of EAZY\footnote{\url{https://github.com/gbrammer/eazy-py}} \citep{2008ApJ...686.1503B}, which we make publicly available\footnote{\url{https://github.com/claragimenez/sed_fit_photometry}}. Like EAZY, our code finds the linear combination of templates that best fits all of the observed photometry of each spatial bin. 

We use a set of four continuum templates and two extra line templates, one with H$\alpha$+[\ion{N}{2}] in a fixed ratio of 0.55, (see \S \ref{section_ha_nii} below) and the other with Pa$\beta$ emission, that allow us to fit the emission lines separately\footnote{We do not fit for additional near-infrared emission lines (e.g., [\ion{S}{0iii}]$\lambda\lambda$9070,9530, HeI$\lambda$10830) that can contribute
up to $\sim$12\% of the signal in the broad F110W filter.}.

Analogous to the stepwise SFH parameterization of the \texttt{Prospector-$\alpha$} software \citep{2017zndo...1116491J, 2017ApJ...837..170L}, we choose four continuum templates to reasonably sample the star formation histories, but without adding excessive flexibility that would lead to the SFHs not being properly constrained by the limited data that we have. The continuum templates are generated with python-fsps (Flexible Stellar Population Synthesis; \citealt{2009ApJ...699..486C}, \citealt{conroy2010}, \citealt{2021zndo...4737461J}), that allows us to specify our own SFHs, for stepwise time bins between 0, 50, 200, 530 Myr, 1.4 Gyr with constant SFR across each bin (see \S \ref{sec:app_sedfit}). However, by deriving the template scaling coefficients using standard least-squares optimization we do not impose any priors on the relative contributions of the step-wise SFH bins. The template normalization coefficients are transformed to emission line fluxes and star formation rates and stellar masses of the continuum-emitting stellar population, and the analytic covariance of the fit coefficients is used to compute the posterior distributions of those derived parameters. This fitting approach is dramatically faster than sampling codes such as \texttt{Prospector}---running $\sim$800 spatial bins of a single target galaxy with \texttt{Prospector} requires 1--2 weeks on a supercomputer on average, whereas we can perform the full fit in the same galaxy in under 5 minutes with our code and a fairly standard laptop computer\footnote{Tested on a 2GHz Quad-Core i5 CPU.}---. In Appendix \ref{sec:app_sedfit} we discuss additional considerations of the choice of templates and the SED-fitting code.

A new implementation with respect to the EAZY machinery is the addition of a reddening grid. Instead of fitting for redshift, which is well-known for the nearby galaxies in our sample, we construct a reddening grid where we redden the continuum templates by a given amount, and fit the normalization coefficients of the two emission lines and reddened continuum templates. The attenuation curve used to redden the continuum templates can be input by the user. In this work we use a \cite{2000ApJ...533..682C} attenuation curve. The introduction of the line templates to directly fit the line fluxes is the main novelty and motivation behind developing our spatially-resolved SED-fitting code. This allows us to constrain on the one hand the attenuation traced by the stellar continuum, which is due to the diffuse interstellar medium (ISM), and on the other hand, we are able to evaluate the attenuation that the gas suffers, traced by the inferred empirical decrement H$\alpha$/Pa$\beta$. As discussed in e.g. \citet{2020MNRAS.495.2305G}, the ``extra'' attenuation experienced by the gas is due to stellar birth clouds, which are \ion{H}{2} regions enshrouded in dust, where dust is clumpier than in the diffuse ISM, and it can be traced with nebular emission lines (H$\alpha$ and Pa$\beta$ in our study). 

Our SED-fitting code finds the best fit solution by minimizing $\chi^2$, defined as in \citet{2008ApJ...686.1503B}:
\begin{equation}
    \chi^2_{A_V} = \sum^{N_{filt}}_{j=1} \frac{(T_{A_V,j}-F_j)^2}{(\delta F_j)^2},
\end{equation}
where $N_{filt}$ is the number of filters, $T_{A_V,j}$ is the synthetic flux of the linear combination of templates in filter $j$ for reddening $A_V$, $F_j$ is the observed flux in filter $j$, and $\delta F_j$ is the uncertainty in $F_j$, obtained combining the observed uncertainty and the template error function (see \citealt{2008ApJ...686.1503B} for more information). Once we find our best fit linear combination, we directly infer the line fluxes H$\alpha$ and Pa$\beta$ from the fitted line emission templates. We do this on a bin-to-bin basis, and the ensemble result is the spatially-resolved fit across the face of the target galaxy. Our SED-fitting code allows user input for a variety of parameters and files, such as the templates that are used to fit both the continuum and the lines, making the code flexible to fit both contributions across all wavelength ranges, as long as the input templates cover the desired interval and line emissions.

In summary, our SED-fitting procedure outputs the line fluxes H$\alpha$ and Pa$\beta$ in erg/s/cm$^2$, as well as various physical properties: the extinction ($A_V$) obtained from the ``empirical'' Balmer-to-Paschen decrement ($A_V($H$\alpha/$Pa$\beta$)) and the $A_V$ inferred from the stellar population ($A_{V}$ or $A_{V,continuum}$, derived with the continuum templates). We also obtain the star-formation rate (SFR) and stellar mass ($M_{*}$) (surface density) in each spatial bin. 

\begin{figure*}[t]
\centering
\includegraphics[width=\textwidth]{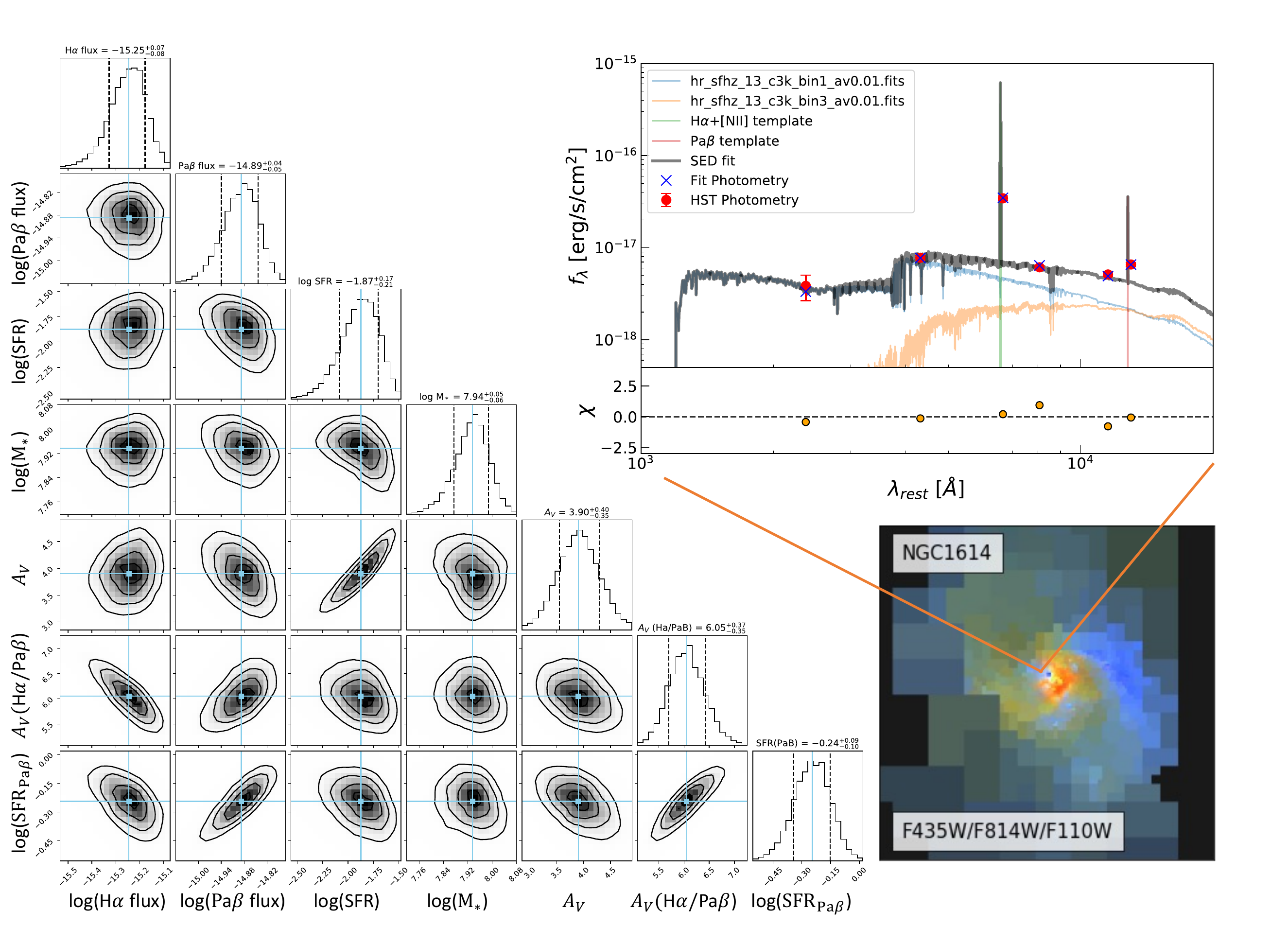} 
\caption{Example of our SED-fitting code applied on the galaxy NGC1614. The bottom right cutout shows the RGB image built combining the F435W, F814W and F110W broad-band filters. The top right plot shows the resulting SED-fit on an example bin from the galaxy. The red points and uncertainties correspond to the original HST photometry. The blue crosses are the resulting synthetic photometry of the best fit. The black curve shows the best fit resulting SED, which is a combination of the different coloured continuum templates and the two line emission templates. The bottom panel shows the residual $\chi$ from the fit. On the left, we display the corner plot for the whole fit parameter space for this example bin, displaying the different correlations and distributions between the physical parameters that our code infers. \label{fig_test_corner_sed}}
\end{figure*}

Additionally, while the SFR from the SED-fit estimates the star formation activity throughout the last $\sim$100 Myr, we derive the instantaneous ($\sim$10 Myr) star formation rate from the H$\alpha$ luminosity, following the \citet{1998ARA&A..36..189K} relation, and dividing by 1.8  \citep[][]{2007ApJ...666..870C,2009ApJ...703.1672K} to convert from a \cite{1955ApJ...121..161S} IMF to \cite{2003PASP..115..763C}:
\begin{equation}
    \textrm{SFR[M}_{\odot}\textrm{yr}^{-1}] = 4.4\times10^{-42} L_{H\alpha,corr} [\textrm{erg s}^{-1}], \label{eq_sfr_ha}
\end{equation}
where $L_{H\alpha,corr}$ is the dust-corrected H$\alpha$ luminosity, calculated as:
\begin{equation}
    L_{H\alpha,corr} = L_{H\alpha,obs}\times10^{0.4 A_{H\alpha}}, \label{eq_3}
\end{equation}
 where $A_{H\alpha}$ is the extinction at the H$\alpha$ wavelength, and can be obtained using the parameterisation of the attenuation $A_{\lambda} = k(\lambda) E(B-V)$, where $k(\lambda)$ is given by an attenuation curve, that we must assume, and $E(B-V)$ is the colour-excess, which we can calculate in terms of the Paschen-$\beta$ and H$\alpha$ observed ratio (e.g. \citealt{1996ApJ...458..132C}):
\begin{equation}
    E(B-V) = \frac{2.5}{k(\textrm{Pa}\beta)-k(\textrm{H}\alpha)} \, \, \textrm{log}\Big(\frac{(\textrm{H}\alpha/\textrm{Pa}\beta)_{obs}}{(\textrm{H}\alpha/\textrm{Pa}\beta)_{int}}\Big),
    \label{eq_color_excess}
\end{equation}
where $(\textrm{H}\alpha/\textrm{Pa}\beta)_{int}$ is the intrinsic ratio (that we set to be 17.56, as explained before). Following \citet{2000ApJ...533..682C} and subsequent work, the emission lines follow the standard MW curve \citep{1999PASP..111...63F}, so $k(\textrm{Pa}\beta)=0.76$ and $k(\textrm{H}\alpha)=2.36$. Equivalently, we can infer the most recent SFR with the Pa$\beta$ emission instead, which later on is analysed in depth and compared with other star-formation tracers in \S\ref{sec4}. The SFR inferred with the Pa$\beta$ luminosity is given by:
\begin{equation}
    \textrm{SFR[M}_{\odot}/\textrm{yr}] = 4.4\times10^{-42} \times\Big(\frac{H\alpha}{Pa\beta}\Big)_{int} \times L_{Pa\beta,corr} [\textrm{erg/s}], \label{eq_sfr_pab}
\end{equation}
where, equivalently to Equation \ref{eq_3}, the dust-corrected Pa$\beta$ luminosity can be obtained with $A_{Pa\beta} = k(Pa\beta)\times E(B-V)$, and we can use Equation \ref{eq_color_excess} to obtain the colour excess.

Finally, we can also infer the visual attenuation $A_V= R_V E(B-V)$, obtaining $E(B-V)$ from the line ratio as before. The Balmer-to-Paschen ratio only provides the actual value of $E(B-V)$ -- and therefore $A_V$ -- if the dust is distributed homogeneously and as a foreground screen, and the attenuation curve employed is appropriate for the case. Following \cite{1994ApJ...429..582C,2000ApJ...533..682C}, we set $R_V$=4.05 for the stellar continuum and $R_V$=3.1 for the nebular lines.

Figure \ref{fig_test_corner_sed} shows an example of the results our spatially-resolved SED-fitting code produces for each individual bin in a galaxy. The code is run on NGC1614, and it produces a best fit SED (black line), shown in the upper right plot for one example bin towards the bulk of the galaxy. We show the original HST available photometry (red points and errorbars), as well as the best fit synthetic photometry (blue crosses). The choice of continuum and line templates that linearly combine to produce the best fit SED are also displayed in different colour curves. The bottom panel shows the residual $\chi$ of the fit (data$-$model). On the left of Figure \ref{fig_test_corner_sed}, we display the corner plot\footnote{We use the visualisation Python package \texttt{corner.py} \citep{corner}} with the parameter space for the selected example bin. We can explore the different correlations between the various physical parameters, as well as seeing the distribution of best fit solutions for each of them. This constitutes an upgrade in the treatment of correlations between parameters and resulting uncertainties in the physical properties that we infer, when compared to previous fast-running codes, and with the advantage of computational speed when compared to other MCMC (Markov Chain Monte Carlo) routines in other codes.

\subsection{Robustness of our Inferred Parameters}

We can conduct some tests and diagnostics to analyse the robustness of the inferred physical properties, estimated with our spatially-resolved SED-fitting scheme. For this, we can use archival observations obtained with different instruments, as well as other SED-fitting codes.

\subsubsection{H$\alpha$ emission line strength} \label{section_ha_nii}

Firstly, we test how reliable our line fluxes are, which is vital if we want robust and trustworthy $A_V($H$\alpha/$Pa$\beta$), SFR$_{H\alpha}$ and SFR$_{Pa\beta}$ estimates. The narrow-band filter that targets H$\alpha$ ($\lambda6564.61$\AA) is wide enough so that we get contamination from the [\ion{N}{2}]$\lambda\lambda6549.86,6585.27$\AA\ doublet, so we need to correct for this. A common approach is to consider a fixed ratio throughout the galaxy; however, the [\ion{N}{2}]/H$\alpha$ ratio can vary significantly not only between galaxies, but also spatially within a galaxy \citep[e.g.,][]{2007ApJ...671..333K, 2013ApJ...779..135W, 2016MNRAS.461.3111B, 2017MNRAS.469..151B}. Multiple processes can be responsible for these variations, e.g. ionisation, shocks, outflows, differences in the electron density and metallicity, amongst others \citep[see e.g.,][for a review]{2019ARA&A..57..511K}. It is common to use a correction of [\ion{N}{2}]/\halpha=0.55 \citep[e.g.,][]{2010ApJS..190..233M,2019ApJS..244...33J}, which we also apply in this work, although the user can choose which correction to implement. We also calculate and account for the corresponding contaminant factor of [\ion{N}{2}], computed as the throughput of the narrow-band filter of [\ion{N}{2}] with respect to H$\alpha$. 

Some of the galaxies in our sample have ground-based archival integral field spectroscopy available (e.g., VLT-MUSE for 6 objects from the GOALS sample), which can be used to measure the spatial variation of the [\ion{N}{0ii}]/H$\alpha$ line ratio, though at somewhat lower spatial resolution than the HST maps. Here we use MUSE integral field cubes of the target MCG-02-01-051, to test the robustness of our method for inferring H$\alpha$ emission line fluxes from the multiband HST images.  

We convolve the HST images with a Moffat kernel \citep[][]{1969A&A.....3..455M}, to match the larger ground-based MUSE PSF (FWHM $\sim 0\farcs6$) and resample them to the MUSE spatial pixel grid. We then recompute the Voronoi bins on the resampled HST images as described above in \S\ref{sec:voronoi} and Appendix \ref{sec:app_voronoi}, and apply the bins to the calibrated MUSE spectral cubes downloaded from the ESO archive. We fit for the fluxes of the H$\alpha$+[\ion{N}{2}] emission lines in the resulting binned spectra with a triple Gaussian model, finding excellent agreement with our MUSE fits and previously published H$\alpha$ maps on our targets (e.g. IRAS13120-5453 from the PUMA Project, \citealt{2021A&A...646A.101P}).

\begin{figure}[!t]
\includegraphics[width=\columnwidth]{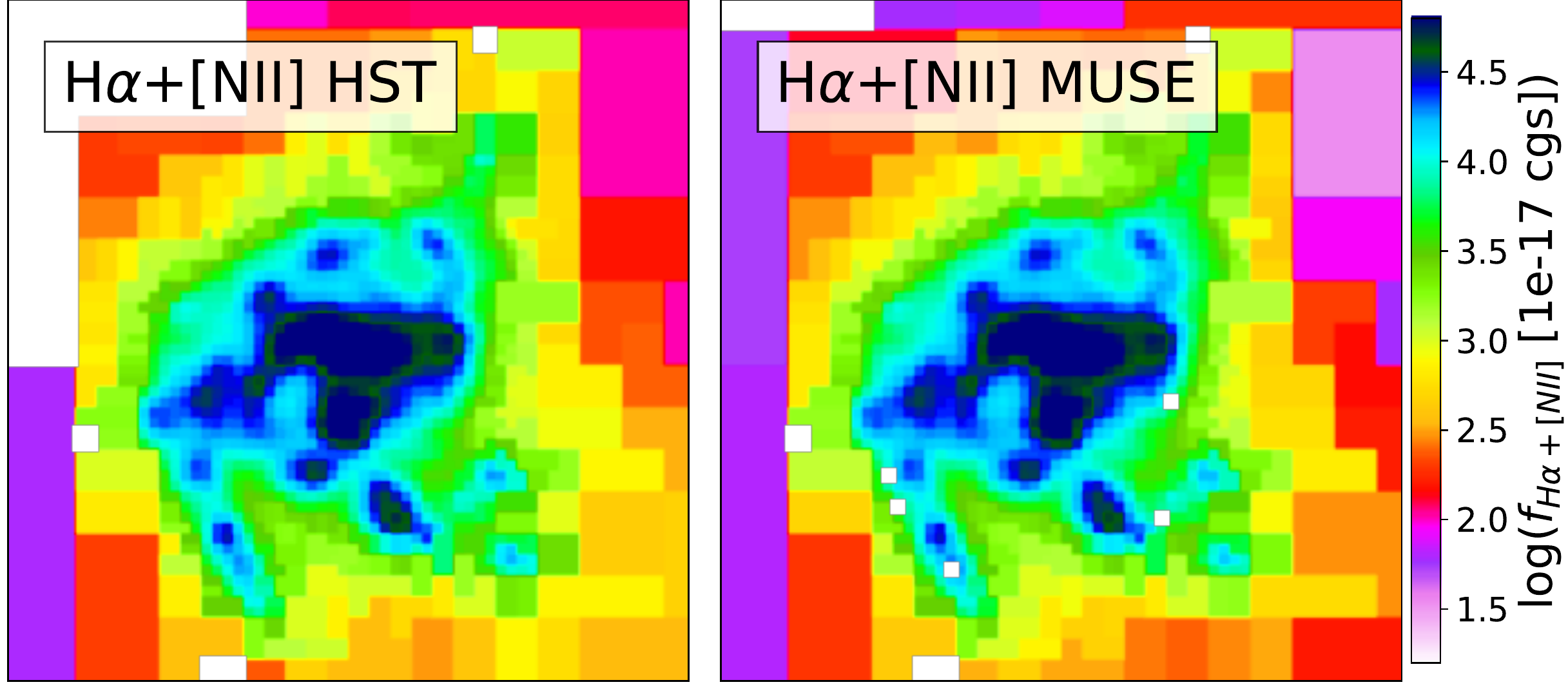}
\caption{Maps of the H$\alpha$+[\ion{N}{2}] line flux for the target MCG-02-01-051 inferred from the HST images with our SED-fitting code (left), and the H$\alpha$+[\ion{N}{2}] emission measured from the MUSE IFU cube (right) of the same source. \label{fig_muse_maps}}
\end{figure}

Figure \ref{fig_muse_maps} shows an example on the target MCG-02-01-051, on the \halpha+[\ion{N}{2}] fit from MUSE compared to our HST SED-fit H$\alpha$+[\ion{N}{2}] estimate. The \halpha\ fluxes derived with our fits to the HST broad- and narrow-band photometry (left) agree remarkably well with the independent measurement from the MUSE spectral cube (right). 

\begin{figure}[!t]
\includegraphics[width=\columnwidth]{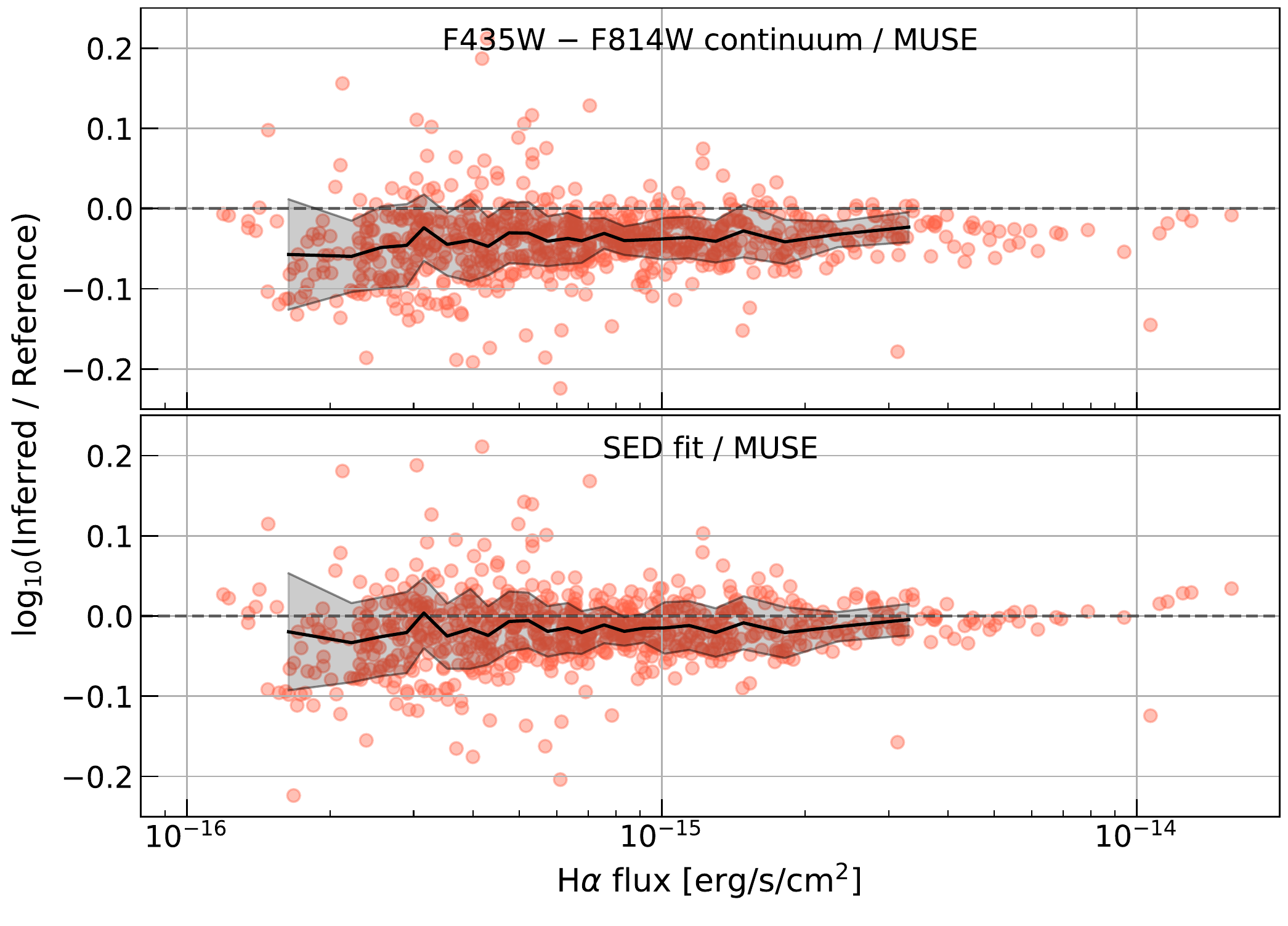}
\caption{Comparison of the resulting H$\alpha$ flux with a simple background subtraction method (top panel, interpolating two broad-band filters) and our full SED machinery (bottom panel), on the galaxy MCG-02-01-051. The vertical axis shows the logarithm of the ratio between each HST subtraction method inferred line estimates, and the reference MUSE H$\alpha$ of this source. The SED-fit H$\alpha$ matches the MUSE one better than the commonly used simple subtraction method. The running median and uncertainty are shown as a solid black line and surrounding shaded region, respectively. \label{fig_subtraction_muse_ha}}
\end{figure}

In Figure \ref{fig_subtraction_muse_ha}, we show a comparison of the resulting H$\alpha$ emission inferred with the simple subtraction method, interpolating two continuum templates and subtracting from the narrow-band that targets \halpha\ (top panel), and the full SED-fitting scheme (bottom panel) for the galaxy MCG-02-01-051. We compare each of these with the direct H$\alpha$ fit from the MUSE cube available for this target. The result clearly shows greater discrepancy between the inferred H$\alpha$ from the simple subtraction method and the MUSE fit (with a mean difference of $-0.04 \pm 0.04$ dex), than the H$\alpha$ from our SED-fit method, which agrees considerably better with the MUSE measurement (with an improved mean discrepancy of only $-0.02 \pm 0.04$ dex). Focusing on the bottom panel, we see that the large majority of the SED-fit \halpha\ fluxes agree well with the MUSE ``ground truth'', with larger scatter at low HST flux that can be explained by the uncertainties. The slight tilt in the comparison likely arises from imperfections in the PSF and alignment matching between the MUSE and HST frames; if we perform the same analysis with the HST F673N (the narrow-band that targets H$\alpha$) versus a synthetic F673N on MUSE (we integrate the MUSE spectra through the filter curve), we obtain the same trend, as well as if we perform the same test on the F814W filter, which does not contain the line. 

We run this study for six targets that have MUSE cubes available, and derive the same conclusions -- that our H$\alpha$+[\ion{N}{2}] fluxes agree considerably well with the H$\alpha$+[\ion{N}{2}] emission measured directly from MUSE spectra, and we even see that our composite SED-fit obtained with the different templates, fits the MUSE spectrum on a bin-to-bin basis, and not only the emission lines.

\begin{figure}[!t]
\centering
\includegraphics[width=0.95\columnwidth]{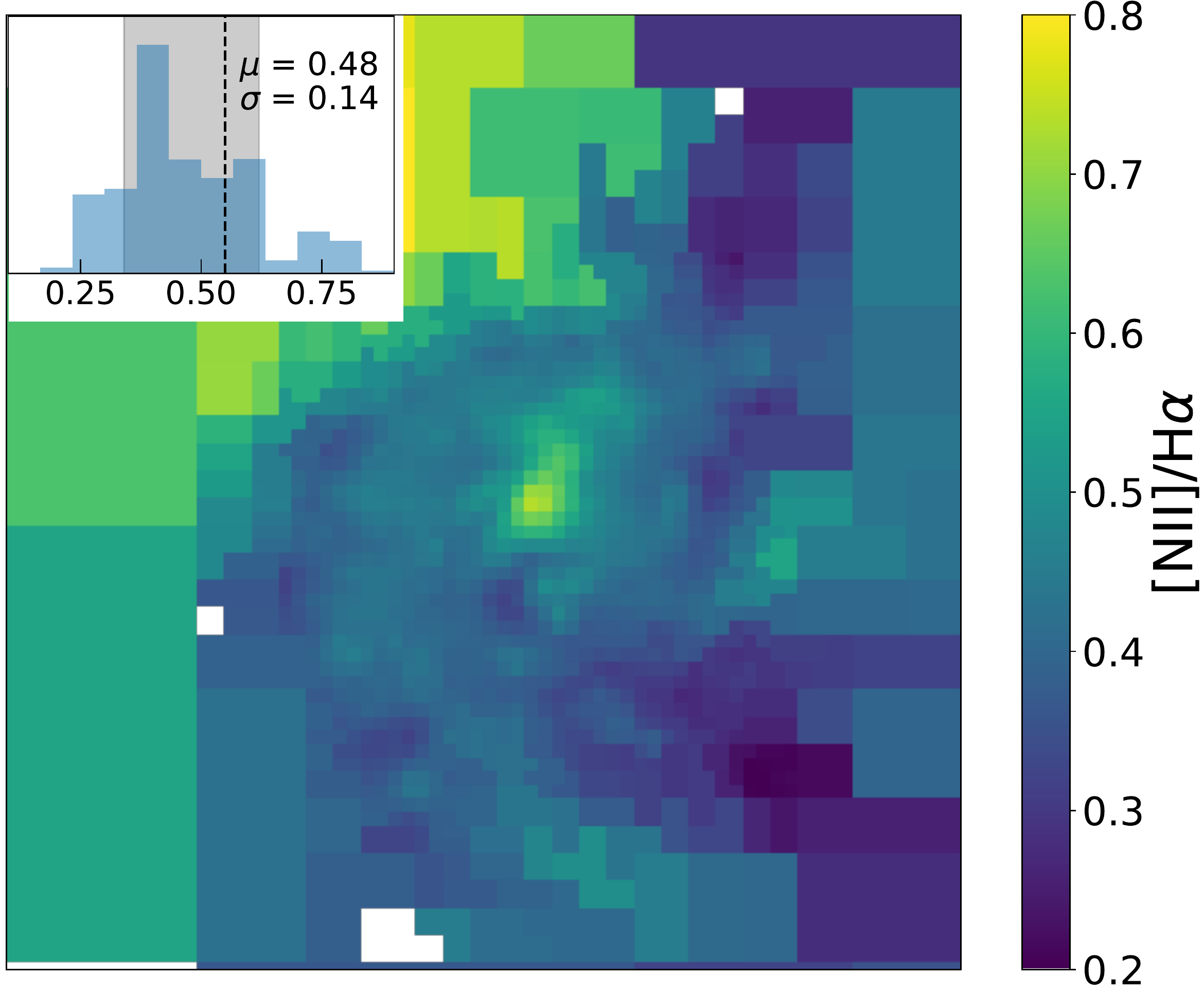}
\caption{Histogram (top left inset plot) and spatial map (main plot) of the [\ion{N}{2}]/\halpha\ ratio for MCG-02-01-051 as measured in the MUSE IFU cube. The value we adopt for the ratio (black dashed line on the histogram) falls within 1$\sigma$ of the mean (grey shaded region). \label{fig_nii_ha_corr}}
\end{figure}

Finally, we can check the [\ion{N}{2}]/\halpha\ spatially resolved correction, inferred from the MUSE IFU cube. Figure \ref{fig_nii_ha_corr} shows the spatial map and histogram of the derived ratio for MCG-02-01-051. We see that the ratio is not uniform across the face of the galaxy, but the constant value that we assume (0.55) --following the assumption that most studies make-- falls within 1$\sigma$ of the mean value for this galaxy. This value is reasonable for most HII-like galaxies and isolated LIRGs that comprise our sample, but we might encounter some targets that can have [\ion{N}{2}]$>$\halpha\ across much of the extent of the galaxy, such as the extreme starburst Arp~220 (see Figure 14 in \citealt{2020A&A...643A.139P}) or galaxies with extended LINER emission \citep[Low-Ionization Nuclear Emission-line Region, e.g.][]{2015MNRAS.449..867B}. Section \ref{section_app_niiha} in the Appendix shows the resulting [\ion{N}{2}]/\halpha\ histograms for the 5 additional targets from the GOALS sample that have MUSE observations available. Naturally, where our assumed constant [\ion{N}{2}]/\halpha\ value differs from the actual ratio, the inferred $A_V$ will be affected directly. For the extreme galaxy in our sample for which we have a MUSE cube, Arp~220, the mean [\ion{N}{2}]/\halpha$=1.92$ ratio observed in the cube corresponds in a systematic shift of 1.3 mag for $A_V$ relative to our assumed value of 0.55 (see Eq. \ref{eq_delta_av}).

\subsubsection{Pa$\beta$ emission line strength} \label{section_pab}

While we do not have an IFU cube to test the \pabeta\ line flux that we infer with our SED-fitting code, we consider a subset of galaxies in our sample where a robust empirical subtraction can be performed with the paired narrow-band images on and off of the emission line.

\begin{figure}[!t]
\includegraphics[width=\columnwidth]{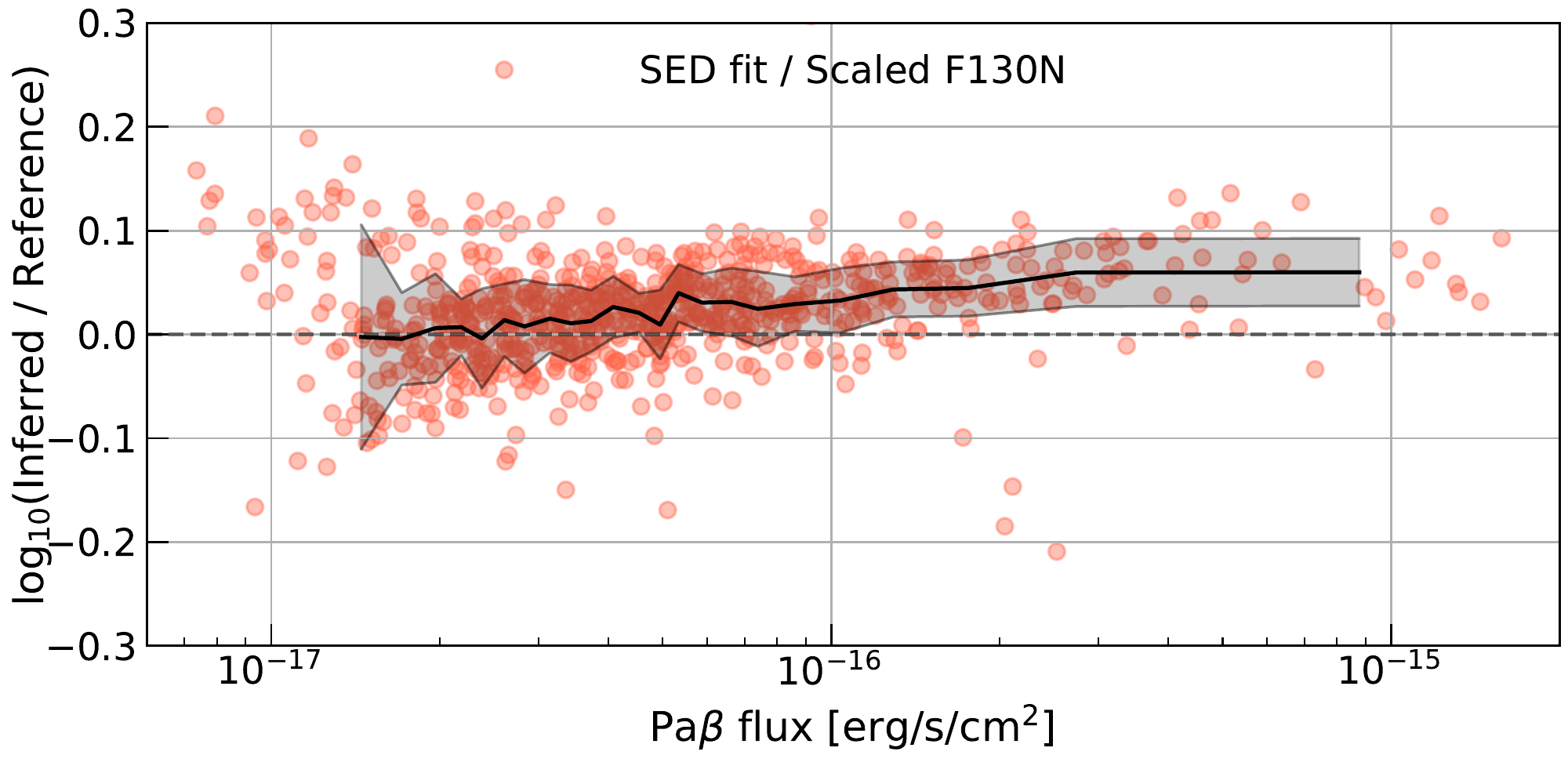}
\caption{Comparison between the \pabeta\ inferred from our SED-fitting scheme and the scaled F130N off-band subtraction for each spatial bin. The running median and scatter are shown as a solid black line and surrounding shaded region, respectively.} \label{fig_subtraction_muse_pab}
\end{figure}

Figure \ref{fig_subtraction_muse_pab} shows the comparison between the \pabeta\ line flux that we infer from our SED fit and the one obtained using the offset narrow-band filter F130N. By scaling the F130N we can perform a simple empirical subtraction from the F132N measurement, and remove the stellar continuum from the narrow-band that targets the recombination line. We see that our \pabeta\ estimate for this galaxy agrees with the empirical simple method (with a mean difference of $0.03 \pm 0.05$ dex). There is a slight flux dependent residual towards the bright end of the \pabeta\ flux, where the discrepancy reaches $\sim0.05$ dex, which is still a good agreement overall.

This and the previous study allow us to show that our line fluxes are within 0.05 dex of their respective reference values, making them reliable and robust on a bin-to-bin basis. Furthermore, the resulting $A_{V,gas}$ inferred from the decrement and the SFRs inferred from both lines are also trustworthy.

\subsubsection{Physical Properties from \texttt{Prospector}}

To prove the robustness of firstly our $A_{V,stars}$ estimates, we use a different SED-fitting code to fit our photometric observations, and compare the inferred $A_V$. For this, we use the SED-fitting code \texttt{Prospector}\footnote{\url{https://github.com/bd-j/prospector}} \citep{2017zndo...1116491J,2017ApJ...837..170L}. Both EAZY (and therefore our code) and \texttt{Prospector} incorporate FSPS \citep[Flexible Stellar Population Synthesis;][]{2009ApJ...699..486C, conroy2010} models, which are composite stellar population models that allow user input.

Until now, SED-fitting techniques of galaxies have broadly adopted basic models to obtain stellar masses, with fixed stellar metallicities, simple parametric star formation histories (SFHs), rigid and simplistic dust attenuation curves, and minimisation of chi-squared. On the other hand, \texttt{Prospector} includes a flexible attenuation curve and metallicity. \texttt{Prospector} uses FSPS, which contains nebular emission and considers both attenuation and re-radiation from dust. It also provides some innovation on flexible SFHs, as well as techniques for sampling parameter distributions. It comprises a 6-component non-parametric star formation history. With such a wide and adjustable range of parameters, \texttt{Prospector} is able to supply realistic uncertainties and unbiased parameters.

\begin{figure*}[t]
\includegraphics[width=\textwidth]{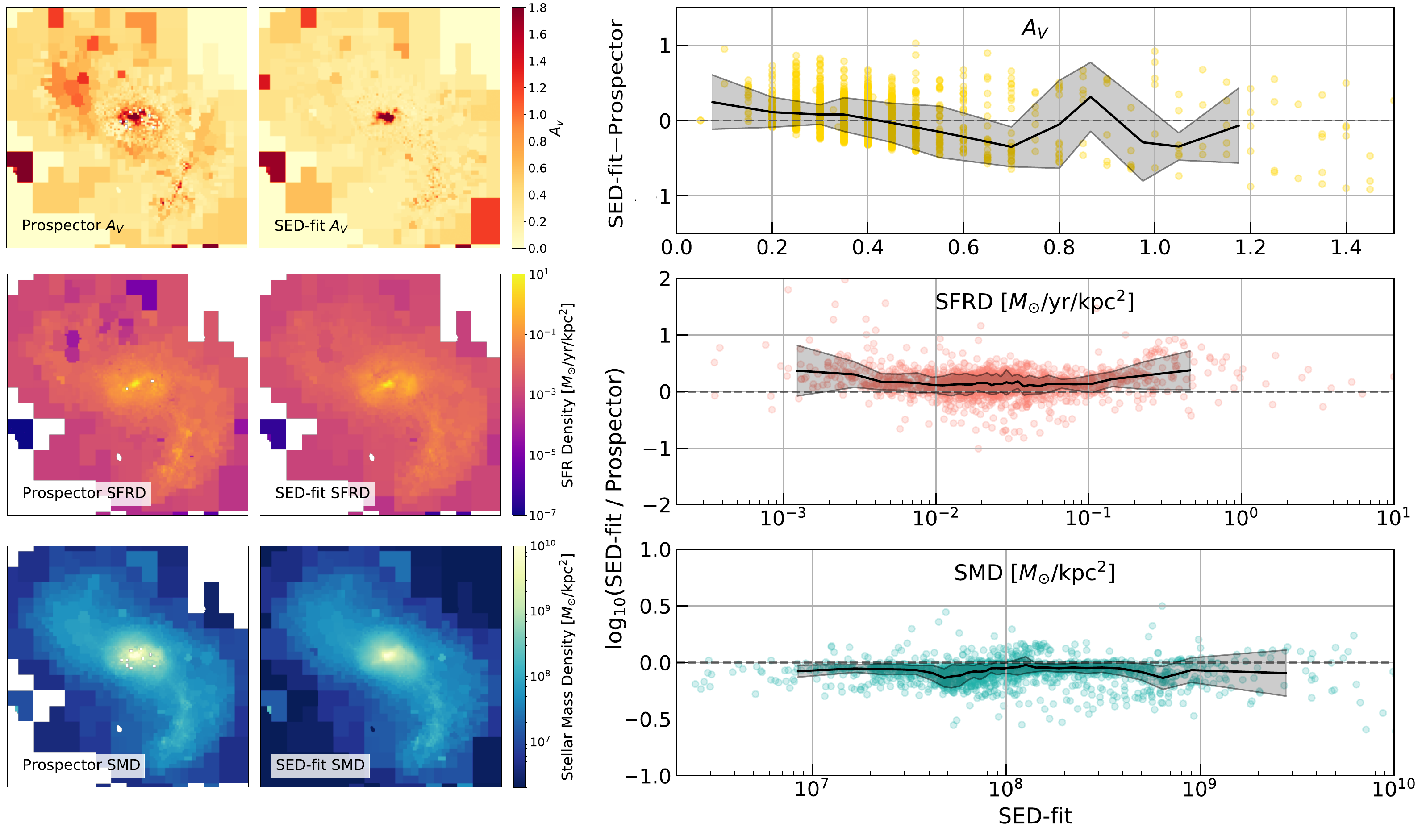} 
\caption{Comparison on IRASF10038-3338 between the physical properties inferred by \texttt{Prospector} and our SED-fitting code. \textbf{Left:} Resulting maps obtained with both codes, of the estimated $A_V$ (top), the star formation rate density (middle) and the stellar mass density (bottom). \textbf{Right:} Scatter plots of each physical property, with our SED-fit inferred parameters on the x axis, and the comparison (difference for $A_V$ and ratio for SMD and SFRD) of the Prospector and SED-fit properties on the vertical axis. Each point corresponds to an individual bin of IRASF10038-3338. The black line and shaded region indicate the running median and standard deviation, respectively. Both qualitatively and quantitatively, the SMD seems the most robust parameter. \label{fig_full_prospector}}
\end{figure*}

We run \texttt{Prospector} on the target IRASF10038-3338 with both a parametric star-formation history and a non-parametric star-formation history. Both yield very similar results. Figure \ref{fig_full_prospector} (top left panel) shows the resulting $A_{V}$ maps inferred from \texttt{Prospector} with a non-parametric SFH (left) and with our SED-fitting code (right). A priori, the maps seem to agree considerably well. Quantitatively (see the top right scatter plot in Figure \ref{fig_full_prospector}), the mean difference between both $A_V$ estimates across the galaxy is 0.06 ($\sim6\%$), with a standard deviation of 0.25, reflecting on the good agreement between the two.

Furthermore, to do a sanity check on our SFR and stellar mass estimates (both inferred from the SED fitting and not the lines alone), we can compare them with the \texttt{Prospector} estimates for the same galaxy, IRASF10038-3338. With \texttt{Prospector}, we obtain the star-formation rate density (SFRD) and stellar mass density (SMD), so to compare one-to-one, we simply divide our estimates by the corresponding physical bin size. At our average redshift of 0.02, for a minimum bin size of 4$\times$4 pixels (0$\farcs$4 per side), this corresponds to 160 pc across, which results in a minimum bin area of 0.0256 kpc$^2$.

Figure \ref{fig_full_prospector} shows the resulting qualitative and quantitative comparison between the two codes, using the non-parametric SFH for \texttt{Prospector}, although the parametric SFH run gives equivalent results. We see that both the SMD and SFRD maps (left middle and bottom panels of the figure, respectively) seem to agree considerably well on a bin-to-bin basis for this galaxy, especially the stellar mass density, which our code samples particularly smooth, even in central bins where the \texttt{Prospector} fit seems to fail. Our code seems to infer a higher star forming bulge than \texttt{Prospector}, most likely driven by the differences in SFH sampling from both codes and the SFH priors used. On a quantitative point of view (middle and bottom scatter plots of Figure \ref{fig_full_prospector}), for the SMD, the mean discrepancy between the two estimates is $-0.07 \pm 0.08$ dex, which is a remarkable agreement. On the other hand, for the SFRD, we obtain a median difference between \texttt{Prospector} and our code of $0.15 \pm 0.19$ dex. We see that the largest difference is obtained at the bright end of the SFRD. Although this may seem like a large discrepancy, the SFR from SED modelling is one of the more uncertain stellar population properties to derive \citep[see e.g.,][]{2009ApJ...701.1839M}, and highly influenced in this comparison by the differences in SFH sampling. Broadly used SED-fitting codes such as EAZY would yield equivalent results when compared to \texttt{Prospector} in terms of the SFRD.

To summarise, our inferred physical properties agree well both qualitatively and quantitatively (see Figure \ref{fig_full_prospector}), which gives us confidence that our estimates are robust and trustworthy.

\section{Results and Discussion} \label{sec4}

For all galaxies in our sample, we have produced H$\alpha$ and Pa$\beta$ maps, as well as extinction maps from the decrement and the stellar continuum, that allow us to measure the typical size scales and extinction levels of the dust obscuration. In this section we present the results. Maps of the various physical properties inferred with our SED-fitting code can be found in the Appendix \ref{app_results} for each individual galaxy in our sample.

\subsection{Case Example}

Before conducting a quantitative analysis on the various inferred line fluxes and physical properties, we present here a case example on our whole methodology applied to one target, NGC1614. As has been discussed in \S\ref{sec:voronoi} (and Appendix \ref{sec:app_voronoi}), the first step with our sample has been to employ a Voronoi tesselation binning scheme. Figure \ref{fig_binning} showed the binning applied on NGC1614 on the bottom panels. We apply the binning generated on F130N to the rest of available filters for this target, which are 6 images from the UV to the NIR.

\begin{figure*}[!t]
\includegraphics[width=\textwidth]{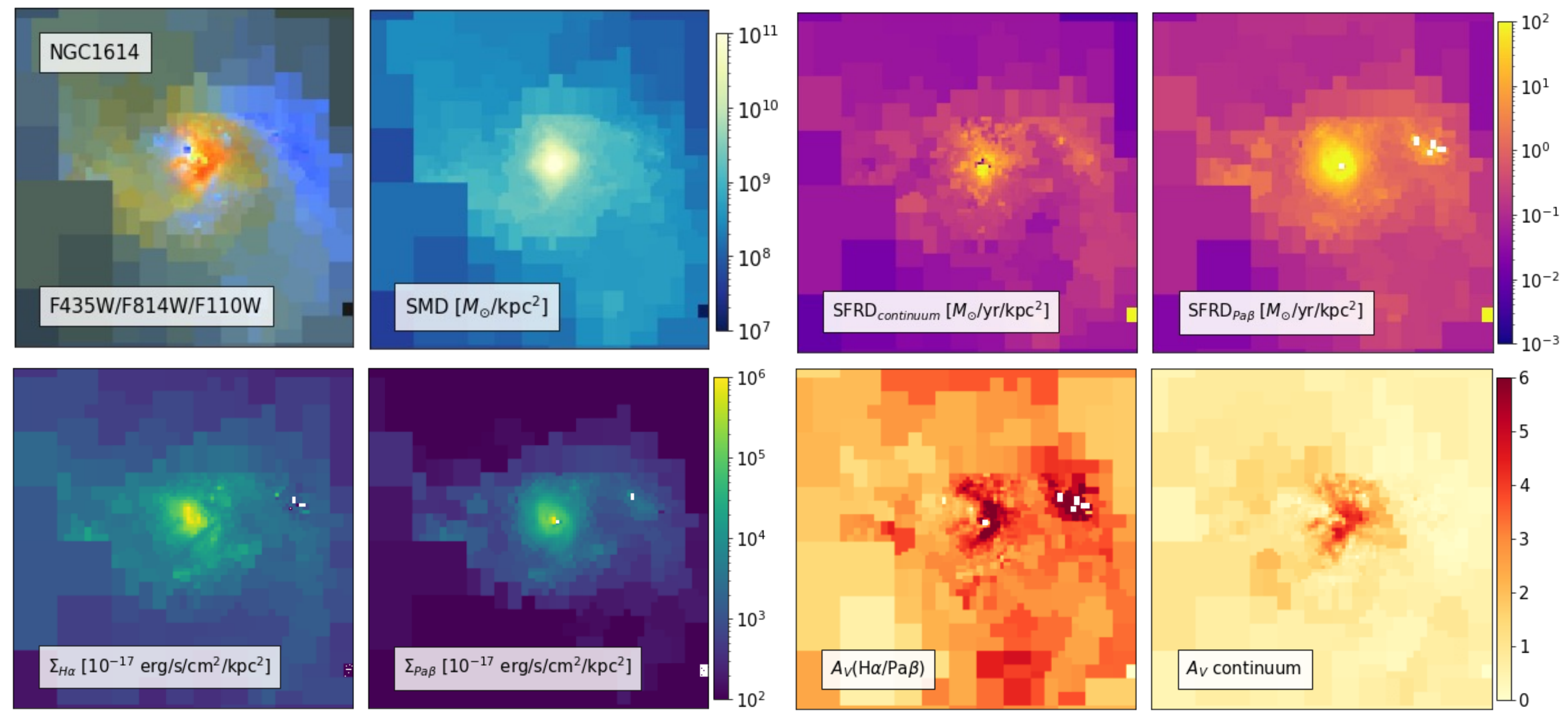}
\caption{Output of the spatially-resolved SED-fit on NGC1614. The top row, from left to right, shows the RGB image (built combining the F435W, F814W and F110W broadband filters), the stellar mass density map, and the SFRD maps inferred with the stellar continuum and with the Pa$\beta$ emission line flux. The bottom row shows the resulting H$\alpha$ and Pa$\beta$ surface density flux maps, as well as the $A_V$ inferred from the empirical Balmer-to-Paschen decrement, and the stellar continuum. \label{fig_example_results2}}
\end{figure*}

Once we have the binned images, we apply our spatially-resolved SED-fitting code (see \S\ref{sec:sedfit} and Appendix \ref{sec:app_sedfit}), and obtain the resulting physical estimates that are shown in Figure \ref{fig_example_results2}. The top row, from left to right, shows the RGB image (built combining the F435W, F814W and F110W broadband filters), the stellar mass density map, and the SFRD maps inferred with the stellar continuum and with the Pa$\beta$ emission line flux. The bottom row shows the resulting H$\alpha$ and Pa$\beta$ surface density flux maps, as well as the $A_V$ inferred from the empirical Balmer-to-Paschen decrement, and from the stellar continuum. We see that most line emission, star-formation, dust content and stellar mass reside in the dusty star-bursty core of NGC1614, which is debated to contain an AGN \citep{2015MNRAS.454.3679P}. Furthermore, we see noticeable differences in the dust maps inferred from the gas (H$\alpha$/Pa$\beta$) and the stellar population ($A_V$ inferred from the continuum). The differential attenuation inferred from each component will be addressed in detail in a forthcoming publication.

\subsection{SFR Indicators}

Accurately determining the star-formation rate is a vital step in understanding galaxy evolution. There are numerous tracers of recent and past star forming activity, and one can infer a comprehensive picture of the star formation history and burstiness in galaxies by comparing different star formation rate indicators \citep{2014ARA&A..52..415M}. 

In this section, we focus on exploiting our Paschen-line observations to trace the obscured star-formation in our sample of nearby star-forming galaxies. First, we compare the optical and NIR SFRs inferred with H$\alpha$, Pa$\beta$ and the stellar continuum emission. Later on, we investigate whether the dust-corrected SFR inferred with Pa$\beta$ can recover the SFR derived from the infrared luminosity and the 24$\mu$m emission observed with MIPS/\textit{Spitzer}. We conduct both spatially-resolved and integrated comparisons for the various tracers. As mentioned before, we conduct this study only on the galaxies of our sample that are also part of GOALS, due to their $L_{IR}$ and 24$\mu$m flux measurements availability.

\subsubsection{Spatially-Resolved Optical and NIR SFRs}

As introduced before, we can infer the SFR over the last $\sim$100 Myr from the stellar continuum emission. On the other hand, hydrogen recombination lines trace the most recent star formation ($\sim$10 Myr), occurring in HII regions throughout the galaxy \citep{2012ARA&A..50..531K}. 

Using Equations \ref{eq_sfr_ha} and \ref{eq_sfr_pab}, we can infer the SFR from the H$\alpha$ and Pa$\beta$ line fluxes, and divide by the physical size to obtain the star formation rate surface densities (in $M_{\odot}$/yr/kpc$^2$). Both SFRDs match once we apply the dust correction inferred from the H$\alpha$/Pa$\beta$ decrement. On the other hand, for the targets where we have MUSE cubes available, we can correct for dust obscuration using the classic Balmer decrement (H$\alpha$/H$\beta$). This allows us to explore whether Pa$\beta$ reveals obscured star formation that is invisible to optical lines, even after applying a dust correction.

\begin{figure}[!t]
\centering
\includegraphics[width=\columnwidth]{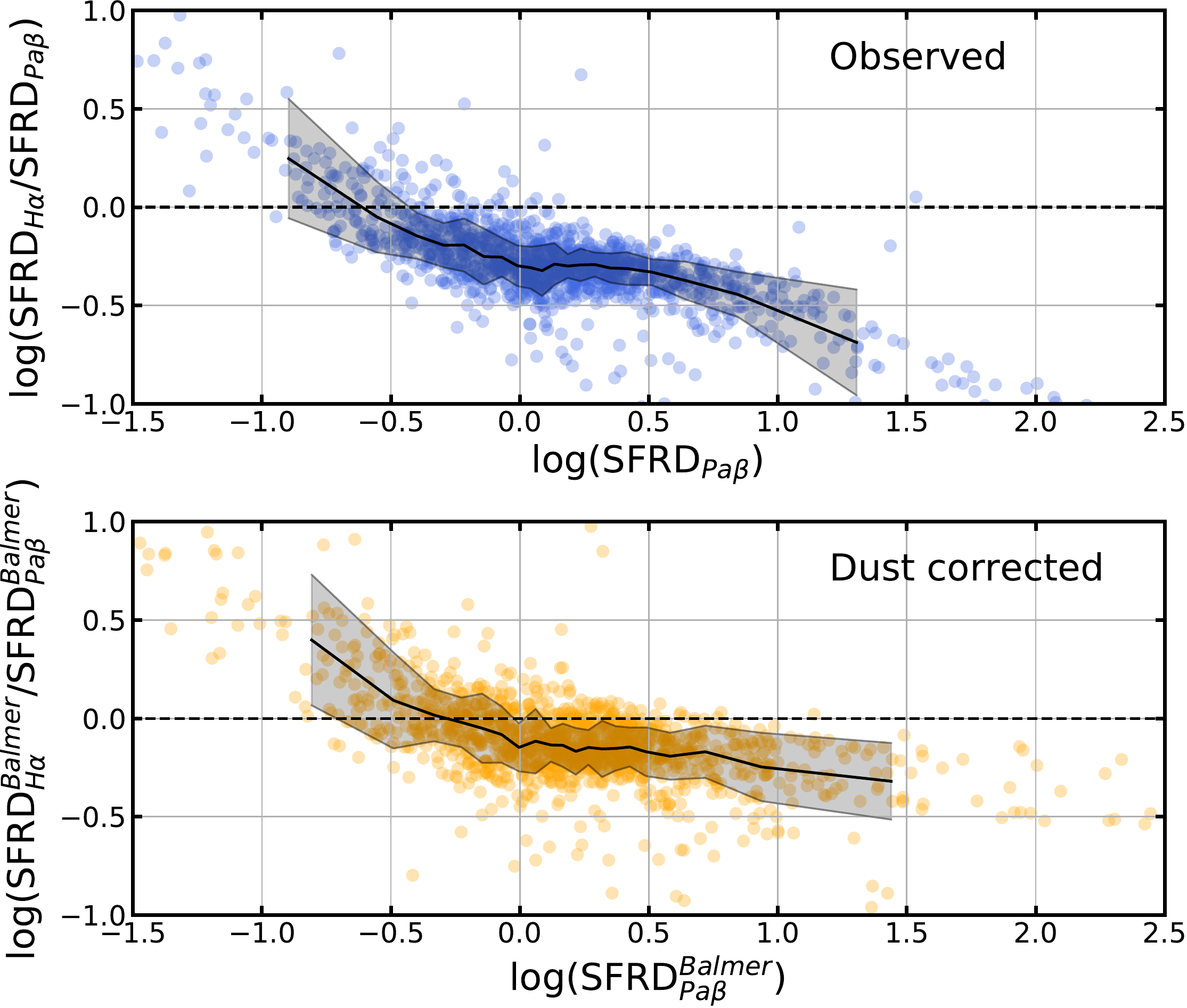}
\caption{Comparison of the spatially-resolved SFRD inferred from H$\alpha$ versus from Pa$\beta$ for the galaxy MCG-02-01-051. The dashed line indicates the one-to-one ratio. The solid black line and shaded region indicate the running median and standard deviation, respectively. \textbf{Top:} SFRDs estimated from the observed H$\alpha$ and Pa$\beta$, without correcting for dust extinction. \textbf{Bottom:} Dust corrected SFRDs, using the Balmer decrement (H$\alpha$/H$\beta$), obtained from the MUSE cube. \label{fig_sfr_indicators}}
\end{figure}

Figure \ref{fig_sfr_indicators} shows the spatially-resolved comparison between SFRD$_{H\alpha}$ and SFRD$_{Pa\beta}$ for the galaxy MCG-02-01-051, without (top panel, blue points) and with (bottom panel, orange points) dust correction prescription applied, using the Balmer decrement obtained from the MUSE cube (indicated by the superscript `Balmer`). In both cases, the SFRD inferred with \pabeta\ is systematically higher. There is a clear trend with the SFRD$_{Pa\beta}$, the disagreement between both estimates is much larger (up to 1 dex) for the bright end of SFRD$_{Pa\beta}$, and less than 0.5 dex at the lower SFRD$_{Pa\beta}$ end. This is what we could expect, since more star-forming regions are also more obscured by dust. The bottom panel illustrates that the Balmer decrement is not enough correction to see the most obscured star forming regions, with an average offset of $-0.1 \pm 0.2$ dex between SFRD$_{H\alpha}^{Balmer}$ and SFRD$_{Pa\beta}^{Balmer}$. Moreover, the discrepancy on the faint end on both panels may arise from differences in the surface brightness sensitivity of the \halpha\ observations relative to the Pa$\beta$, and we would require deeper \pabeta\ observations in the faint regions to study this regime better.

\begin{figure}[!t]
\centering
\includegraphics[width=\columnwidth]{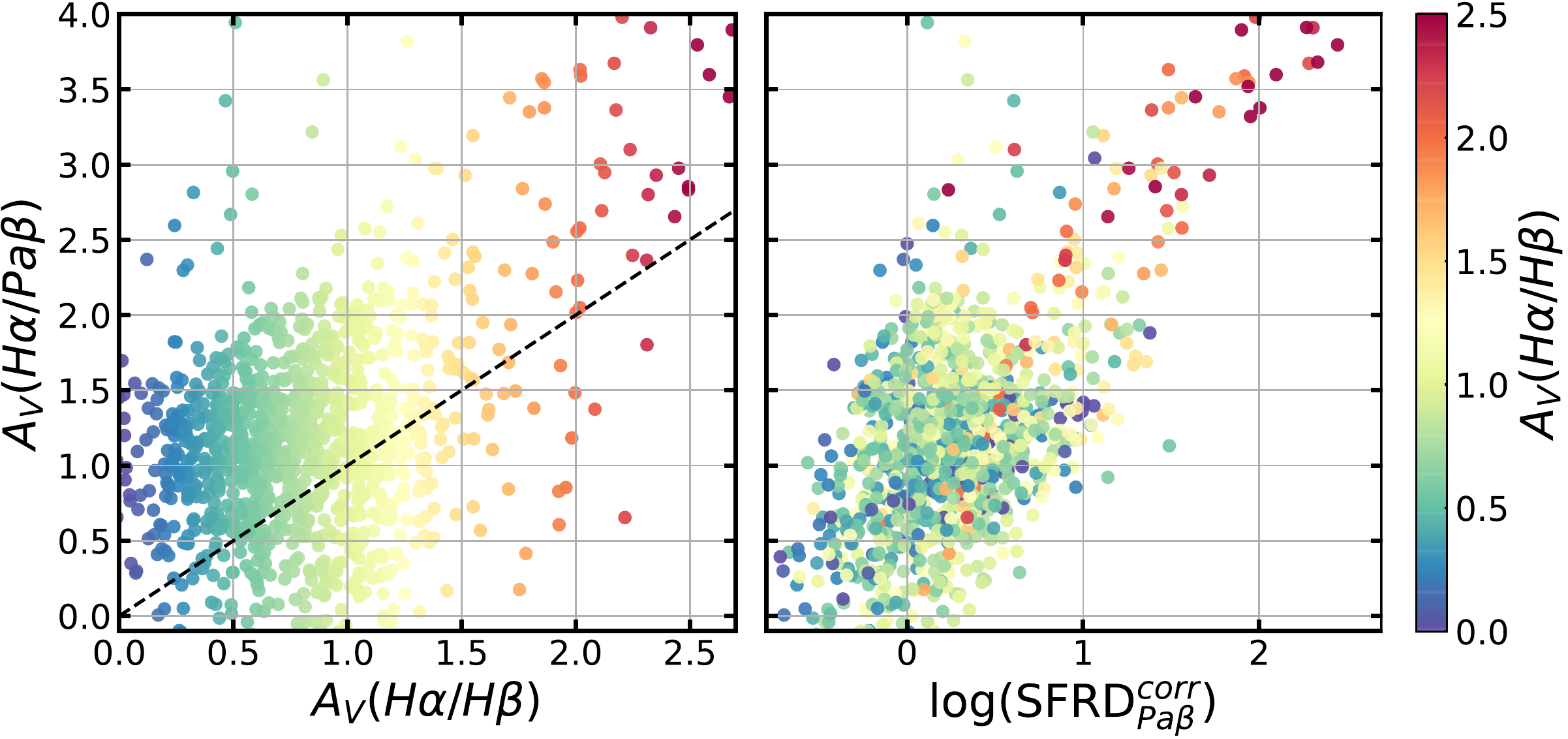}
\caption{Spatially-resolved visual extinction inferred from the Balmer decrement (left) and SFRD inferred from the dust-corrected Pa$\beta$ (right) versus the visual extinction from the Balmer-to-Paschen decrement. The colour-code indicates the $A_V$ inferred with the classic Balmer decrement. \label{fig_sfr_pab_av_lines}}
\end{figure}

Furthermore, we can directly see the relationship between the dust-corrected SFRD$_{Pa\beta}$ (using \halpha/\pabeta), and the visual extinction inferred with the Balmer-to-Paschen decrement. Figure \ref{fig_sfr_pab_av_lines} (right panel) shows the relation between the two for the spatially-resolved data of MCG-02-01-051, colour-coded according to the $A_V$ inferred with the classic Balmer decrement. We see that both quantities correlate, indicating that higher star-forming activity will also be more obscured. On top of this, on the left panel we compare the $A_V$ inferred with each decrement, and we clearly see that the $A_V$ inferred with Paschen lines probes larger optical depths than $A_V$(\halpha/H$\beta$) by $>1$ mag, agreeing with works such as L13. 

These results agree with previous comparisons with Paschen-line SFRs such as \cite{2016A&A...590A..67P}, where they find that the SFR$_{Pa\alpha}$ is on average a factor $\times3$ larger than SFR$_{H\alpha}$, even after applying extinction corrections. They conduct this study on a sample of local U/LIRGs, using Brackett and Paschen lines to correct SFR$_{Pa\alpha}$, and the classic Balmer decrement to dust-correct SFR$_{H\alpha}$. \cite{2015ApJS..217....1T} also report that SFR$_{Pa\alpha}^{corr}$ is systematically larger than SFR$_{H\alpha}^{corr}$ in a sample of nearby star-forming galaxies (33 of them from the IRAS Revised Bright Galaxy Sample catalog), which implies that Paschen-lines can reveal star-formation that is otherwise obscured for H$\alpha$, reaching dustier regions than the optical Balmer lines. Albeit with spectroscopic data and non spatially-resolved observations, \cite{2020arXiv200900617C} have Pa$\beta$ and H$\alpha$ measurements, and report the same conclusions found in this work.

\begin{figure}[!t]
\centering
\includegraphics[width=\columnwidth]{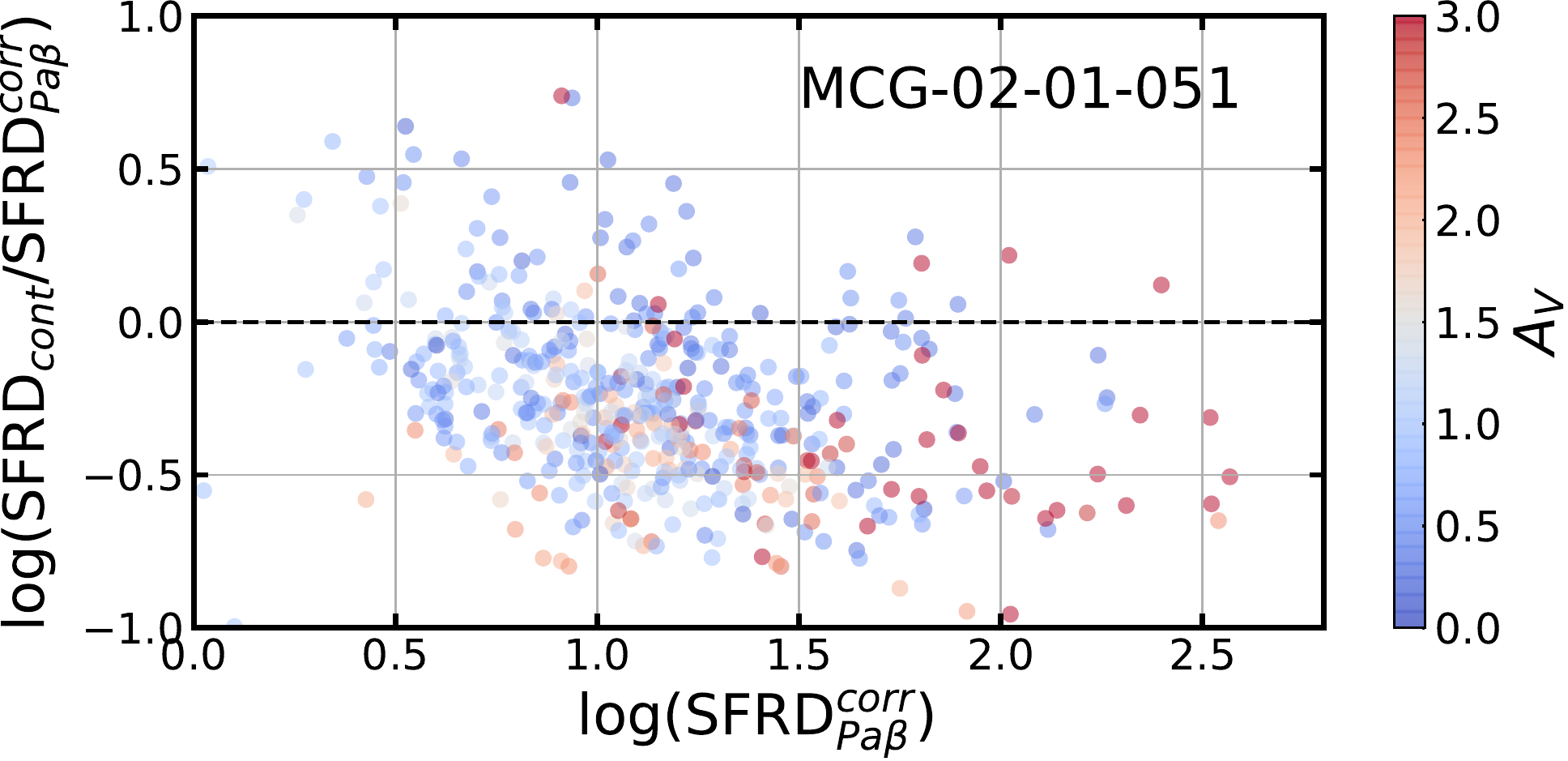}
\includegraphics[width=\columnwidth]{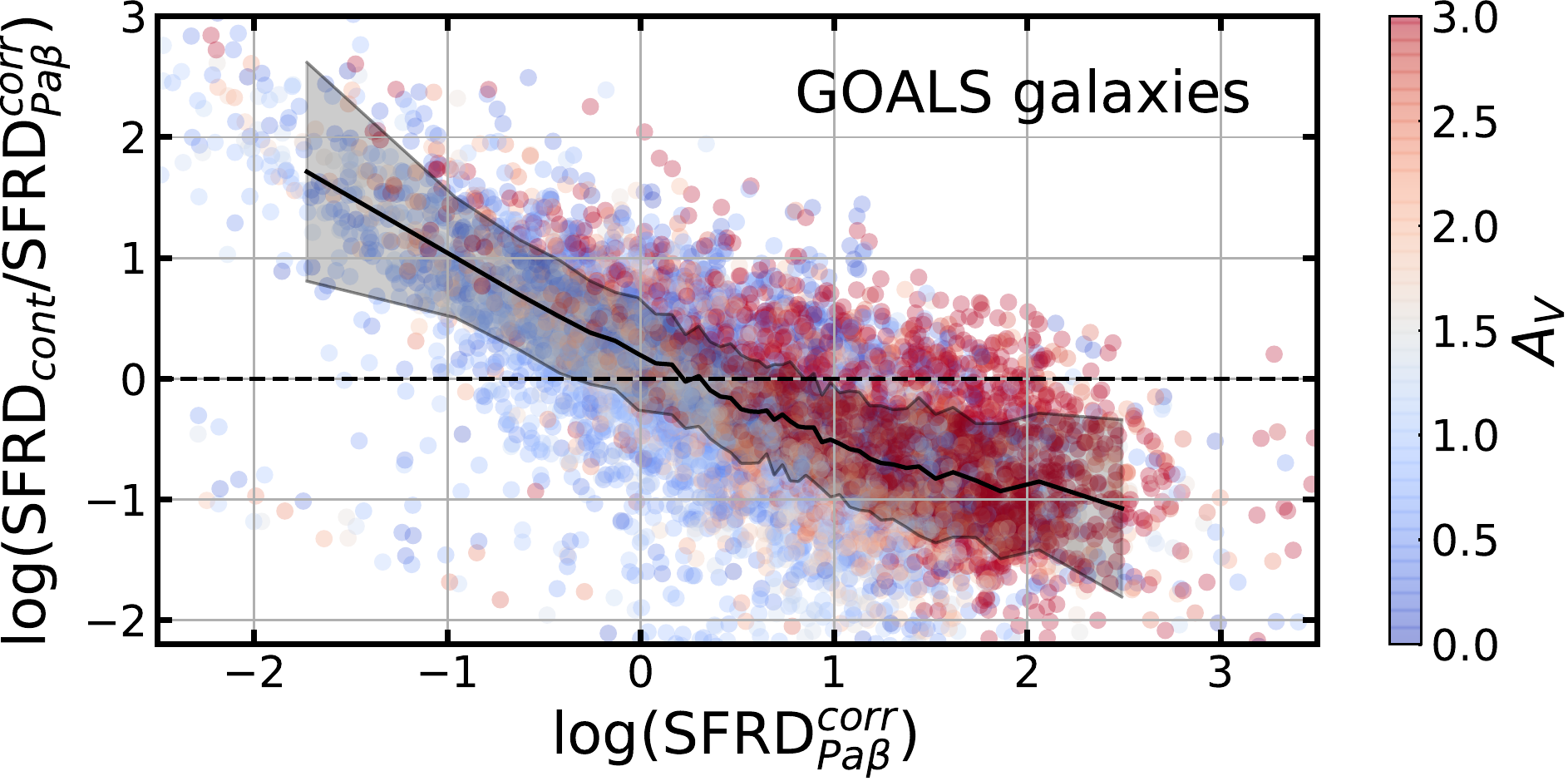}
\caption{Comparison of the spatially-resolved SFRDs inferred from stellar continuum versus from the dust-corrected Pa$\beta$. The dashed line indicates the one-to-one ratio. The colour-code indicates the stellar continuum $A_V$. \textbf{Top:} For the target MCG-02-01-051. \textbf{Bottom:} For the targets in our sample that are part of GOALS. The black line and shaded region indicate the running median and standard deviation, respectively. \label{fig_sfr_pab_cont}}
\end{figure}

Besides the nebular lines, we can also study whether Pa$\beta$ 'sees' more star formation than the SFR inferred from the stellar continuum (estimated with our spatially-resolved SED-fitting code, which is computed directly as the normalisation of the two youngest continuum templates, where the reported SFR is averaged over 100 Myr). Figure \ref{fig_sfr_pab_cont} shows in the vertical axis the difference between these two indicators, as a function of SFRD$_{Pa\beta}^{corr}$, dust-corrected with the Balmer-to-Paschen ratio, for the galaxy MCG-02-01-051 (top panel) and for the whole sample of our galaxies that belong to GOALS (bottom panel). The individual bins are colour-coded according to their $A_V$ inferred with the SED-fitting code, which applies to the stellar continuum templates. For MCG-02-01-051, we clearly see that the SFRD inferred with Pa$\beta$ is systematically higher than with the stellar continuum for the wide majority of bins, with a mean discrepancy of $-0.3\pm0.3$ dex. There is a slight decrease in the difference towards lower SFRD$_{Pa\beta}^{corr}$. On top of this, the bins with higher obscuration traced by the stellar continuum ($A_V$) seem to yield bigger difference between both SFR indicators. Besides the fact that the NIR line is less obscured than the optical continuum, the discrepancy between both indicators can also be explained by the different star formation timescales that they both trace. This could be the case particularly for the plot of the whole sample (bottom panel in Figure \ref{fig_sfr_pab_cont}), where the mean discrepancy between the SFR tracers is $-0.3\pm0.7$ dex. For small SFRD$_{Pa\beta}^{corr}$ values, the continuum yields higher star formation. This could be due to the recent star formation being lower than the star formation over the last 100 Myr, or it could be the same effect observed in Figure \ref{fig_sfr_indicators} (where both indicators trace identical timescales), resulting from the depth of the images. On the other end of high SF activity, where the mean discrepancy goes beyond 1 dex, dust might dominate the difference between the two indicators, as we can see by the enhanced $A_V$, since we see that the most obscured bins infer larger SFRD with the nebular line than with the stellar continuum. This leads us to conclude that near-IR lines are able to see star formation that is invisible to the stellar continuum, reaching further depths than any optical component.

Furthermore, the attenuation experienced by the stars is due to the diffuse ISM, whereas the nebular lines tracing the gas are obscured also by birth clouds, where dust is clumpier and more patchy \citep{2020MNRAS.495.2305G}. \cite{2018ApJ...862L..22C} illustrate obscured starburst as an optically thick core with an enclosing layer, emitting NIR and optical nebular lines, which explains both the larger SFR and extinction inferred by NIR lines when compared to optical tracers \citep{2017ApJ...838L..18P}. The combination of all of these factors results in having differential attenuation and SFRs inferred by nebular lines and the stellar continuum.

With upcoming observations of JWST of Paschen-line emission in galaxies, our results suggest that we will be able to reveal star-formation activity that is otherwise currently obscured for optical observations, and it will allow us to paint a more complete picture of the evolution of galaxies and their star formation activity across cosmic time.

\subsubsection{Integrated IR SFR Indicators} 
We have found that Paschen-lines can reveal obscured star formation invisible to optical indicators, but now it remains to be explored whether these lines can recover the star formation activity that we can infer with mid- and far-IR indicators, arising from re-emitted absorbed UV starlight. Infrared and far-IR luminosities have been shown to be good star-formation tracers in dusty starburst galaxies \citep[e.g.,][]{1998ARA&A..36..189K,2002AJ....124.3135K}. Past studies have found a tight agreement between the SFR inferred from dust corrected Paschen-lines and SFR$_{24\mu m}$, derived from the MIPS/\textit{Spitzer} 24 $\mu$m observations (e.g. \citealt{2016A&A...590A..67P}). On top of this, a reasonable linear correlation has also been found between Paschen star formation rates and the SFR inferred from the infrared luminosity (e.g. \citealt{2015ApJS..217....1T}).

The light that dust absorbs in the UV range of the spectrum, is later re-emitted at longer wavelengths, in the IR regime. Therefore, IR indicators of reprocessed stellar light have been broadly employed to infer SFRs. Here we focus on the $L_{IR}$ and 24$\mu$m emission. We can derive the SFR$_{IR}$ using the calibration by \cite{1998ARA&A..36..189K}, for a \cite{2003PASP..115..763C} IMF:
\begin{equation}
    SFR_{IR} [M_{\odot} \textrm{yr}^{-1}] = 2.5 \times 10^{44} \times L_{IR} [\textrm{erg s}^{-1}] \label{eq_ir}
\end{equation}
where $L_{IR}$ is defined as the luminosity in the spectral range $8\mathrm{-}1000\,\mu\mathrm{m}$. Since we do not have multiple available measurements along the IR part of the SED, we cannot infer a spatially-resolved analysis with our code. Therefore, we conduct these comparisons in an integrated manner.

Firstly, we compare SFR$_{Pa\beta}^{corr}$ and SFR$_{IR}$. The SFR$_{Pa\beta}^{corr}$ is inferred from the Paschen-beta line flux following Equation \ref{eq_sfr_pab}, and applying a dust correction using the Balmer-to-Paschen decrement (H$\alpha$/Pa$\beta$) (calculated on a bin-to-bin basis and then integrating over the whole galaxy). As introduced before, previous studies have found that redder hydrogen recombination lines than H$\alpha$ and H$\beta$, reach further depths than the common Balmer decrement (e.g. \citealt{1996ApJ...458..132C},  \citealt{2013ApJ...778L..41L}), and therefore infer higher obscuration. In agreement with this, in the previous section we have found that Pa$\beta$ systematically traces more star-formation than H$\alpha$, even when the Balmer decrement is used for applying a dust correction. For the next comparisons, we use the integrated SFR$_{Pa\beta}^{corr}$ and we obtain the infrared luminosity values from the GOALS survey measurements \citep{2009PASP..121..559A}, that can be found on Table \ref{tab_sample} in this work.

\begin{figure}[!t]
\centering
\includegraphics[width=\columnwidth]{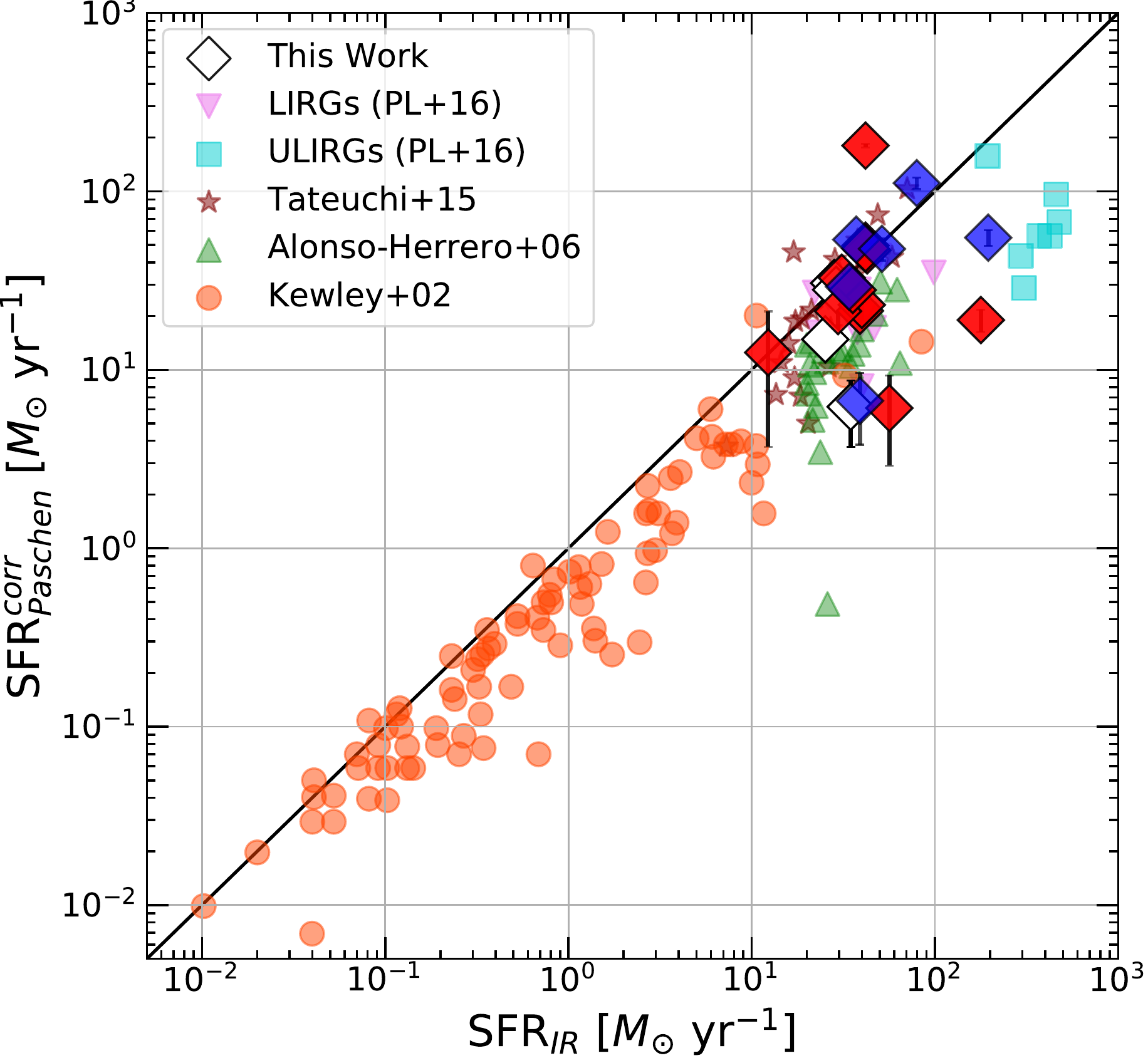}
\caption{Extinction corrected SFR$_{Paschen}^{corr}$ as a function of the SFR inferred from the infrared luminosity. The various datasets are detailed in the main text. Our sample is indicated by the diamond symbol, blue coloured if they host an AGN and red coloured if not. \label{fig_sfr_ir}}
\end{figure}

Figure \ref{fig_sfr_ir} shows the comparison between the extinction-corrected SFR inferred with Paschen lines and SFR$_{IR}$ for various studies. The targets from our sample are indicated with the diamond symbol, and following the colour-coding from Figure \ref{fig_sample}, where blue means that the galaxy hosts and AGN and red that it does not. Our data is plotted inferring the y-axis SFR with \pabeta . On the other hand, to put our results into the context of previous studies, we include other works that infer the SFR with Pa$\alpha$ instead. Some discrepancy between both Paschen lines SFRs might be encountered, but the effect of dust between the two tracers should not be as significant as for example the previous study between \pabeta\ and \halpha . Figure \ref{fig_sfr_ir} includes data from various samples. The turquoise squares (ULIRGs) and violet downward triangles (LIRGs) are data from the \cite{2016A&A...590A..67P} sample (PL+16), observed with VLT-SINFONI, using Brackett and Paschen lines to dust-correct SFR$_{Pa\alpha}$ and a \cite{2000ApJ...533..682C} attenuation curve with a 0.44 ratio between the ionized gas and stellar continuum applied. The green upward triangles correspond to local LIRGs from \cite{2006ApJ...650..835A}, obtained measuring the Pa$\alpha$ fluxes and comparing them with their \halpha\ and Br$\gamma$ measurements, respectively, and estimating the extinction to the gas using the \cite{1985ApJ...288..618R} extinction law and a foreground dust screen model. The orange circles are measurements of normal galaxies from the Nearby Field Galaxy Survey (NFGS, \citealt{2002AJ....124.3135K}). For these galaxies the SFR is inferred with \halpha\ and dust-corrected with the classic Balmer decrement. The maroon stars correspond to HII galaxies from the sample of local LIRGs presented in \cite{2015ApJS..217....1T}, where they use the H$\alpha$/Pa$\alpha$ ratio to correct for the extinction their Pa$\alpha$ measurements and a \cite{2000ApJ...533..682C} attenuation curve with $R_V=4.05$.

Overall, we find a considerable agreement between the SFR inferred from Paschen-lines and the IR luminosity, with some targets having higher SFR$_{IR}$, as also encountered and discussed in e.g. \cite{2016A&A...590A..67P}. \cite{2015ApJS..217....1T} also find a systematic offset of $-$0.07 dex of SFR$_{Pa\alpha}^{corr}$ with respect to SFR$_{IR}$, and a scatter of 0.27 dex. The targets presented in this work seem to agree well with all previous published results. The discrepancy between both indicators for our galaxies is $-0.04 \pm 0.23$ dex, with the majority of targets lying on the one-to-one relation. We also find some extreme cases where both indicators considerably disagree. This could be due to very high obscuration, or other extreme environments within the galaxy, such as is the case for Arp~220, a known extreme starbursty ULIRG, which is the biggest outlier in Figure \ref{fig_sfr_ir}. Previous studies such as \cite{2018ApJ...862L..22C}, report that SFRs inferred with tracers from the UV to the NIR are systematically underestimated when compared to SFR$_{IR}$ in ULIRGs, which we also find here.

The discrepancy that we find could also be due to the choice of radii within which we integrate the Pa$\beta$ flux, and its difference with respect to the GOALS choice to integrate $L_{IR}$ in these targets. Furthermore, ULIRGs infrared luminosity may originate from very compact regions \citep{2021A&A...651A..42P}. 

On the other hand, the SFR$_{IR}$ might trace star formation occurring in longer timescales than the most recent star formation \citep{2018ApJ...862L..22C}, which is traced by hydrogen recombination lines arising from HII regions, such as we explored before when comparing the SFR from the optical stellar continuum and the lines. This could be the case for example for the NFGS sample \citep{2002AJ....124.3135K}, that appears to have systematically lower SFR$_{Pa\alpha}$ than SFR$_{IR}$, while our results are closer to the 1:1 line, on average. This might be due to their SFRs being inferred with \halpha\ and dust corrected with the Balmer decrement, instead of redder Paschen lines. Or it could also be explained by an age effect, i.e. we see a recent star formation that was a systematic factor lower than the star formation over the last 100 Myrs. This is mainly due to the fact that old stellar populations can still heat the dust that emits in the far-IR, a known effect that has been observed at the level of a factor of $\sim2$ in the SFRs difference.

\cite{2015ApJS..217....1T} attribute the discrepancy between the inferred SFRs to high dust obscuration, IR cirrus component or the presence of AGNs in the galaxies, which could be possible sources of systematic differences. If we focus on our sample, the targets that are known to host an AGN do not exhibit a particular trend in Figure \ref{fig_sfr_ir}. The presence of an AGN could explain the discrepancy for 2/6 targets, but the other four are on top of the 1:1 agreement. Therefore, we cannot report a systematic trend on the SFR$_{Pa\beta}^{corr}$ and SFR$_{IR}$ relationship due to the presence of an AGN in our galaxies.

\begin{table}[t]
\centering
\begin{tabular}{l c c c c}
\hline
\hline
Target & SFR$_{Pa\beta}^{corr}$ & SFR$_{IR}$ & SFR$_{24\mu m}$ & log($\frac{M_*}{M_{\odot}}$)\\ 
\hline
Arp~220 & 19.0 & 178.2 & 75.8\textsuperscript{$\dagger$} & 10.8 \\
ESO550-IG025 & 32.8 & 31.0 & 23.4\textsuperscript{$\dagger$} & 11.0 \\
IRAS03582+6012 & 14.8 & 25.2 & 36.5\textsuperscript{*} & 10.3 \\
IRAS08355-4944 & 21.4 & 39.0 & 88.0\textsuperscript{*} & 10.0 \\
IRAS12116-5615 & 180.4 & 41.8 & 47.6\textsuperscript{*} & 10.9 \\
IRAS13120-5453 & 55.0 & 195.4 & 163.3\textsuperscript{*} & 11.3 \\
IRAS18090+0130 & - & 41.8 & 39.1\textsuperscript{*} & 11.1 \\
IRAS23436+5257 & 6.2 & 34.7 & 50.0\textsuperscript{*} & 10.8 \\
IRASF10038-3338 & 6.1 & 56.4 & - & 10.5 \\
IRASF16164-0746 & 6.7 & 39.0 & - & 10.5 \\
IRASF16399-0937 & 23.1 & 39.9 & - & 11.0 \\
MCG-02-01-051 & 30.6 & 28.2 & 48.0\textsuperscript{$\dagger$} & 10.6 \\
MCG+12-02-001 & 21.3 & 29.6 & 49.7\textsuperscript{*} & 10.5 \\
NGC1614 & 49.6 & 41.8 & 95.7\textsuperscript{$\dagger$} & 10.7 \\
NGC2146 & 12.5 & 12.3 & 9.5\textsuperscript{*} & 10.2 \\
NGC2623 & 53.6 & 37.2 & 27.6\textsuperscript{$\dagger$} & 10.4 \\
NGC5256 & 29.2 & 34.0 & 39.1\textsuperscript{$\dagger$} & 10.9 \\
NGC5331 & 46.7 & 42.7 & 37.4\textsuperscript{*} & 11.2 \\
NGC6090 & 28.0 & 35.6 & 49.2\textsuperscript{$\dagger$} & 10.6 \\
NGC6240 & 111.1 & 79.6 & 122.8\textsuperscript{*} & 11.3 \\
NGC6670 & 47.9 & 41.8 & 49.7\textsuperscript{*} & 11.0 \\
NGC6786 & 28.1 & 28.9 & 51.7\textsuperscript{*} & 10.7 \\
NGC7592 & - & 23.5 & 29.9\textsuperscript{$\dagger$} & - \\
VV340A & 47.6 & 51.4 & 27.9\textsuperscript{$\dagger$} & 11.2 \\
\hline
\multicolumn{4}{l}{\textsuperscript{*}\footnotesize{SFR from the 25$\mu$m flux from \cite{2003AJ....126.1607S}.}}\\
\multicolumn{4}{l}{\textsuperscript{$\dagger$}\footnotesize{SFR from the 24$\mu$m flux from \cite{2012ApJS..203....9U}}}.\\
\end{tabular}
\caption{Integrated measurements of the SFR inferred with different tracers for the targets in our sample that are part of GOALS \citep{2009PASP..121..559A}. All SFRs are in units of $M_{\odot}$/yr. The integrated stellar mass is obtained with our SED-fitting code.}
\label{tab_sfr_ir_24}
\end{table}

Besides using the $L_{IR}$ to infer the SFR, we can also use another mid-IR indicator, namely the emission from the continuum at 24 $\mu$m. We use Equation 10 presented in \cite{2009ApJ...692..556R}, to infer the SFR from the 24$\mu$m luminosity:

\begin{equation}
    SFR_{24\mu m} [M_{\odot} \textrm{yr}^{-1}] = 7.8 \times 10^{-10} \times L_{24\mu m} [L_{\odot}] \label{sfr_fir}
\end{equation}

\begin{figure}[!t]
\centering
\includegraphics[width=\columnwidth]{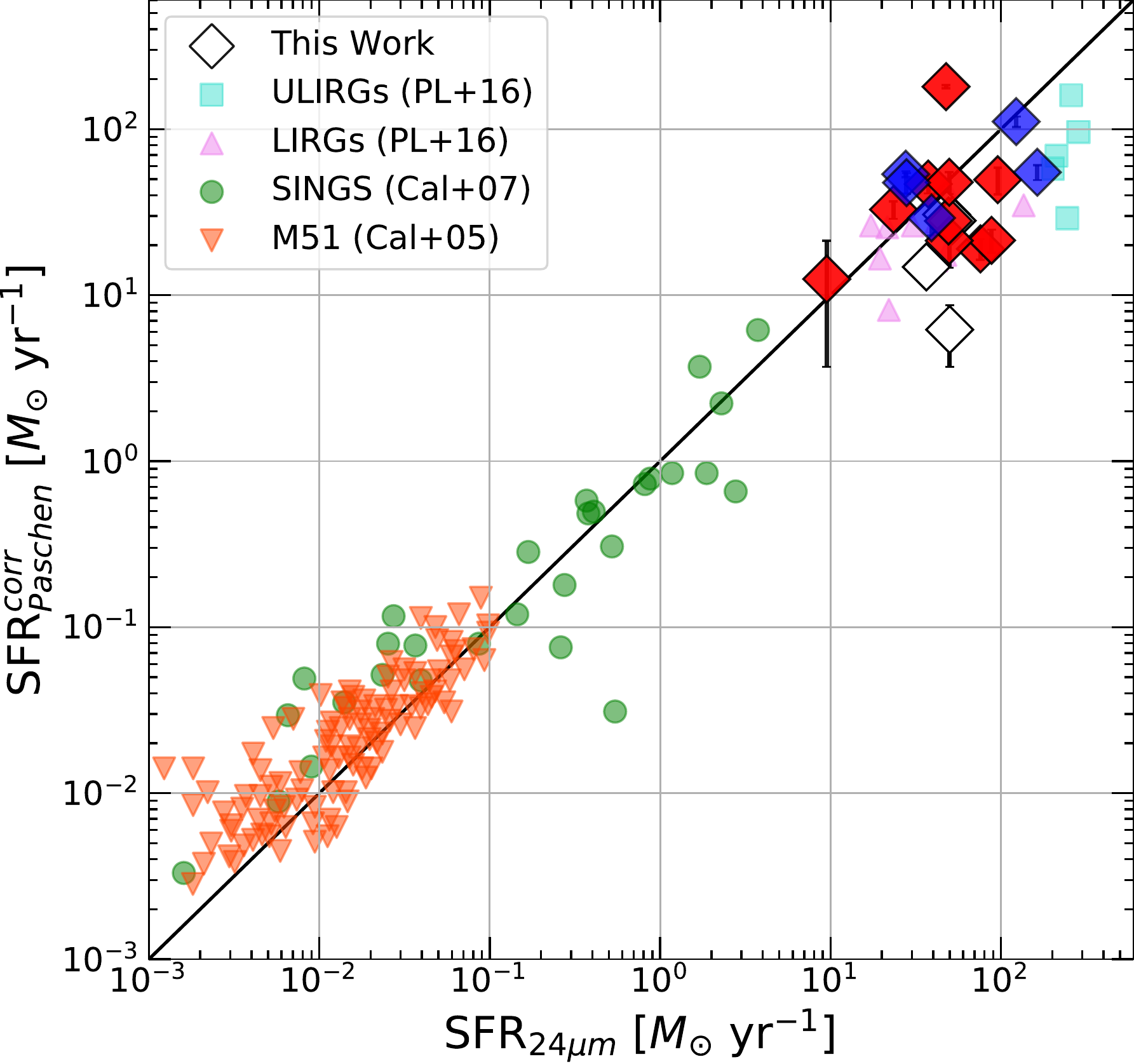}
\caption{Extinction corrected $SFR_{Pa\alpha}^{corr}$ as a function of the SFR inferred from the 24$\mu$m continuum emission. The various datasets are detailed in the main text. Our sample is indicated by the diamond symbol, blue coloured if they host an AGN and red coloured if not. \label{fig_sfr_24}}
\end{figure}

We can now compare SFR$_{Pa\beta}^{corr}$ and SFR$_{24\mu m}$. Figure \ref{fig_sfr_24} shows the resulting comparison. We retrieve a similar linear correlation to the previously discussed with SFR$_{IR}$. For our targets, the discrepancy between both indicators is $-0.14\pm0.32$ dex, so the 24$\mu$m emission and \pabeta\ are providing consistent SFR estimates. Previous studies have found similar conclusions, as well as deviations at the high-luminosity end \citep[e.g.,][]{2006ApJ...650..835A,2009ApJ...692..556R,2016A&A...590A..67P}. Figure \ref{fig_sfr_24} also includes results from previous studies, to be able to place our conclusions into context. As for the SFR$_{IR}$ comparison displayed before, these studies use Pa$\alpha$ to infer SFRs instead of \pabeta\ for our study. In Figure \ref{fig_sfr_24}, the turquoise squares (ULIRGs) and violet upward triangles (LIRGs) are data from PL+16. The green circles correspond to SINGS (SIRTF Nearby Galaxy Survey, \citealt{2003PASP..115..928K}) galaxies from \cite{2007ApJ...666..870C}. The orange downward triangles are measurements from individual star-forming regions of M51 from \cite{2005ApJ...633..871C}. The available 24$\mu$m galaxies from our sample are displayed with white diamonds, and can be seen in Figure \ref{fig_sfr_24} perfectly agreeing with the other works, as well as falling on top of the one-to-one line, on average. As before, we do not see a trend in the SFRs relationship due to the presence of an AGN. It is quite remarkable to see the agreement over almost six orders of magnitude in SFR$_{24\mu m}$, and covering sizes from individual star forming regions in M51 to integrated LIRGs and ULIRGs. On the other hand, we would expect both SFRs to agree if the 24$\mu$m emission was entirely due to star formation alone, since SFR$_{24\mu m}$ is calibrated from Paschen recombination lines (e.g. \citealt{2007ApJ...666..870C}). The contribution from evolved older stellar populations could also explain the deviations from the one-to-one trend in both comparisons in this section. Various studies imply that SFRs inferred from IR tracers are overestimated due to the emission from the dust heated by old stars (see e.g. Fumagalli el al. 2014). On top of this, the uncertainty in the dust temperature determination and possible evolution with redshift, leads to considerable uncertainties on the inferred physical galaxy properties, such as the obscured SFR and $L_{IR}$ (e.g. Sommovigo et al. 2020, 2022), which is an important factor to consider particularly at higher redshifts. All tabulated values from our sample used for Figures \ref{fig_sfr_ir} and \ref{fig_sfr_24} can be found on Table \ref{tab_sfr_ir_24}.

In summary, we report that the SFR inferred with dust-corrected Pa$\beta$ emission can recover the star formation estimated via mid-IR tracers, therefore being enough to estimate the SFR for each individual galaxy, in an integrated sense. A good agreement is found both with the MIPS 24 $\mu$m emission and the SFR$_{IR}$, following previously published results. Other works in the literature indicate that SFRs derived from MIPS 24$\mu$m + UV emission may overestimate the SFR when compared to the SFR derived from modelling the entire UV-to-IR SEDs, especially at the low-SFR end \citep[e.g.,][]{2019ApJ...882...65M,2019ApJ...877..140L}. MIPS 24$\mu$m derived SFRs may be robust for starbursting galaxies, but this might not be the case for all galaxies, in particular those with low specific SFRs and evolved stellar populations.

\section{Summary and Conclusions} \label{sec6}

In this work we have presented a sample of 53 local star-forming galaxies, covering the SFR range from normal star-forming spirals to LIRGs and up to ULIRGs, 24 belonging to the GOALS survey \citep{2009PASP..121..559A}. For each target we have HST narrow-band observations targeting the H$\alpha$ and Pa$\beta$ emission lines, as well as nearby broad-band continuum filters. This allows us to constrain the dust obscuration in these objects with the Balmer-to-Paschen decrement, which has been demonstrated to reach further depths and therefore obscuration than the commonly used Balmer decrement (H$\alpha$/H$\beta$). This provides the first systematic study on a large sample of very high quality observations, with this level of high-resolution and for the first time, both sub-kpc resolved  \halpha\ and \pabeta\ measurements.

For 24 of our targets that are also part of the GOALS survey of local luminous and ultra-luminous infrared galaxies, we have presented a methodology to treat spatially resolved observations, as well as obtaining robust parameters from local or high-z photometric observations with our SED-fitting technique. Our code provides trustworthy emission line fluxes from photometric measurements, that match those obtained from direct spectra on the same targets (within 0.05 dex). Our inferred physical parameters match those estimated with other SED-fitting codes, at a much smaller computational cost when compared to i.e. \texttt{Prospector}, and help us analyse in detail the nature of our targets.

We have performed an in depth analysis of the different SFR indicators that we can infer to probe the star formation in our galaxies at various wavelengths and timescales. We have demonstrated in a spatially-resolved manner, that the SFRD inferred with Pa$\beta$ is systematically higher than with H$\alpha$, even when a dust correction is applied using the Balmer decrement, with a mean offset of $-0.1\pm0.2$ dex. Our results agree with previous published works, although this had not been tested on a bin-to-bin basis before. Furthermore, Pa$\beta$ seems to recover more star-formation than the optical stellar continuum in the most obscured parts of each galaxy. This could also indicate a bursty nature of our sources, as the SFR inferred with the stellar continuum provides the star formation convolved over a larger period of time. Besides this, the emission from the stellar continuum and the nebular lines come from different regions within the galaxies. 

On top of this, we have tested whether SFR$_{Pa\beta}$ can recover the star formation inferred by IR indicators. We find consistent SFR estimates both with the emission traced by the 24$\mu$m continuum (with a ratio $-0.14\pm0.32$ dex) and with the SFR$_{IR}$ (with a ratio $-0.04\pm0.23$ dex). This agrees with previous studies on the different SFR tracers, and gives us confidence in using Paschen-lines to paint a more complete picture of star formation at all redshifts.

We obtain high resolution and robust extinction maps for all targets, probing unusually high $A_V$ inferred by the gas, especially towards the core of the galaxies. Forthcoming publications will analyse this in detail, in particular the differential attenuation experienced by the gas (traced by the Balmer-to-Paschen decrement) and the stars (traced by the stellar continuum). Accurately inferring large obscuration in galaxies could yield very exciting new avenues, such as discussed in \cite{2018ApJ...862L..22C}, where it is introduced that one could identify mergers by their extreme obscuration levels. At high redshift, one cannot rely in the morphological signs to identify merging galaxies, and severe values of $A_V$ have not been able to be explained by other mechanisms. Upcoming JWST observations of NIR lines could help us test this idea, and possibly find a new way of identifying high-z mergers.

Our study paves the way to very exciting upcoming works, that will benefit from the first high quality spatially-resolved observations at higher redshifts to be obtained with JWST. This telescope will be able to observe Paschen-line emission in high-z galaxies at $z>1$, breaking the ice for extending studies and methodologies like the ones presented in this work to higher redshifts. This will aid us in our pursuit to understand how galaxies form and evolve across cosmic time, helping us constrain their star formation activity by seeing through the dust like we have not been able before. Paschen lines will also provide a better estimate of the star formation history in the $z\sim1-3$ Universe produced in dusty star forming galaxies.

\acknowledgments

The authors thank the anonymous referee for the helpful and constructive comments received. The Cosmic Dawn Center is funded by the Danish National Research Foundation (DNRF) under grant $\#$140. DM and DLV acknowledge the very generous support by HST-GO-14095, provided by NASA through a grant from the Space Telescope Science Institute, which is operated by the Association of Universities for Research in Astronomy, Incorporated, under NASA contract NAS5-26555. LC acknowledges support by grant No. MDM-2017-0737 Unidad de Excelencia ‘‘Maria de Maeztu’’-Centro de Astrobiología (INTA-CSIC) by the Spanish Ministry of Science and Innovation/State Agency of Research MCIN/AEI/10.13039/501100011033, and grant  PID2019-106280GB-I00. MP is supported by the Programa Atracci\'on de Talento de la Comunidad de Madrid via grant 2018-T2/TIC-11715, and acknowledges support from the Spanish Ministerio de Econom\'ia y Competitividad through the grant ESP2017-83197-P, and PID2019-106280GB-I00.

\vspace{5mm}
\facilities{HST (ACS), HST (WFC3)}
\software{Astropy \citep{2013A&A...558A..33A}, Matplotlib \citep{Hunter:2007}, NumPy \citep{numpy}, SciPy \citep{scipy}, Grizli \citep{grizli}}

\clearpage
\appendix \label{sec:appendix}

\section{Voronoi Binning} \label{sec:app_voronoi}

As stated in \S\ref{sec:voronoi}, we develop an adaptation of the Voronoi binning procedure from \citet{2003MNRAS.342..345C} to be able to perform the binning in large datasets and in regimes of low S/N per original data point. 

Our hybrid block-averaging approach starts by defining a reference image that is used in the binning, and selecting the desired S/N threshold. We choose this to be the narrow-band image that targets Pa$\beta$, usually being F130N or F132N, since it is important for our further use of the line fluxes to obtain robust S/N line detections in the narrow-band images, particularly Pa$\beta$, being usually fainter than H$\alpha$.
The code starts by creating a block averaged image, where each block has 32$\times$32 original pixels. We mask any pixels where these large bins have S/N less than a minimum threshold, which is one of our binning parameters. Then, we run \citet{2003MNRAS.342..345C} optimal Voronoi 2D-binning algorithm on this large blocked image. We adopt as a Voronoi bin anything with more than one individual (blocked) pixel ($N>1$), and these pixels are now "frozen" in the binning. Individual pixels that satisfy the target S/N (i.e., $N=1$ blocked bins) are sent to the next step, which consists in reducing the block size by a factor of two (e.g., 16$\times$16 at the second stage). Pixels in blocks where there is not enough S/N to reach the target S/N, and therefore the Voronoi code bins multiple blocks together, are not sent to the next step and remain in that assigned Voronoi bin. The code repeats these steps until the desired minimum bin size is achieved (e.g., if the minimum possible size is wanted, it stops when it reaches 1$\times$1, i.e., native pixels). For our study we stop at 2$\times$2 pixels as the minimum size of a bin, reducing the number of points to be able to perform further analyses, but at the same time not losing drastically the high resolution in our images. For example, for a 2$\times$2 pixels bin, which would be 0\farcs2 across, we still achieve a resolution of 80 pc at $z\sim0.02$.

Once we create the Pa$\beta$ narrow-band binned image, we apply the derived binning to the other filter images available for that galaxy. However, this means that for some bins the target S/N criteria is not necessarily met in all bands, which could later on affect the robustness of the derived physical parameters in that bin. By imposing the S/N threshold on the Pa$\beta$ narrow-band image, which is usually the band with least S/N across the image, we reduce this issue. Nonetheless, it is important to have the same binning in all images to be able to perform different studies on them. Specific information on how to run the binning code can be found on the Github repository\footnote{\url{https://github.com/claragimenez/voronoi}}.

\section{SED-Fitting}  \label{sec:app_sedfit}
\begin{figure}[h]
\centering
\includegraphics[width=\columnwidth]{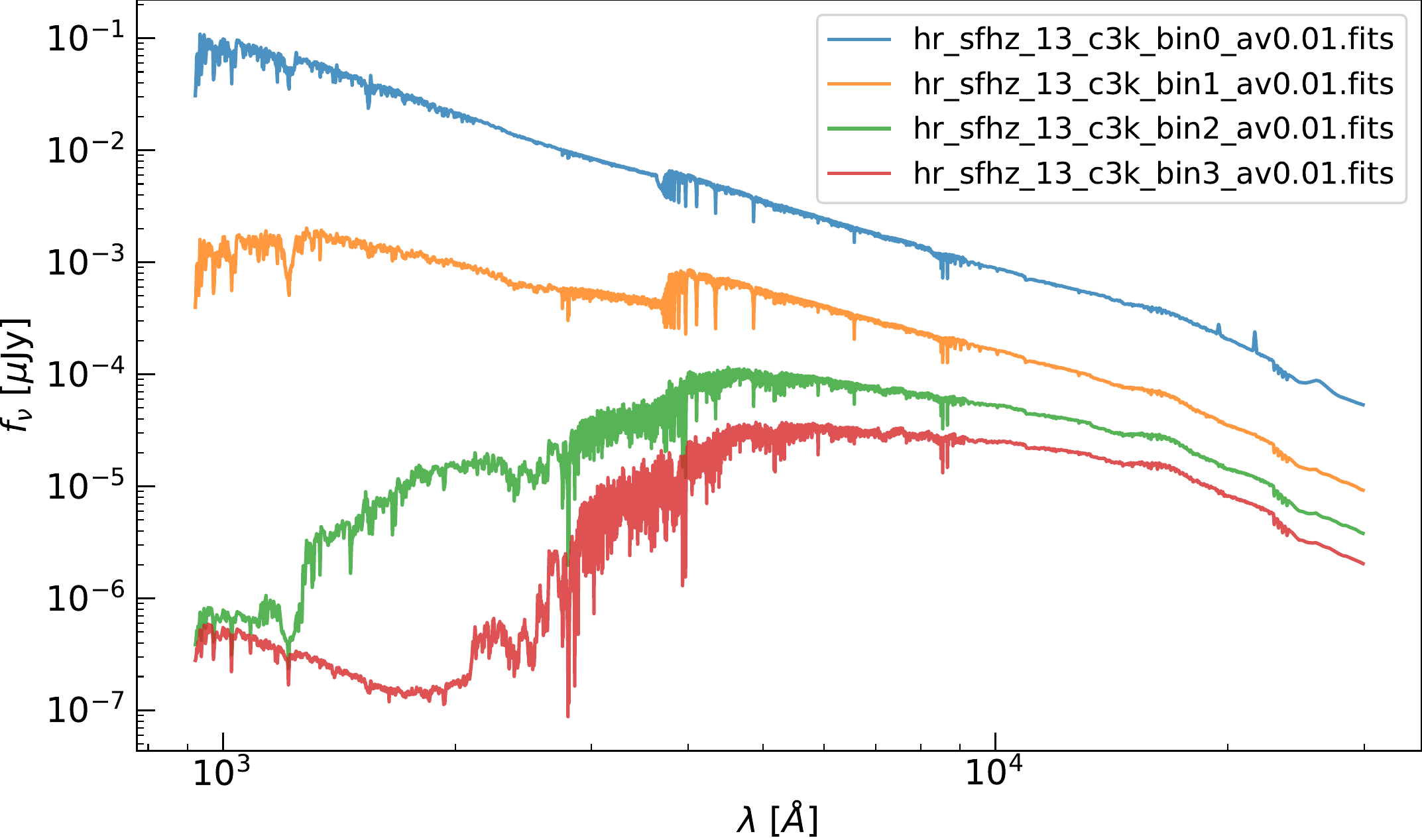}
\includegraphics[width=\columnwidth]{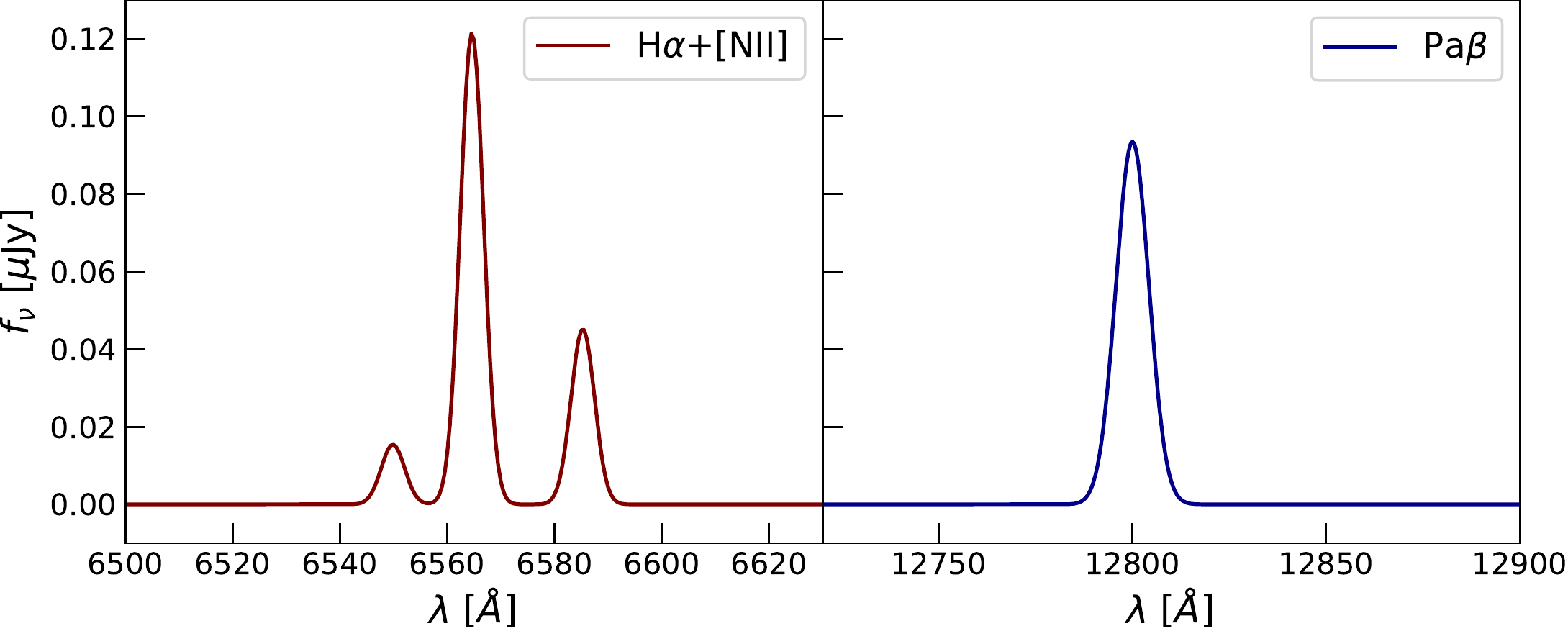}
\caption{\textbf{Top:} Set of continuum templates that we use in our SED-fitting code. \textbf{Bottom:} H$\alpha$+[\ion{N}{2}] (left) and Pa$\beta$ (right) pure line emission templates. \label{fig_templates}}
\end{figure}

We develop an adaptation of the pythonic version of EAZY\footnote{\url{https://github.com/gbrammer/eazy-py}} \citep{2008ApJ...686.1503B}, which includes an extra-reddening grid, which allows 4 continuum templates to fit $A_V$ from 0.0 to 5.0, and two emission line templates $-$H$\alpha$+[\ion{N}{2}] and Pa$\beta-$. The set of templates that we use are shown in Figure \ref{fig_templates}. Analogous to the step-wise SFH parameterization of Prospector \citep{2017ApJ...837..170L}, the continuum templates are generated for stepwise time bins between 0, 50, 200, 530 Myr, 1.4 Gyr with constant SFR across each bin. The line templates are created as Gaussian peaks with a continuum set to zero, and a sigma width of 100 km/s. We set the [\ion{N}{2}]/H$\alpha$ ratio to 0.55. We note that these templates are not identical to those that EAZY uses for photometric redshift estimates. Here we fit explicitly for dust reddening, whereas in EAZY the reddening is a property of the linear combination of basis templates that have diversity of SFH and dust reddening \citep{2008ApJ...686.1503B}.

Our code receives a catalog as input (which can have multiple targets, or in our study it has multiple bins to study spatially-resolved properties), performs the fit, and outputs the inferred physical estimates. The SED-fitting code works as follows. Firstly, it imports the templates (that can be defined by the user, as can multiple other features that follow) and the photometric measurements. For this, it follows how EAZY reads in catalogues given by the user. The reddening grid is then defined, and the templates are integrated through the filter band-passes to produce synthetic photometry. Then the fit at each point of the grid and bin is performed, calculating the resulting $\chi^2$ between the observed data and synthetic photometry and saving it at each step. After this, the covariance matrix is calculated, and we draw 1000 random coefficients, that we later use to better sample our uncertainties. We can directly infer the line fluxes from the combined draws and reddening grid that is built, as well as other physical properties such as $A_V$(H$\alpha$/Pa$\beta$), SFR, stellar mass, etc. We then implement a prior, requiring that the stellar mass map is smooth. From the $\chi^2$ grid we can calculate the normalised probability density function (PDF), that we modify with the mass prior. Once we have the prior normed PDF, we can calculate the uncertainties and values of our inferred physical properties by calculating the percentiles, interpolating the 16\%, 50\% and 84\% with the cumulative sum of the normed PDF and each parameter evaluated in the full grid space. From the 0.5 percentile we get an estimate of the parameter value in each bin $-$ as well as saving the full sampling of each parameter $-$, and from the half difference of the 0.84 and 0.16 percentiles we estimate its uncertainty. Finally, we can estimate the $\chi^2$ of the fit (as well as the coefficients and "best" model) where the sampling grid has the maximum posterior distribution value.

Our SED-fitting code outputs the line fluxes H$\alpha$ and Pa$\beta$, as well as various physical properties: the $A_V$ obtained from the 'empirical' Balmer-Paschen decrement ($A_V($H$\alpha/$Pa$\beta$)), the $A_V$ inferred from the stellar population ($A_{V,stars}$). We also obtain the star-formation rate (SFR) and stellar mass ($M_{*}$) per individual bin. We can also derive the most recent star formation with the H$\alpha$ luminosity, as well as the better tracer for obscured star formation, the SFR derived from the Pa$\beta$ luminosity, dust-corrected with the Balmer-to-Paschen decrement.

\section{MUSE Archival Data} \label{section_app_niiha}

\begin{figure}[h!]
\centering
\includegraphics[width=\columnwidth]{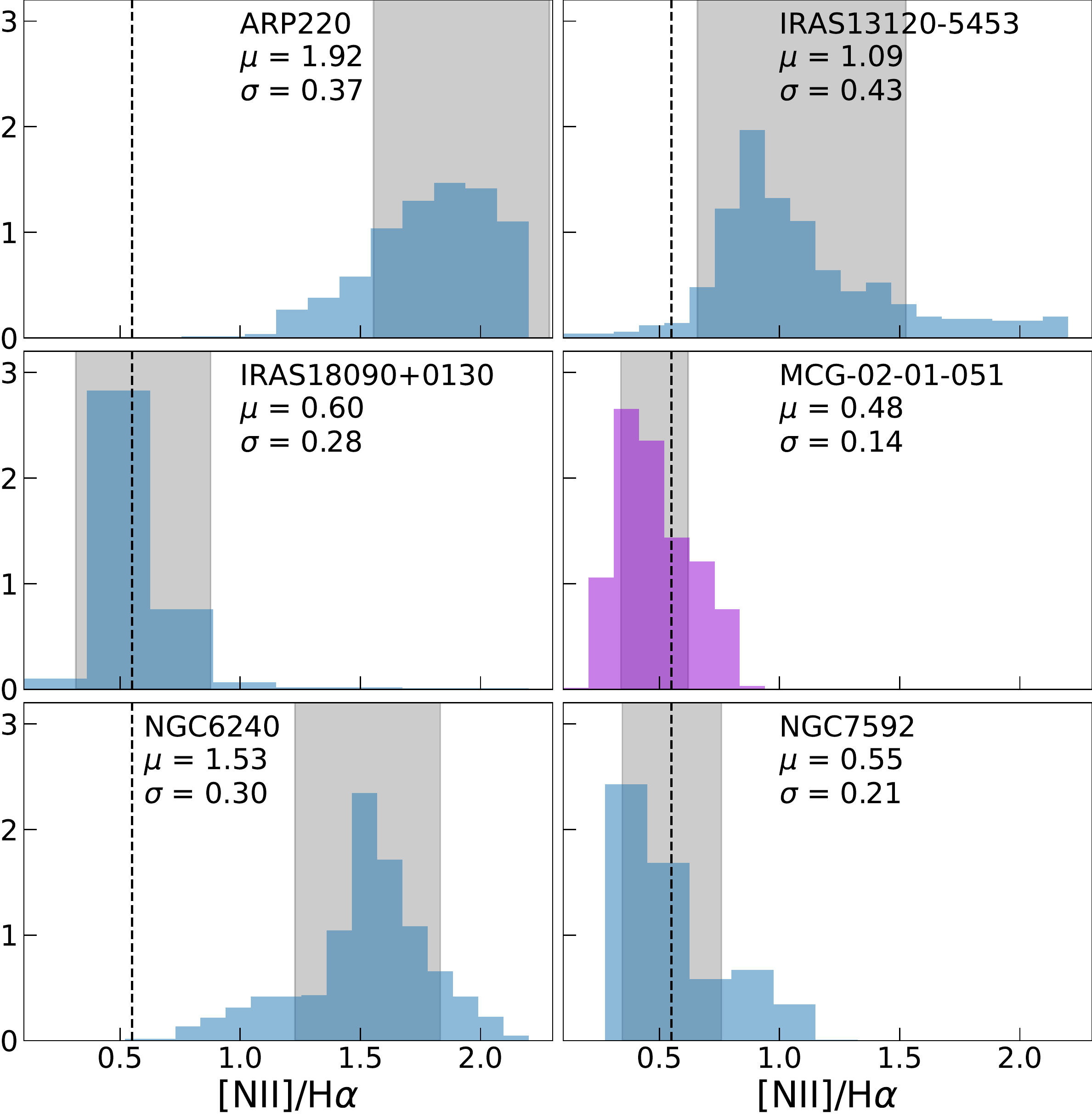}
\caption{Histograms of the [\ion{N}{2}]/\halpha\ ratio for the 6 galaxies that are part of the GOALS survey, as measured in the MUSE IFU cube. The value we adopt for the ratio is indicated with a black dashed line on the histogram, and the grey shaded region shows the 1$\sigma$ interval around of the mean. The y-axis is normalised.} \label{fig_app_niiha}
\end{figure}

As discussed in \S\ref{section_ha_nii}, 6 of the galaxies in our sample have archival MUSE observations that we use to test the reliability of our inferred \halpha\ fluxes from the HST photometric data. Here we present the resulting [\ion{N}{0ii}]/H$\alpha$ histograms for the 5 additional targets that are not presented in \S\ref{section_ha_nii}. Figure \ref{fig_app_niiha} shows the histograms for the targets Arp~220, IRAS13120-5453, IRAS18090+0130, MCG-02-01-051 (presented also in the main text, displayed with a different colour below), NGC6240 and NGC7592. Half of the targets agree with our adopted value of 0.55 within 1$\sigma$ of the mean, whereas the other 3 show quite extreme [\ion{N}{0ii}]/H$\alpha$, with the majority or all bins having [\ion{N}{0ii}]$>$H$\alpha$ across the galaxy, as found in other works (e.g. \citealt{2021A&A...646A.101P}).

The systematic shift in $A_V$, $\Delta A_V$, caused by using two different [\ion{N}{0ii}]-to-H$\alpha$ correction factors, $f^{corr}_i$, will be given by:
\begin{equation}
    \Delta A_V = C_1 \log\Big(\frac{f^{corr}_1}{f^{corr}_2}\Big)
\label{eq_delta_av}
\end{equation}

Where $C_1$ is the following constant:
\begin{equation}
    C_1 = R_V \times \frac{2.5}{k(\textrm{Pa}\beta)-k(\textrm{H}\alpha)}
\end{equation}

And the correction factor is computed as follows:

\begin{equation}
    f^{corr} =\frac{\textrm{H}\alpha/[\textrm{\ion{N}{0ii}}]}{\textrm{H}\alpha/[\textrm{\ion{N}{0ii}}]+1}
\end{equation}
So that the resulting \halpha\ flux that is corrected for [\ion{N}{0ii}] contamination is given by $f_{H\alpha}$ = $f^{corr}$$\times$ $f_{H\alpha+\textrm{[\ion{N}{0ii}]}}$, where the last term is the measured combined [\ion{N}{0ii}]+H$\alpha$ flux.

\section{Filter Coverage} \label{section_filters}

Here we present a test of how the specific filter coverage available for each object can affect the emission line maps derived with our method. We want to test if the different filter widths or number of available filters affect our ability to recover the physical properties that we infer, such as the H$\alpha$ line flux. For this, we make use of the archival MUSE cube for MCG-02-01-051 (see also  \S\ref{section_ha_nii}). We can integrate the MUSE spectra through various filter bandpasses to produce synthetic images. For MCG-02-01-051, we have 6 HST bands available (F673N, F130N, F132N narrow and F435W, F814W, F110W broad). We add synthetic observations from the MUSE cube of the HST ACS/WFC F555W, F606W and F775W bandpasses, to better sample the optical continuum. We then run our binned SED-fitting software with the same set-up as before with the observed HST and synthetic MUSE images.

\begin{figure}[h!]
\centering
\includegraphics[width=\columnwidth]{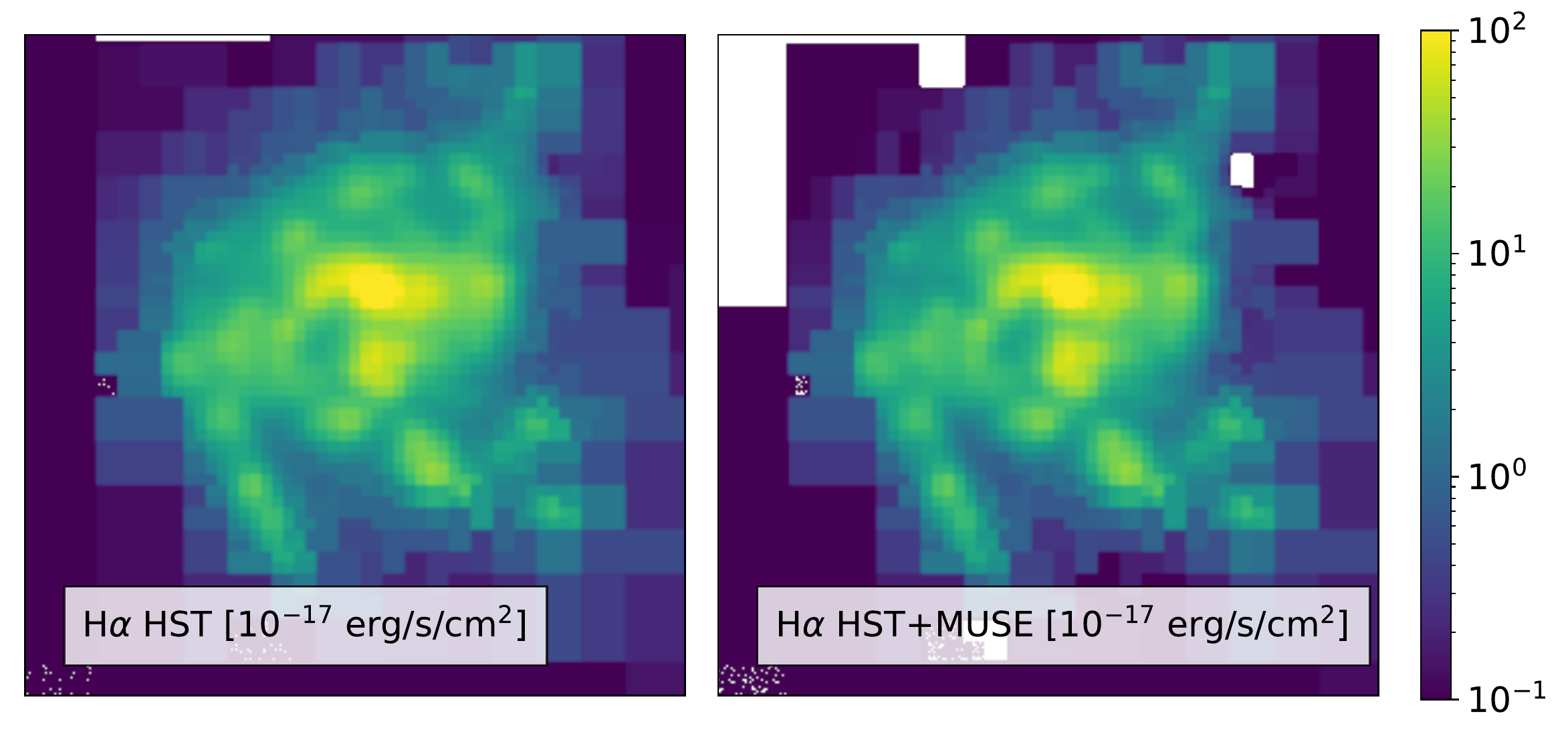}
\includegraphics[width=\columnwidth]{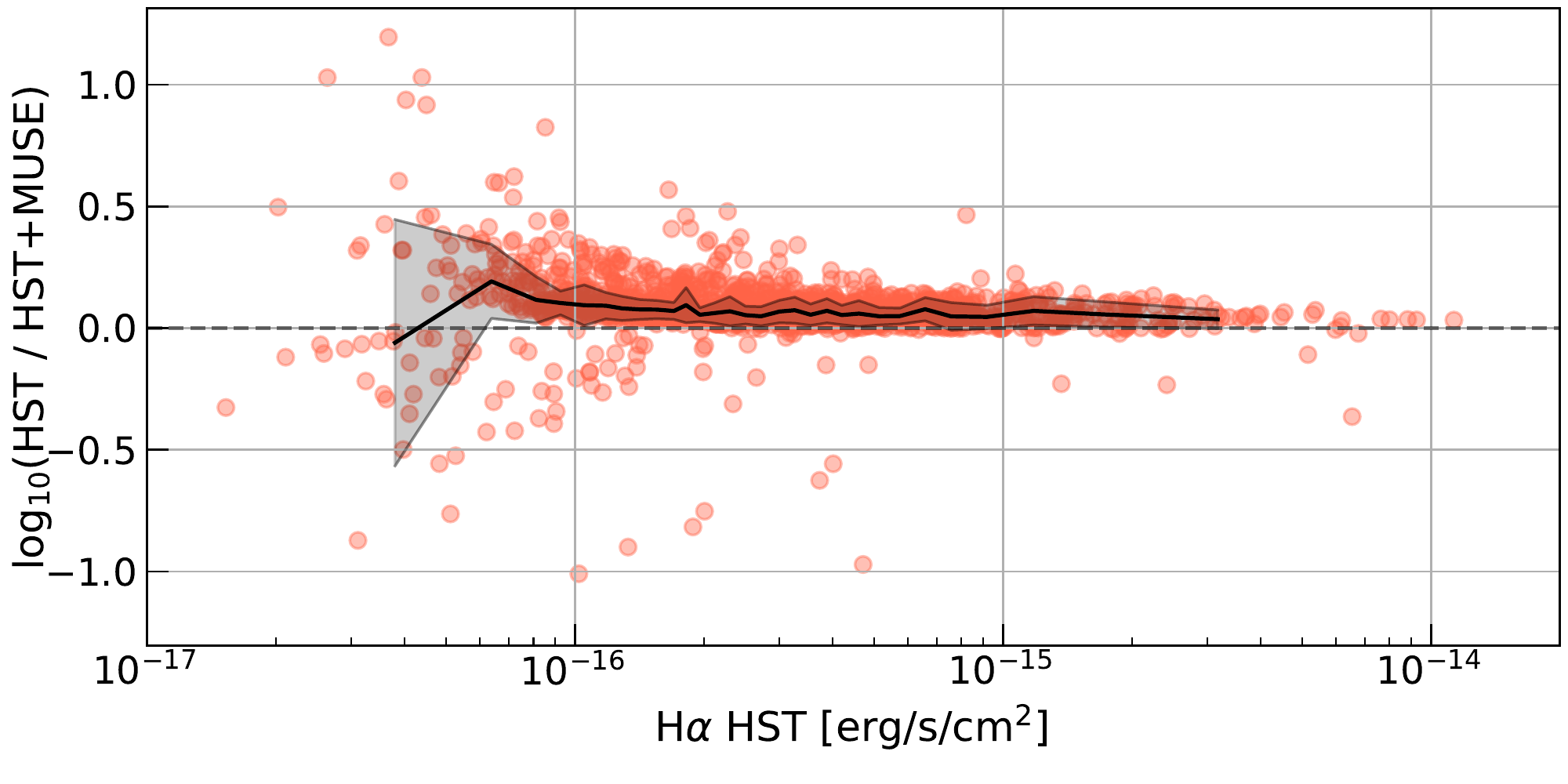}
\caption{\textbf{Top:} Maps of the H$\alpha$ line flux for the target MCG-02-01-051 inferred with only HST photometry (left), and with HST and extra synthetic photometry from MUSE (right). \textbf{Bottom:} Ratio of both H$\alpha$ estimates in each spatial bin. The running median and scatter are shown as a solid black line and surrounding shaded region, respectively.} \label{fig_filter_coverage}
\end{figure}

Figure \ref{fig_filter_coverage} displays the comparison of the resulting H$\alpha$ flux inferred with both estimates, in maps (top panels) and as a scatter plot (bottom panel). We find that the emission line map derived from the limited available HST data is fully consistent with the best-case scenario with additional broad-band optical filters (with a median log ratio of 0.06 $\pm$ 0.07 dex). When considering the rest of inferred physical parameters, both runs are consistent, having less than 0.3 dex off between them when inferring the SFR, the stellar mass or the $A_V$.

In order to fully constrain the SFH and dust obscuration of the continuum, we prefer to have a broader wavelength coverage. Nonetheless, we conclude that with the average filter coverage of our sample, we can adequately infer physical properties with our SED-fitting code.

\section{Individual galaxies} \label{app_results}

\begin{figure*}[!th]
\centering
\includegraphics[width=\textwidth]{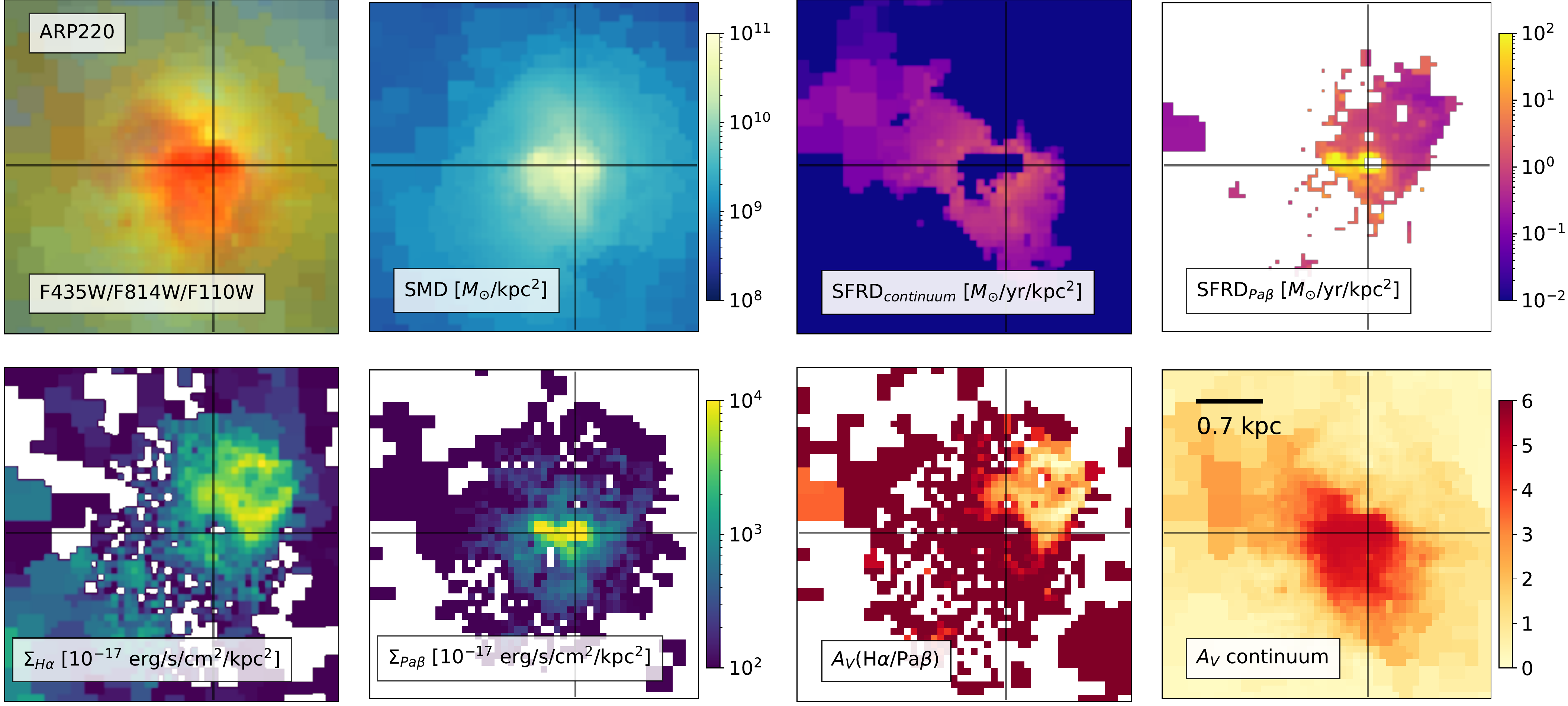}
\caption{Output of the spatially-resolved SED-fit on Arp~220. The top row, from left to right, shows the RGB image (built combining three broadband filters), the stellar mass density map, and the SFRD maps inferred with the stellar continuum and with the \pabeta\ emission line flux. The bottom row shows the resulting \halpha\ and \pabeta\ surface density flux maps, as well as the $A_V$ inferred from the empirical Balmer-to-Paschen decrement, and the stellar continuum. The cross is centered on the brightest F110W pixel. The physical scale is indicated on the $A_V$ continuum panel (bottom right).} \label{appendix_arp220}
\end{figure*}

Figures \ref{appendix_arp220}-\ref{appendix_vv340a} display the three-colour composites, H$\alpha$ and Pa$\beta$ emission line maps, extinction maps inferred from the gas and the stellar population, as well as SFR and stellar mass maps for each galaxy presented in this paper.

\begin{figure*}[!h]
\centering
\includegraphics[width=\textwidth]{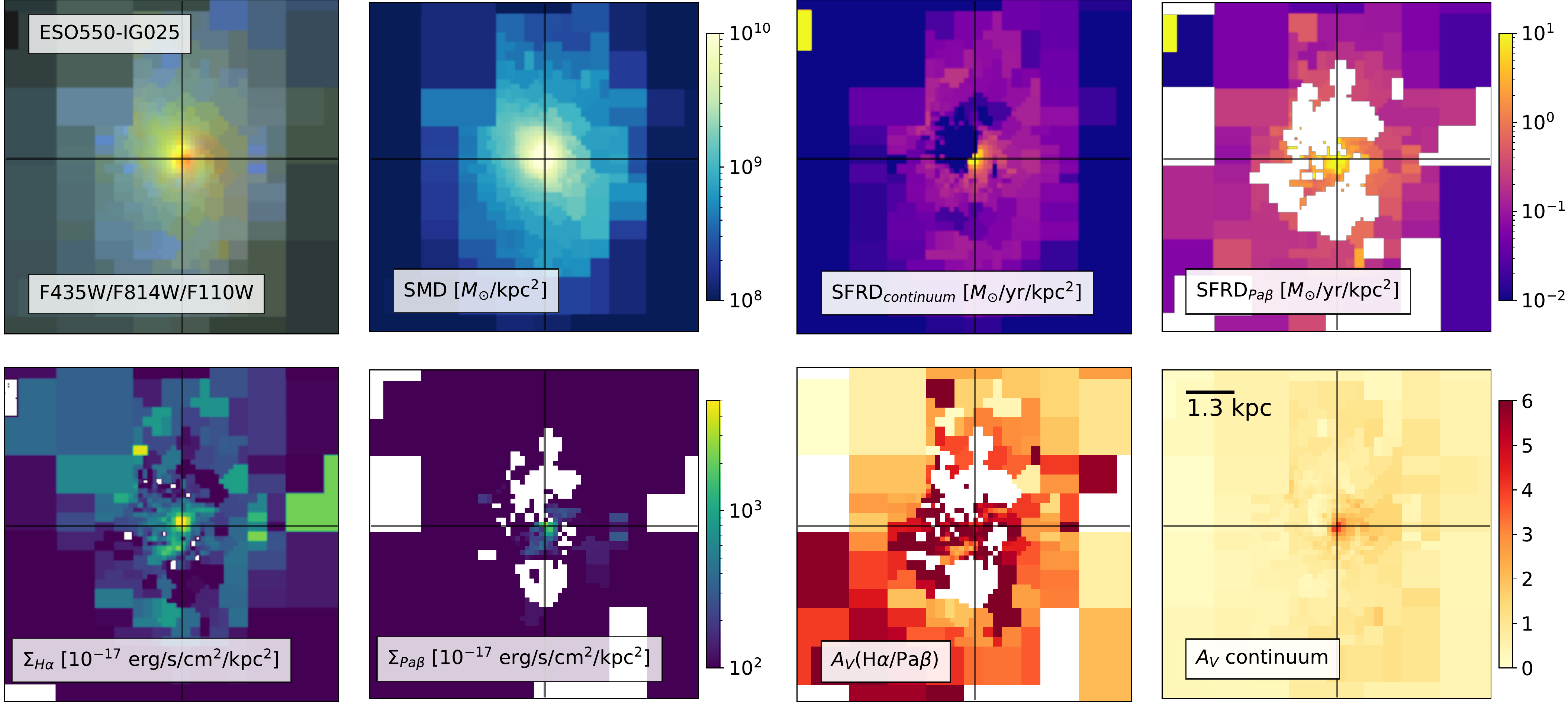}
\caption{Maps of the physical properties inferred with our SED-fitting code for ESO550-IG025. See Figure \ref{appendix_arp220} for more details.}
\end{figure*}

\begin{figure*}[!th]
\centering
\includegraphics[width=\textwidth]{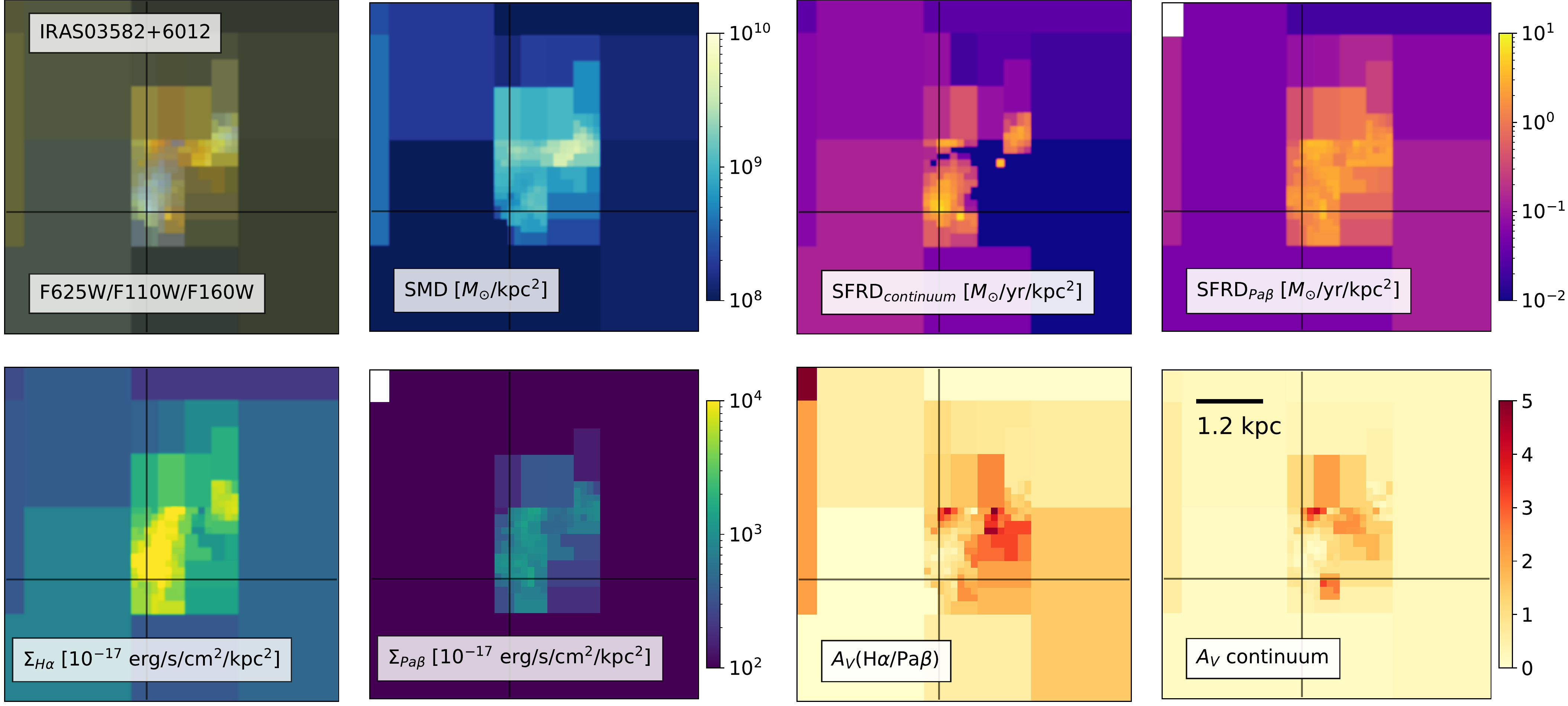}
\caption{Maps of the physical properties inferred with our SED-fitting code for IRAS03582+6012. See Figure \ref{appendix_arp220} for more details.}
\end{figure*}

\begin{figure*}[!h]
\centering
\includegraphics[width=\textwidth]{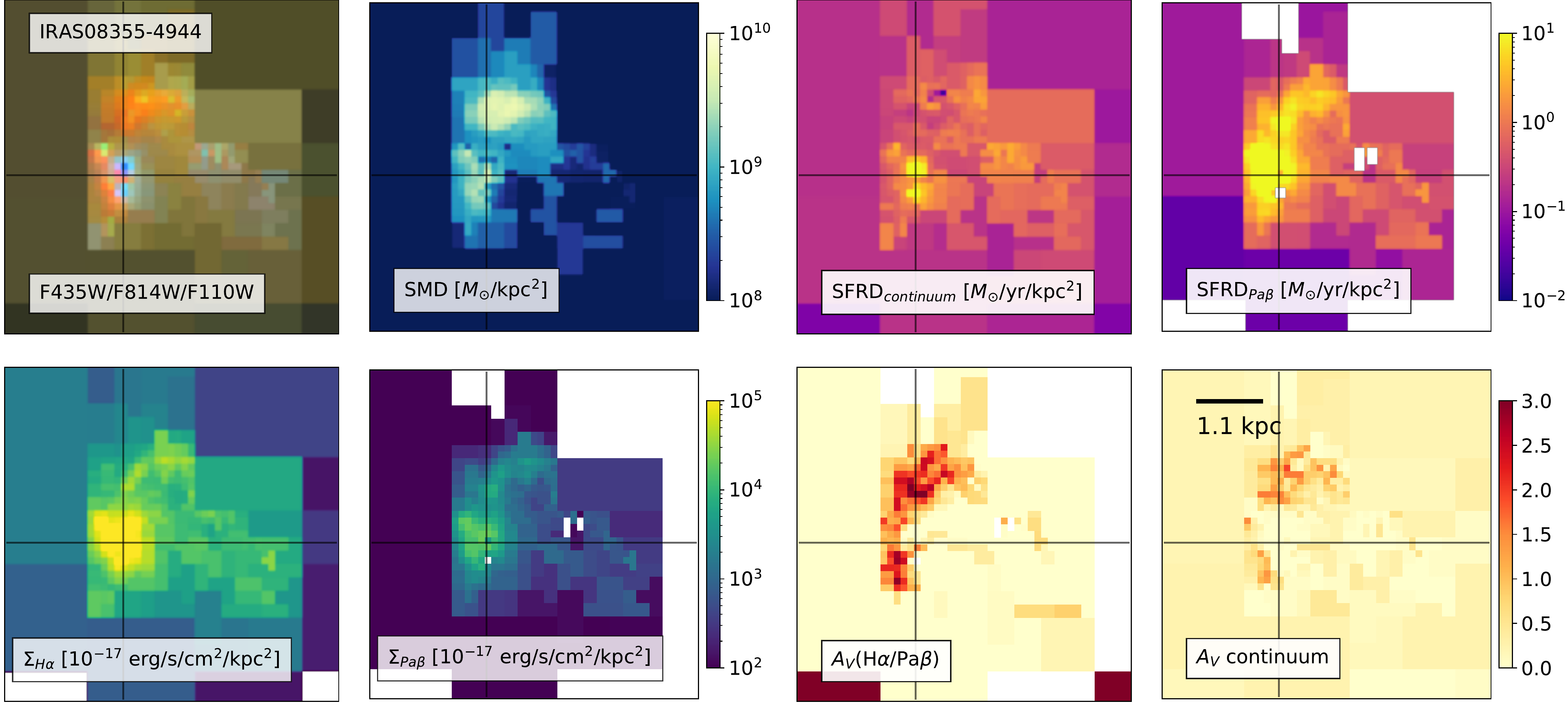}
\caption{Maps of the physical properties inferred with our SED-fitting code for IRAS08355-4944. See Figure \ref{appendix_arp220} for more details.}
\end{figure*}

\begin{figure*}[!th]
\centering
\includegraphics[width=\textwidth]{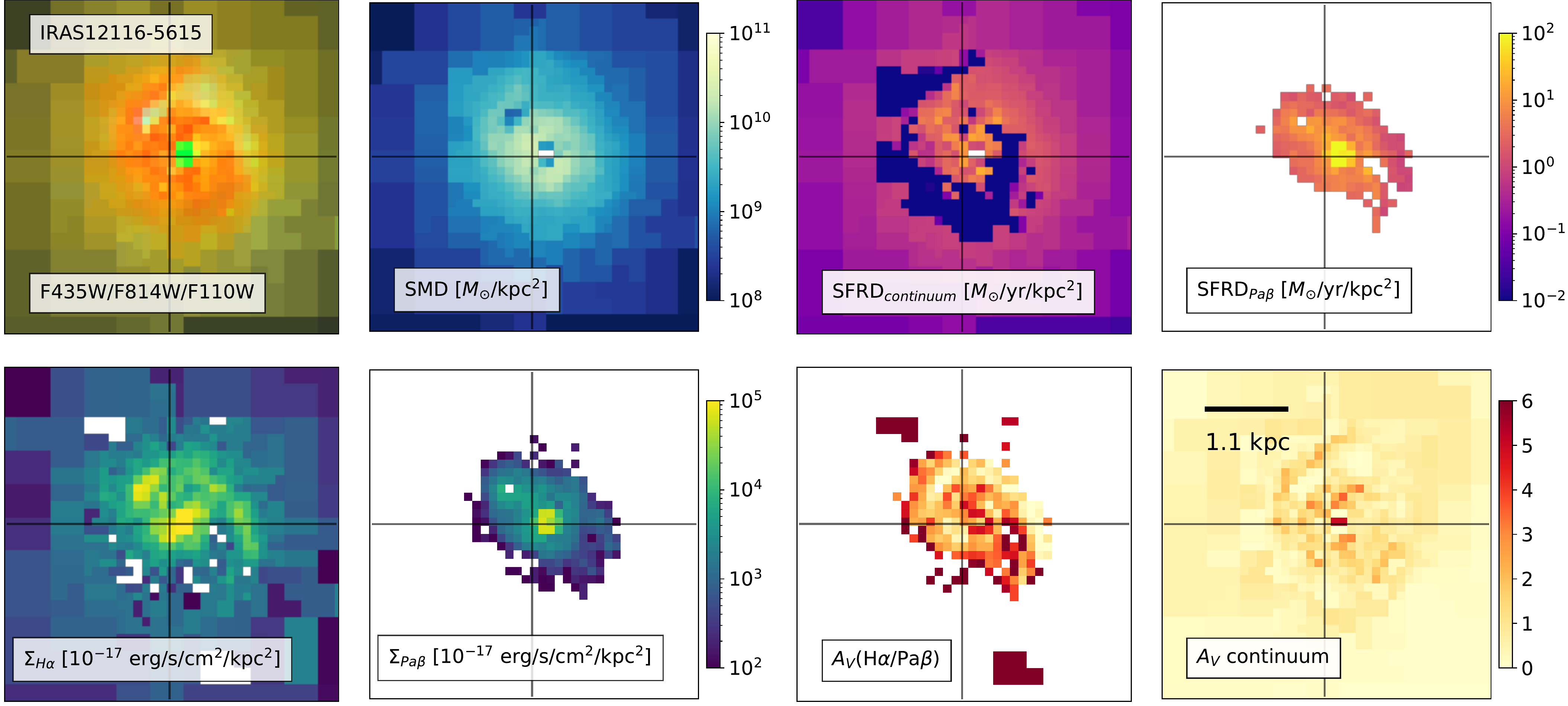}
\caption{Maps of the physical properties inferred with our SED-fitting code for IRAS12116-5615. See Figure \ref{appendix_arp220} for more details.}
\end{figure*}

\begin{figure*}[!h]
\centering
\includegraphics[width=\textwidth]{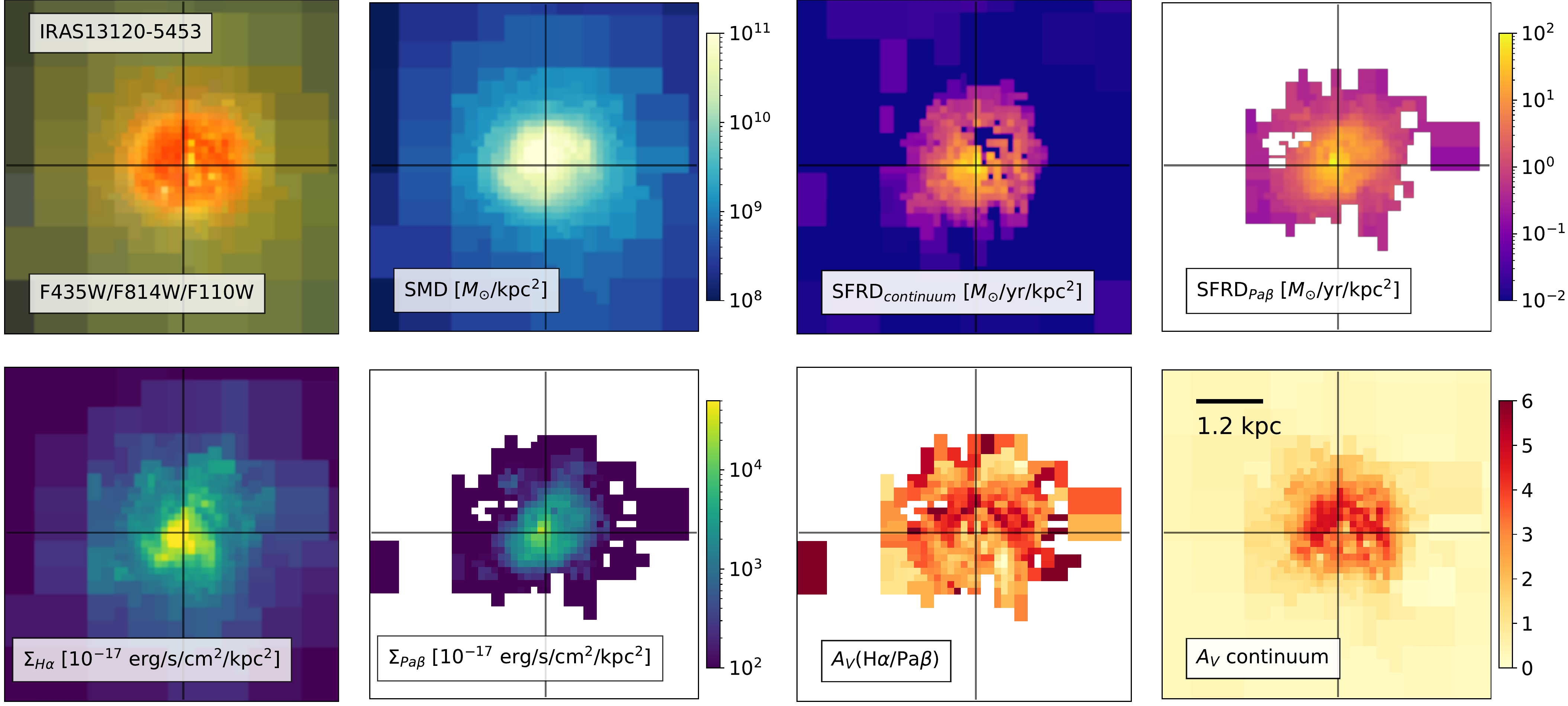}
\caption{Maps of the physical properties inferred with our SED-fitting code for IRAS13120-5453. See Figure \ref{appendix_arp220} for more details.}
\end{figure*}

\begin{figure*}[!th]
\centering
\includegraphics[width=\textwidth]{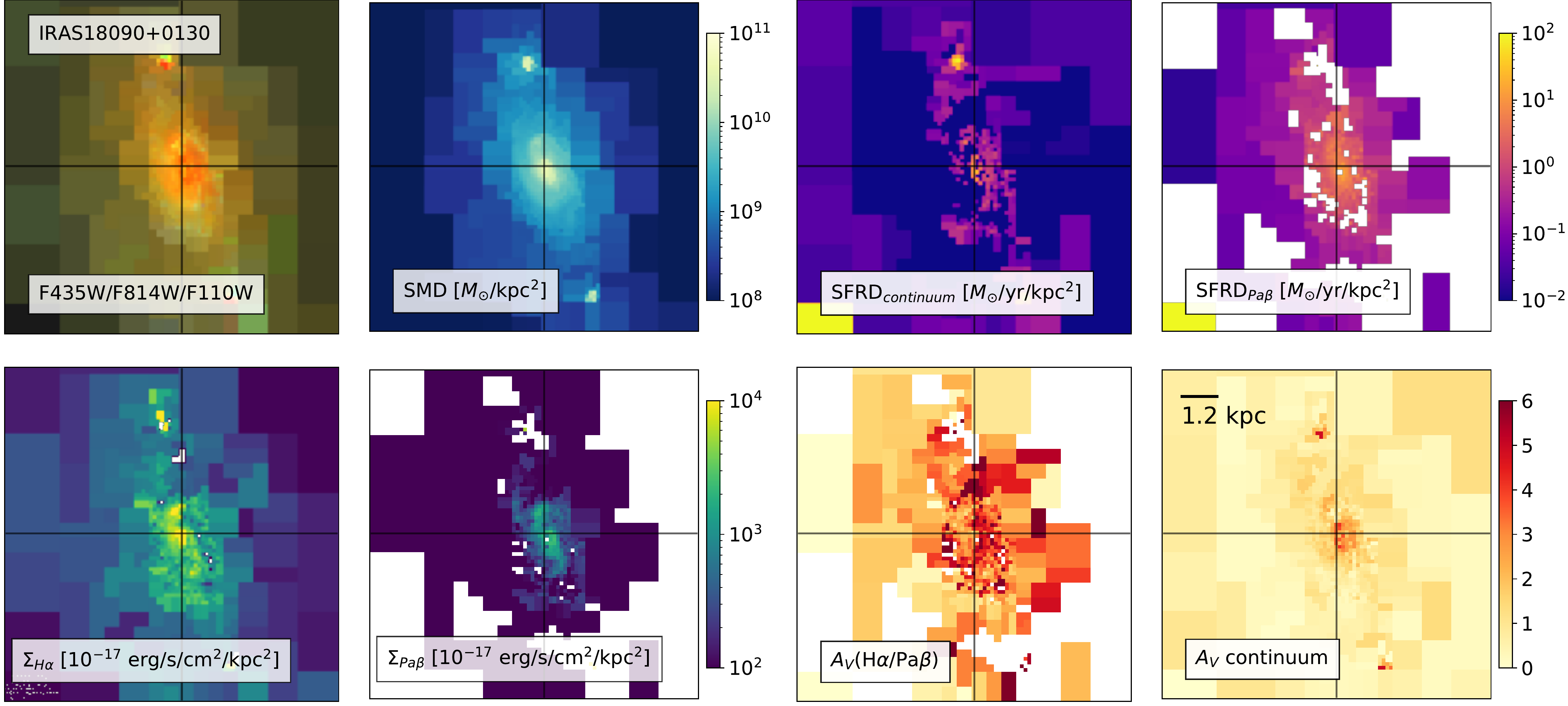}
\caption{Maps of the physical properties inferred with our SED-fitting code for IRAS18090+0130. See Figure \ref{appendix_arp220} for more details.}
\end{figure*}

\begin{figure*}[!h]
\centering
\includegraphics[width=\textwidth]{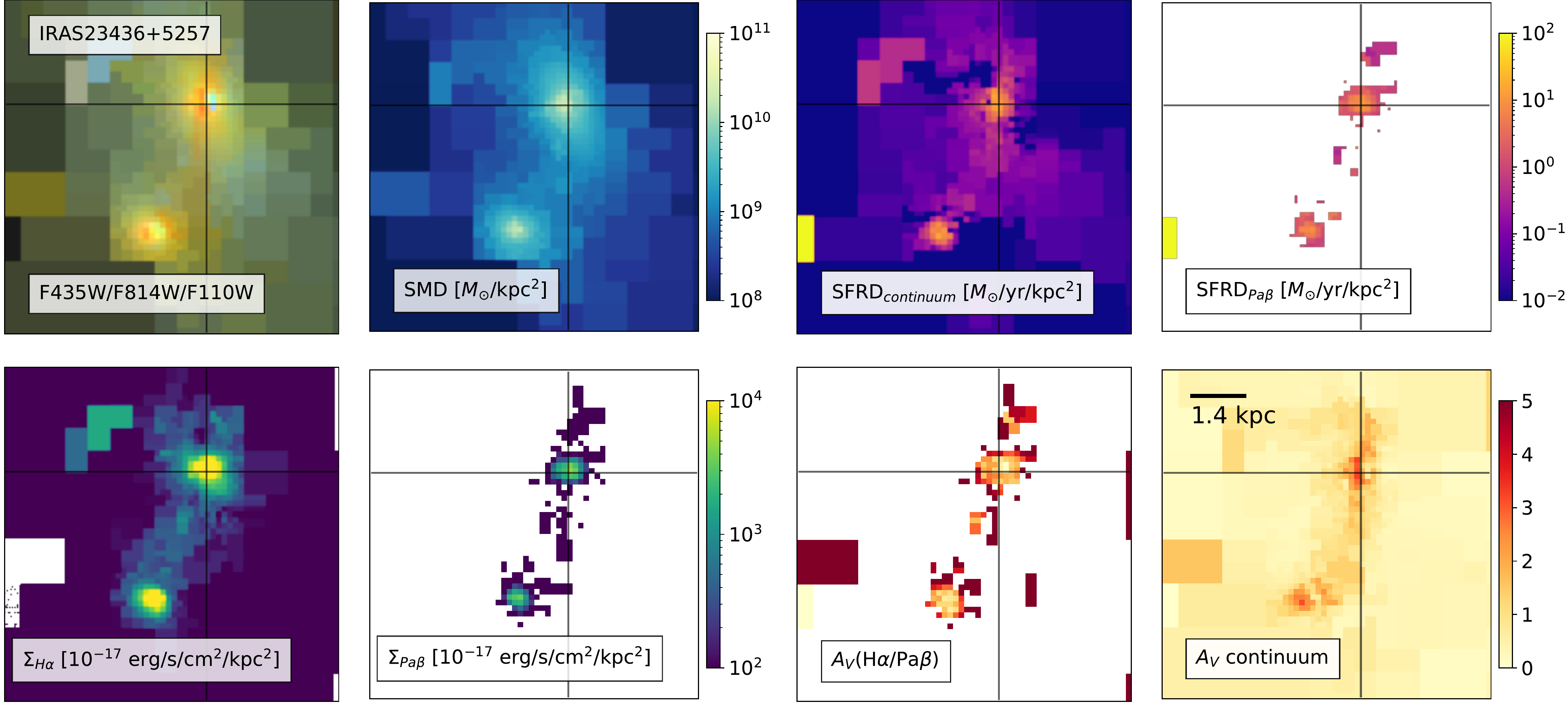}
\caption{Maps of the physical properties inferred with our SED-fitting code for IRAS23436+5257. See Figure \ref{appendix_arp220} for more details.}
\end{figure*}

\begin{figure*}[!th]
\centering
\includegraphics[width=\textwidth]{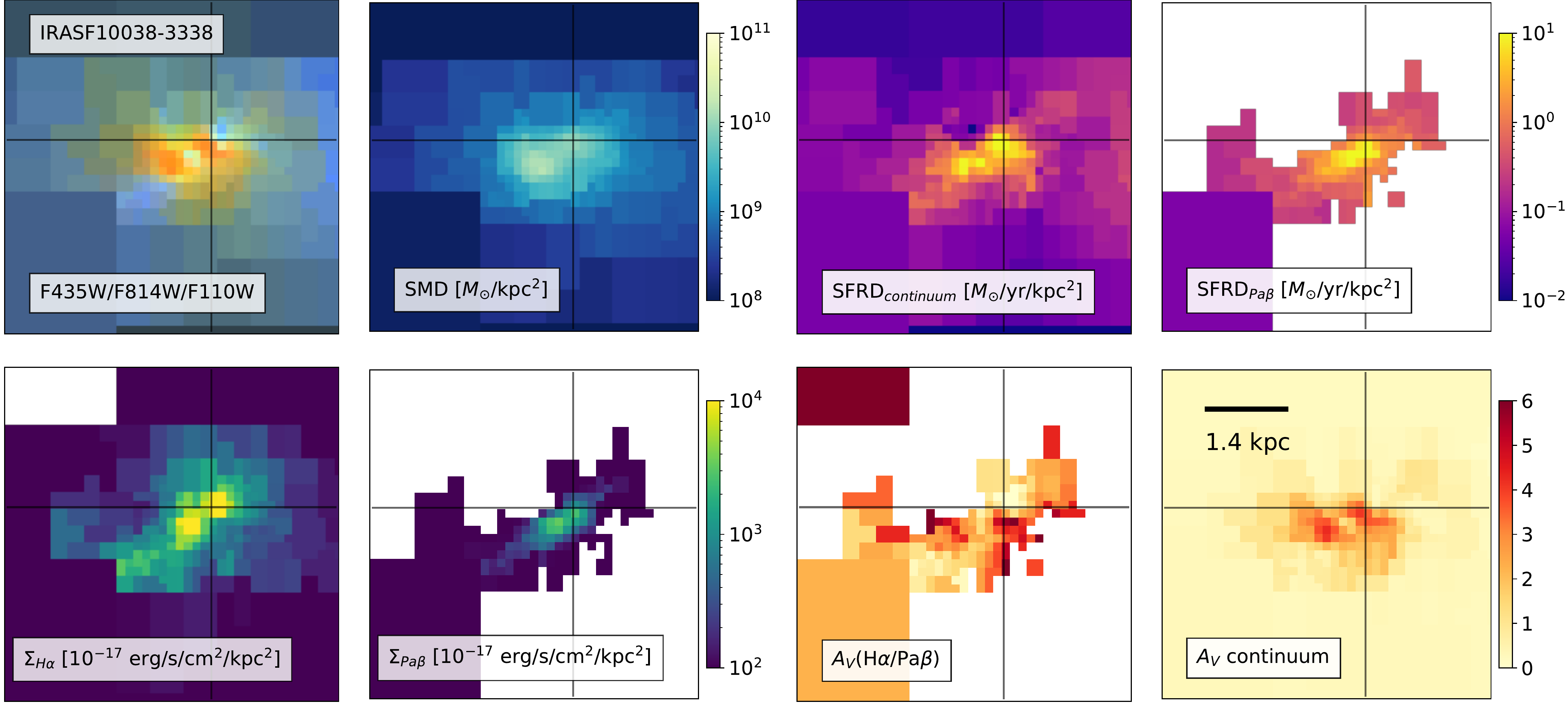}
\caption{Maps of the physical properties inferred with our SED-fitting code for IRASF10038-3338. See Figure \ref{appendix_arp220} for more details.}
\end{figure*}

\begin{figure*}[h]
\centering
\includegraphics[width=\textwidth]{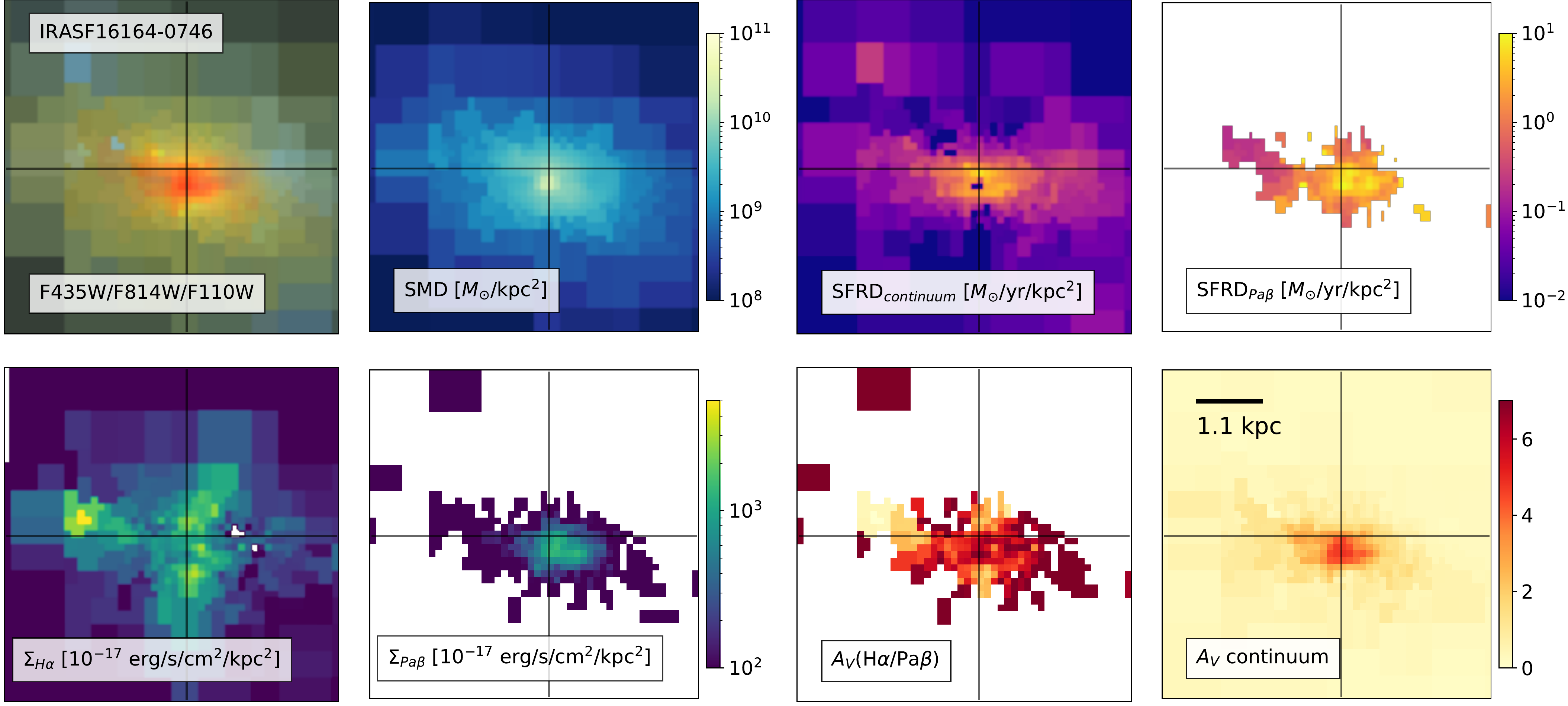}
\caption{Maps of the physical properties inferred with our SED-fitting code for IRASF16164-0746. See Figure \ref{appendix_arp220} for more details.}
\end{figure*}

\begin{figure*}[h]
\centering
\includegraphics[width=\textwidth]{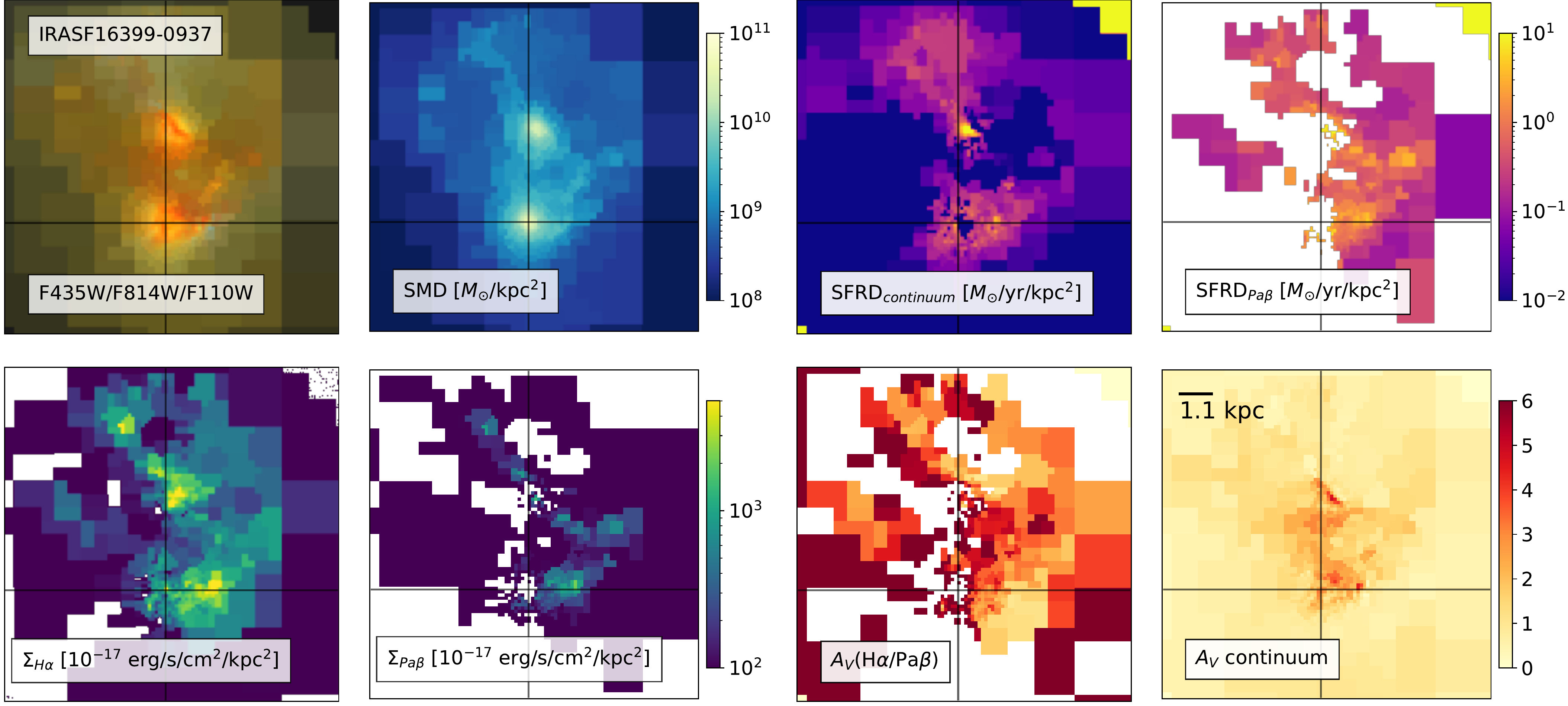}
\caption{Maps of the physical properties inferred with our SED-fitting code for IRASF16399-0937. See Figure \ref{appendix_arp220} for more details.}
\end{figure*}

\begin{figure*}[h]
\centering
\includegraphics[width=\textwidth]{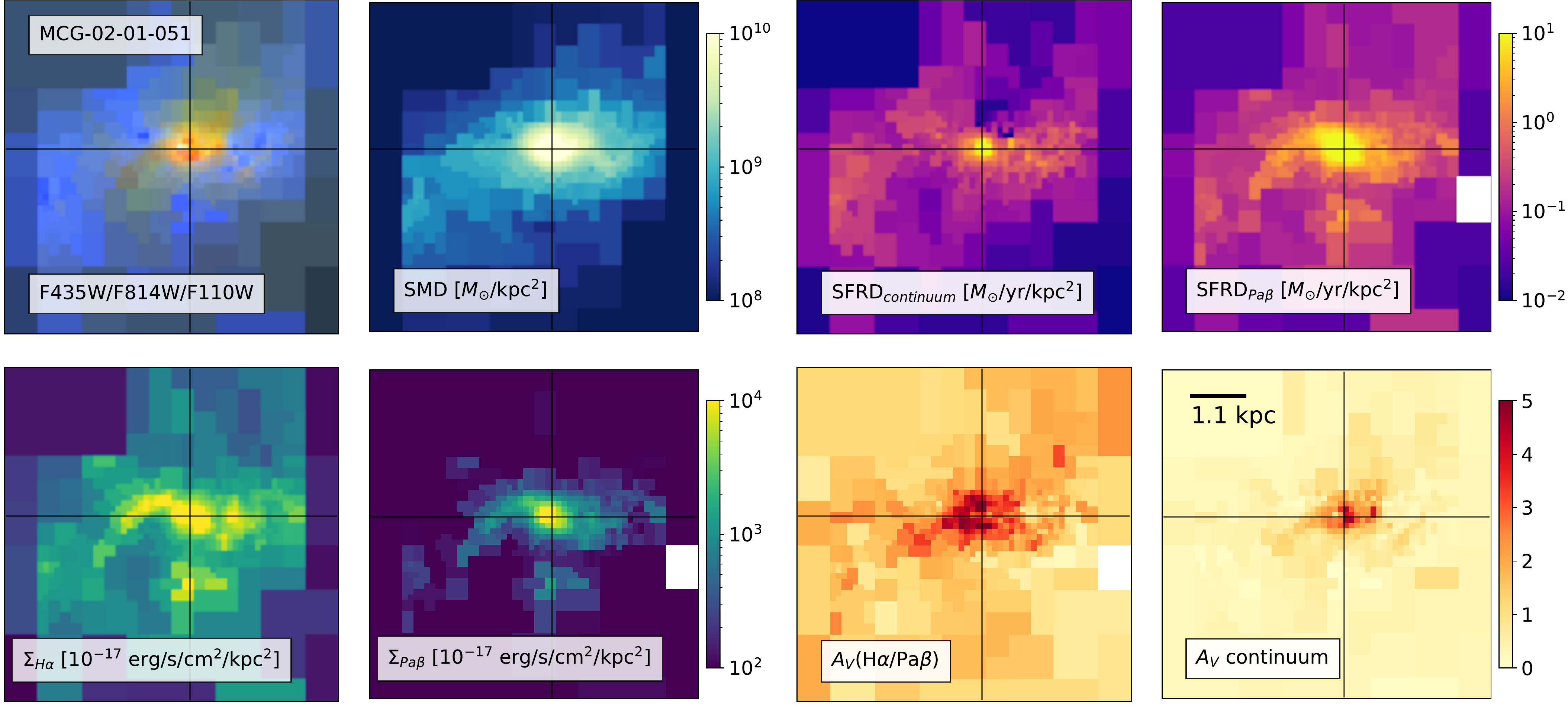}
\caption{Maps of the physical properties inferred with our SED-fitting code for MCG-02-01-051. See Figure \ref{appendix_arp220} for more details.}
\end{figure*}

\begin{figure*}[h]
\centering
\includegraphics[width=\textwidth]{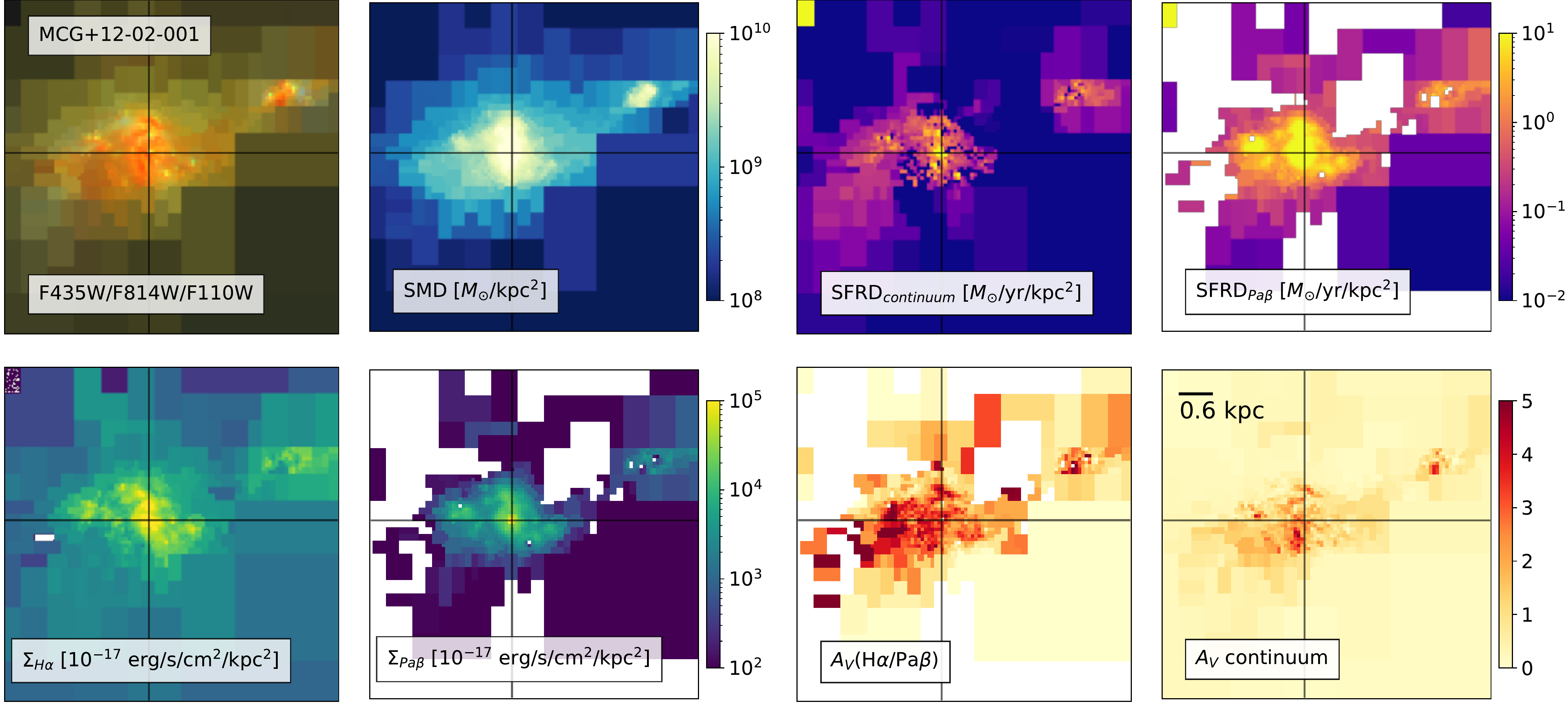}
\caption{Maps of the physical properties inferred with our SED-fitting code for MCG+12-02-001. See Figure \ref{appendix_arp220} for more details.}
\end{figure*}

\begin{figure*}[h]
\centering
\includegraphics[width=\textwidth]{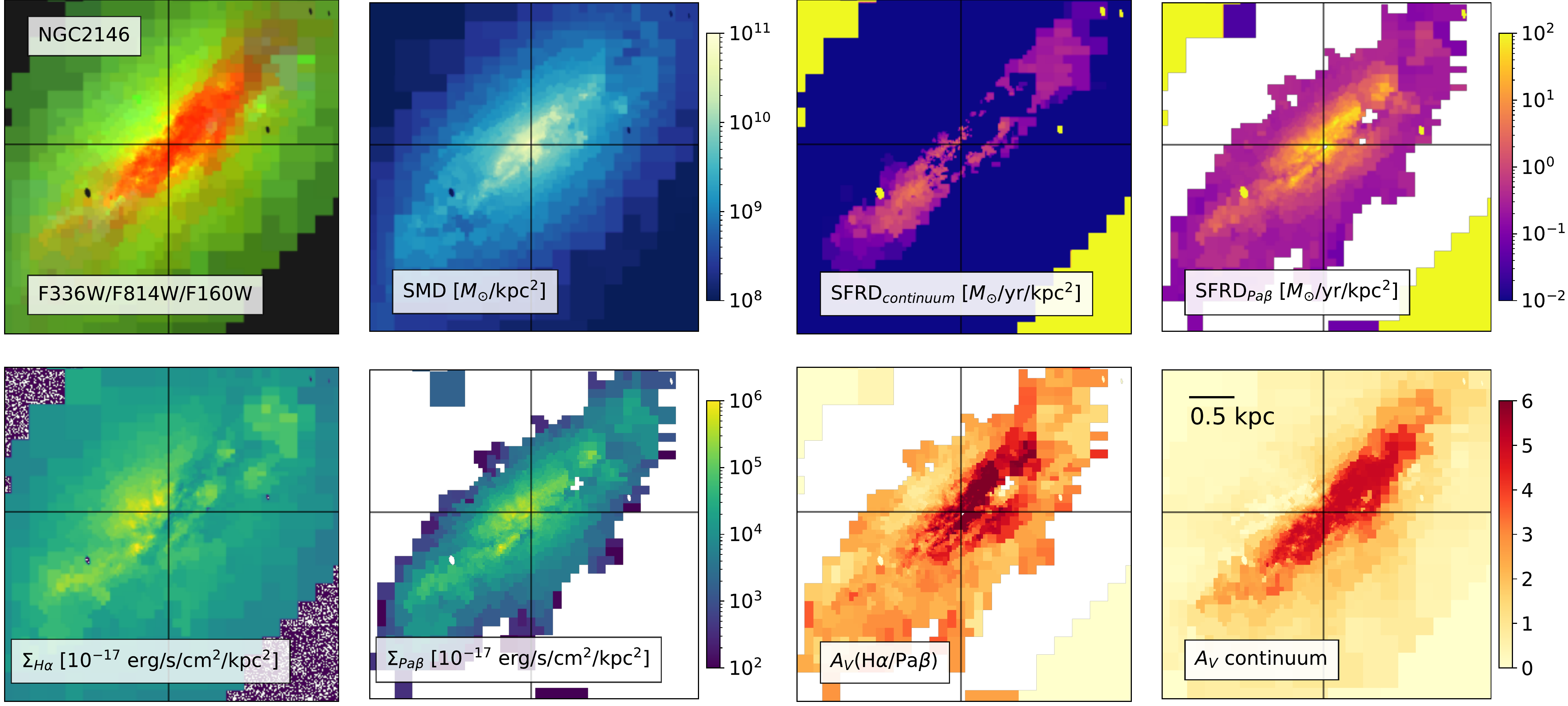}
\caption{Maps of the physical properties inferred with our SED-fitting code for NGC2146. See Figure \ref{appendix_arp220} for more details.}
\end{figure*}

\begin{figure*}[h]
\centering
\includegraphics[width=\textwidth]{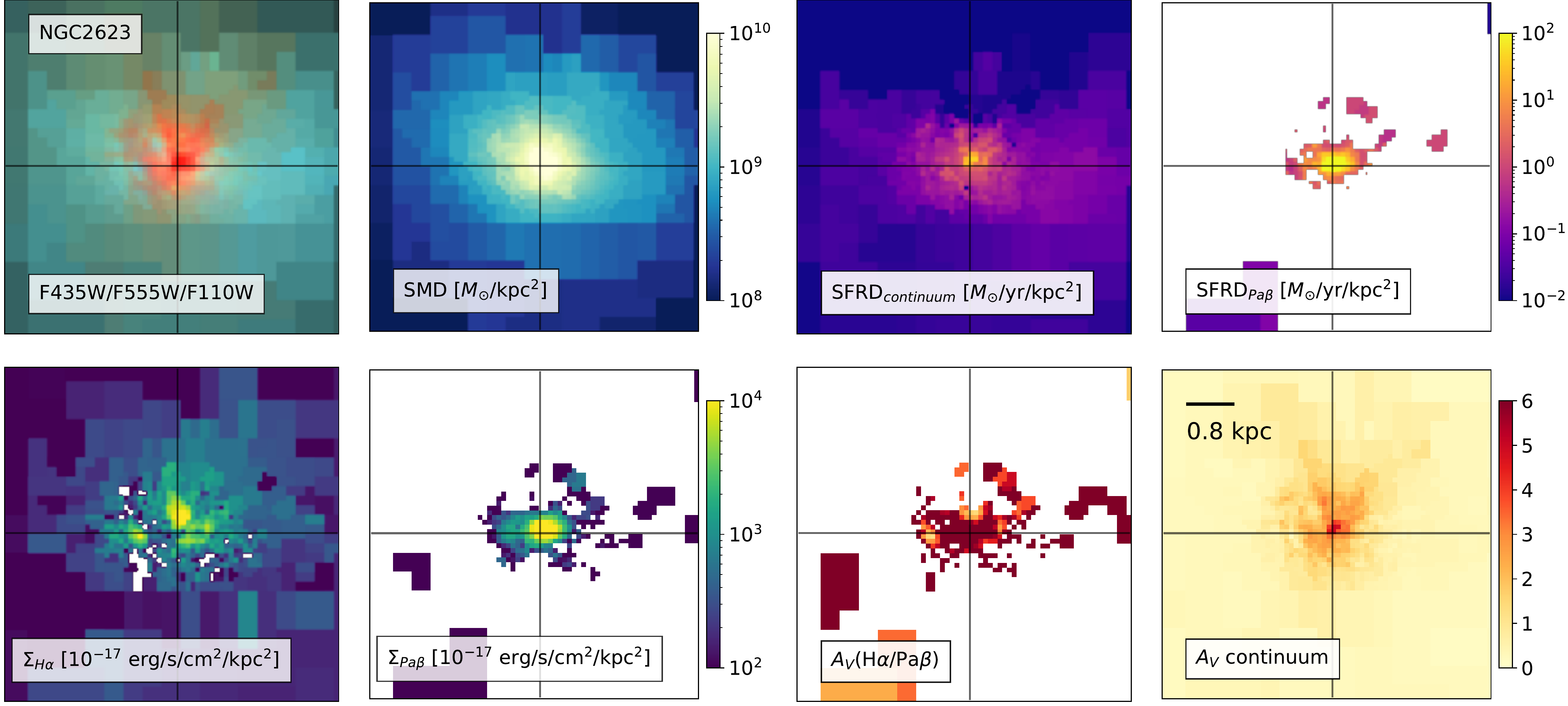}
\caption{Maps of the physical properties inferred with our SED-fitting code for NGC2623. See Figure \ref{appendix_arp220} for more details.}
\end{figure*}

\begin{figure*}[h]
\centering
\includegraphics[width=\textwidth]{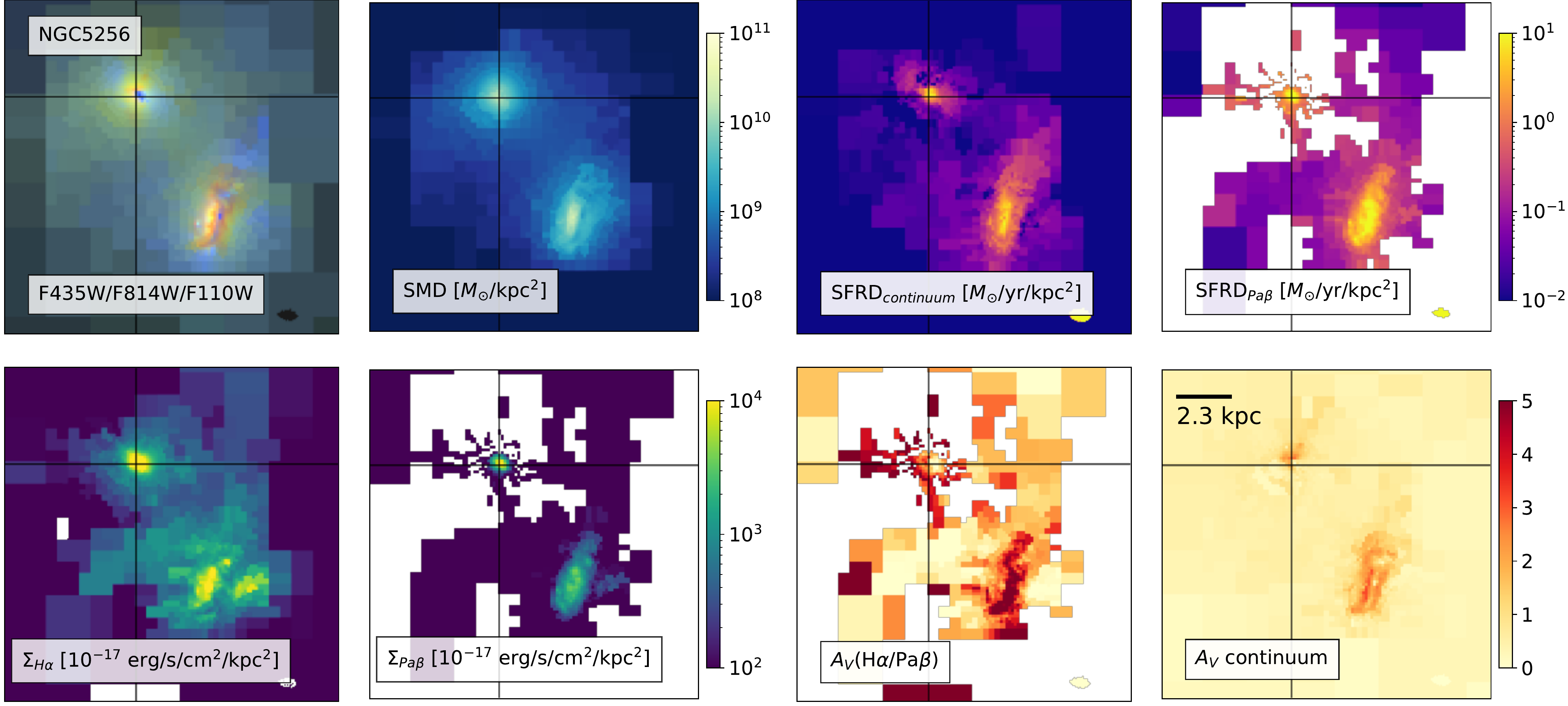}
\caption{Maps of the physical properties inferred with our SED-fitting code for NGC5256. See Figure \ref{appendix_arp220} for more details.}
\end{figure*}

\begin{figure*}[h]
\centering
\includegraphics[width=\textwidth]{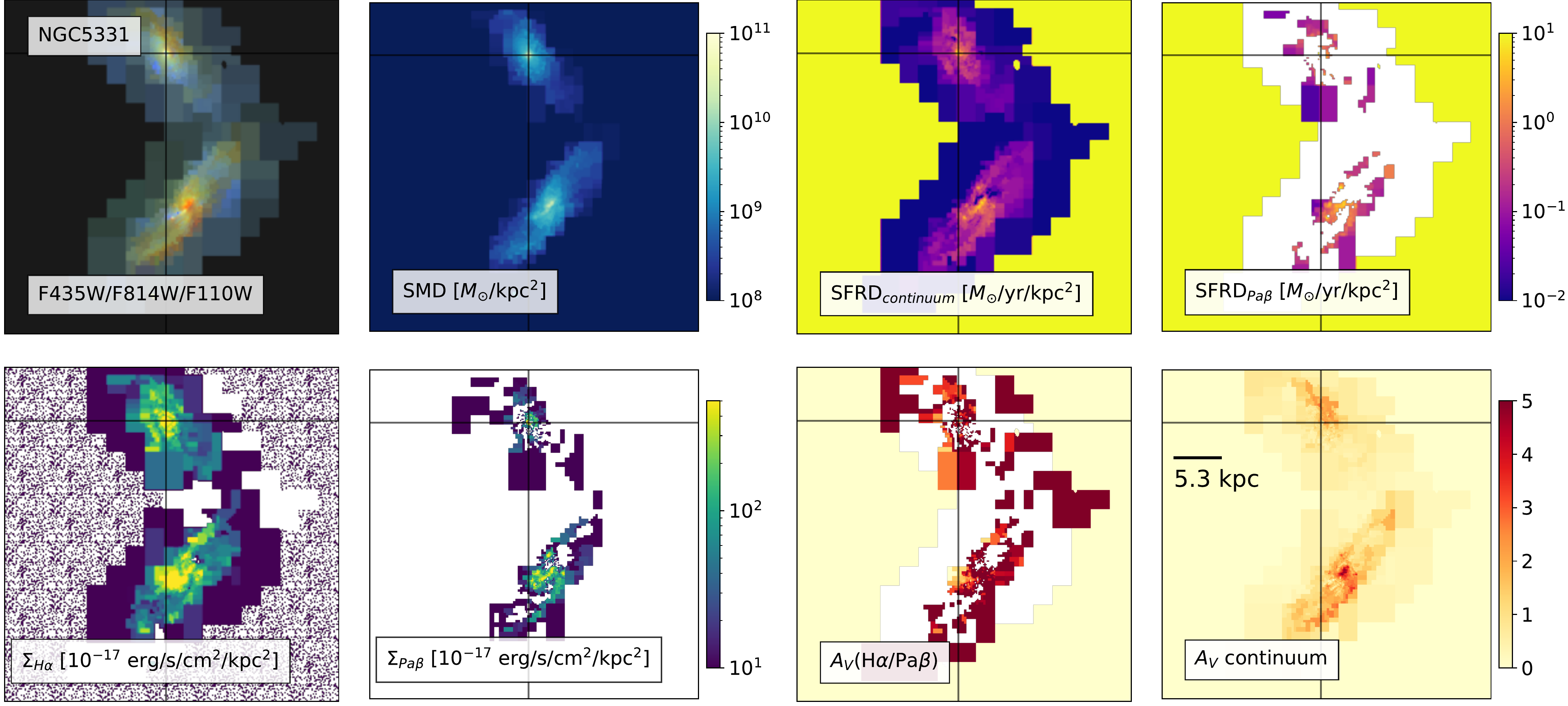}
\caption{Maps of the physical properties inferred with our SED-fitting code for NGC5331. See Figure \ref{appendix_arp220} for more details.}
\end{figure*}

\begin{figure*}[h]
\centering
\includegraphics[width=\textwidth]{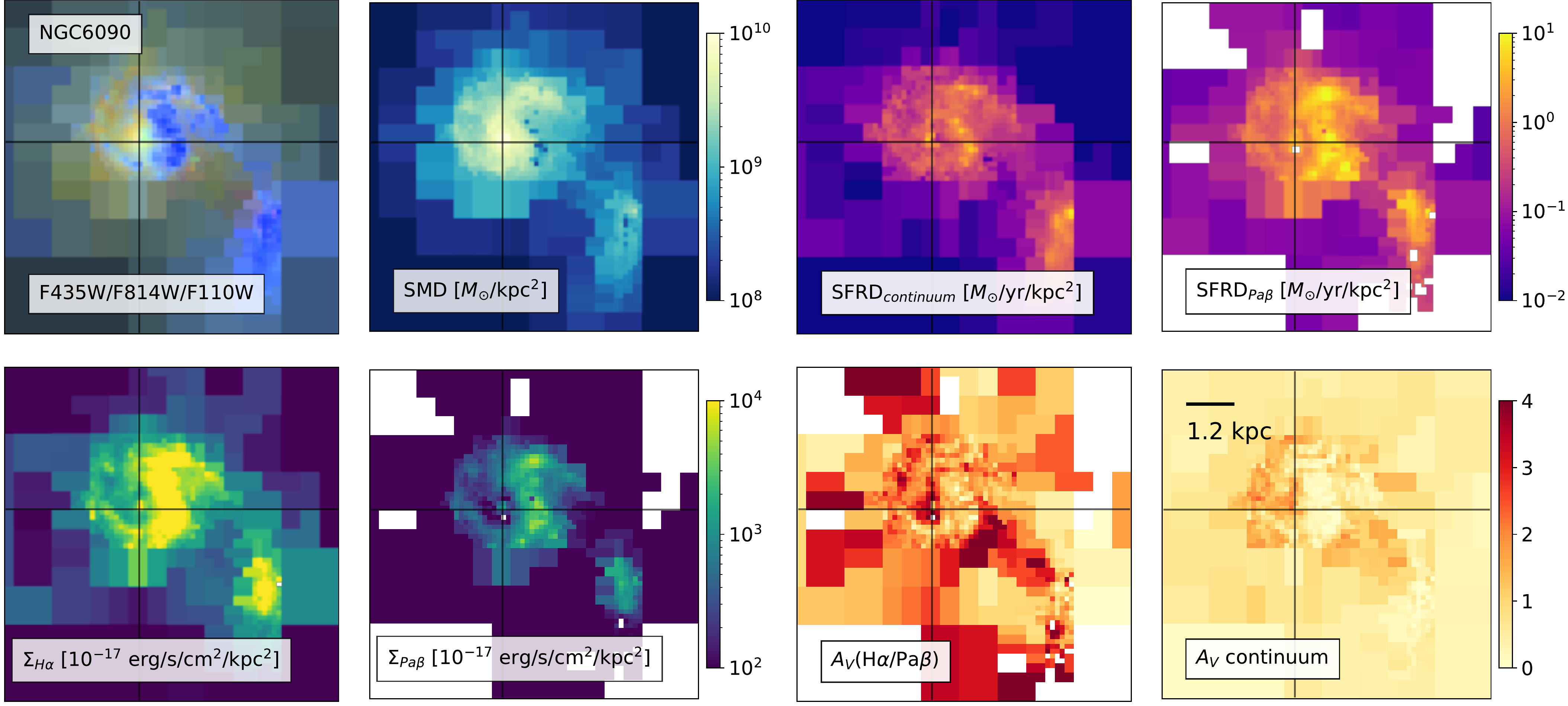}
\caption{Maps of the physical properties inferred with our SED-fitting code for NGC6090. See Figure \ref{appendix_arp220} for more details.}
\end{figure*}

\begin{figure*}[h]
\centering
\includegraphics[width=\textwidth]{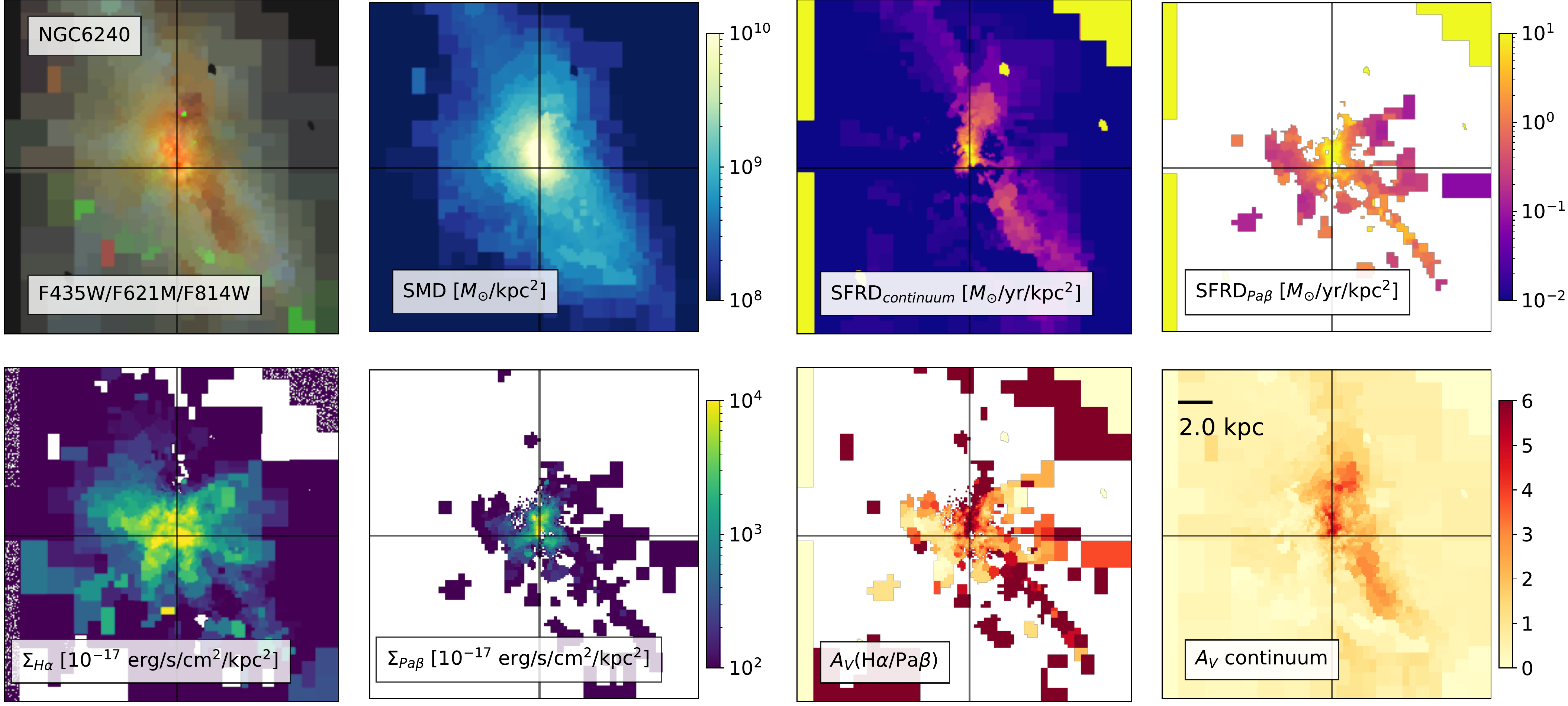}
\caption{Maps of the physical properties inferred with our SED-fitting code for NGC6240. See Figure \ref{appendix_arp220} for more details.}
\end{figure*}

\begin{figure*}[h]
\centering
\includegraphics[width=\textwidth]{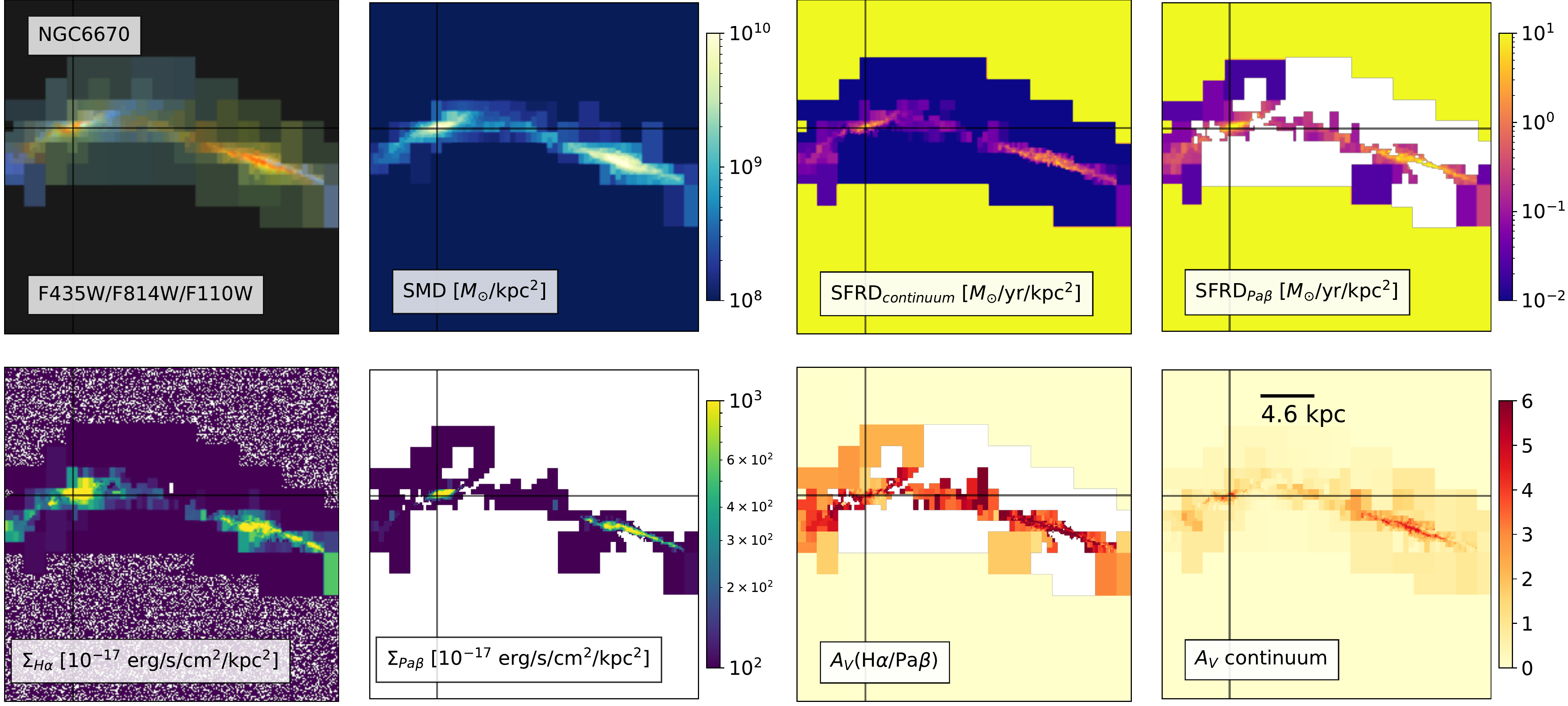}
\caption{Maps of the physical properties inferred with our SED-fitting code for NGC6670. See Figure \ref{appendix_arp220} for more details.}
\end{figure*}

\begin{figure*}[h]
\centering
\includegraphics[width=\textwidth]{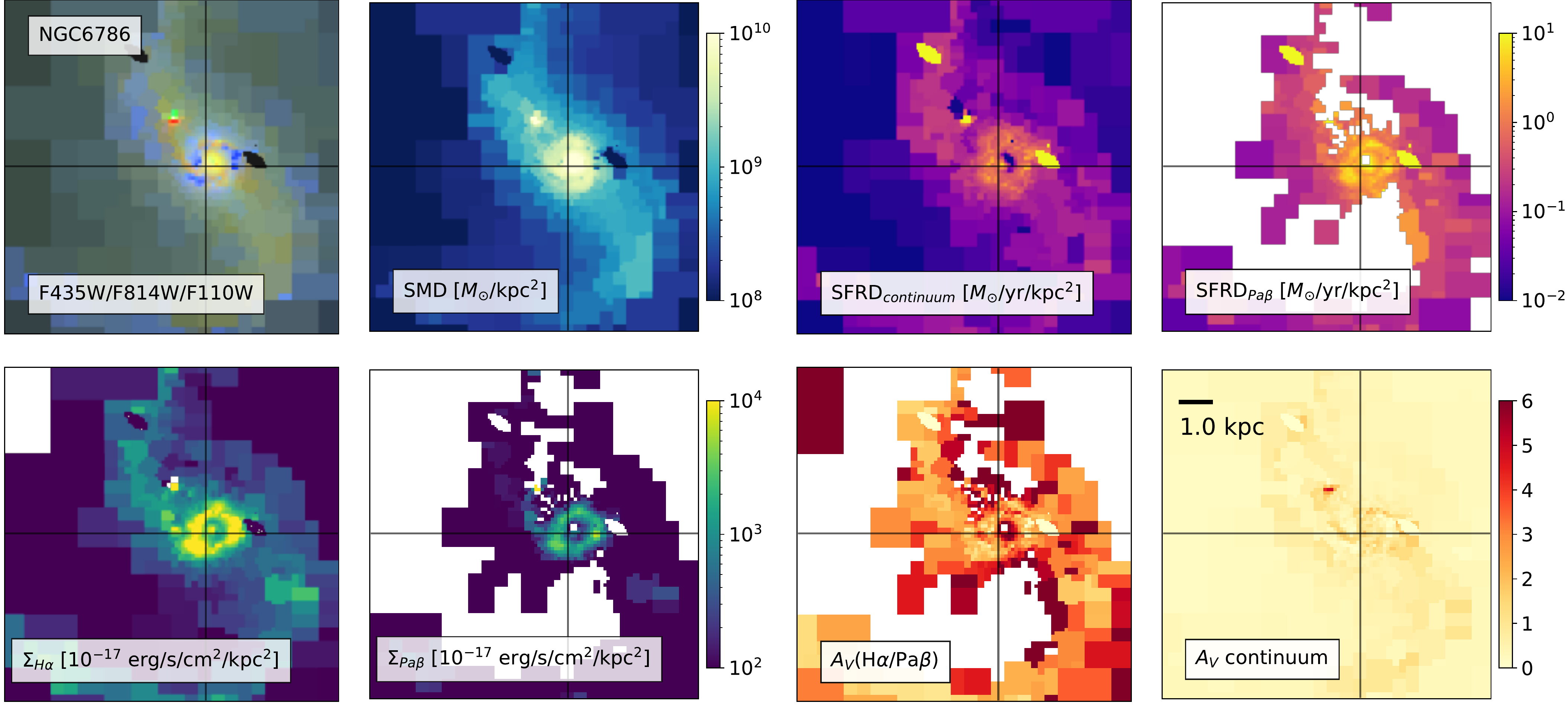}
\caption{Maps of the physical properties inferred with our SED-fitting code for NGC6786. See Figure \ref{appendix_arp220} for more details.}
\end{figure*}

\begin{figure*}[h]
\centering
\includegraphics[width=\textwidth]{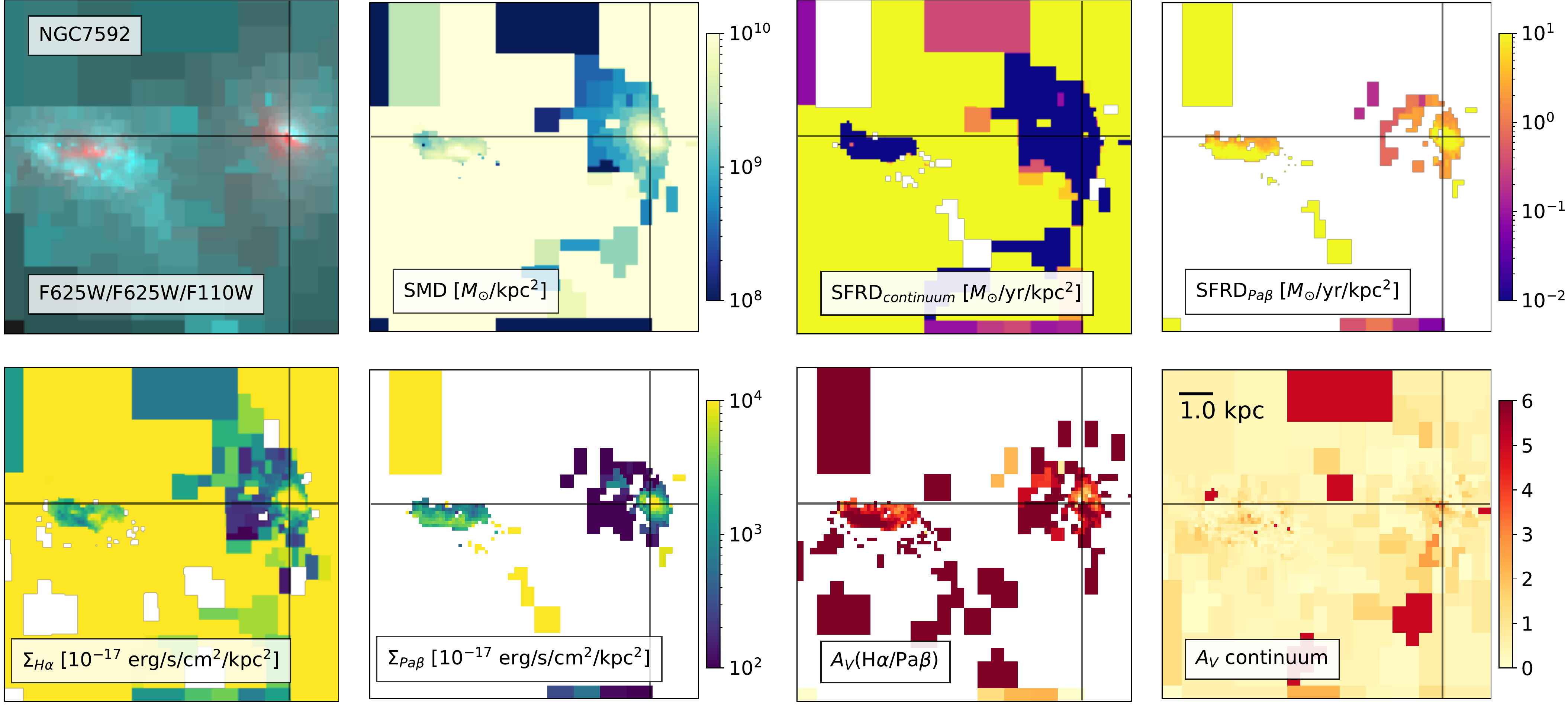}
\caption{Maps of the physical properties inferred with our SED-fitting code for NGC7592. See Figure \ref{appendix_arp220} for more details.}
\end{figure*}

\begin{figure*}[!t]
\centering
\includegraphics[width=\textwidth]{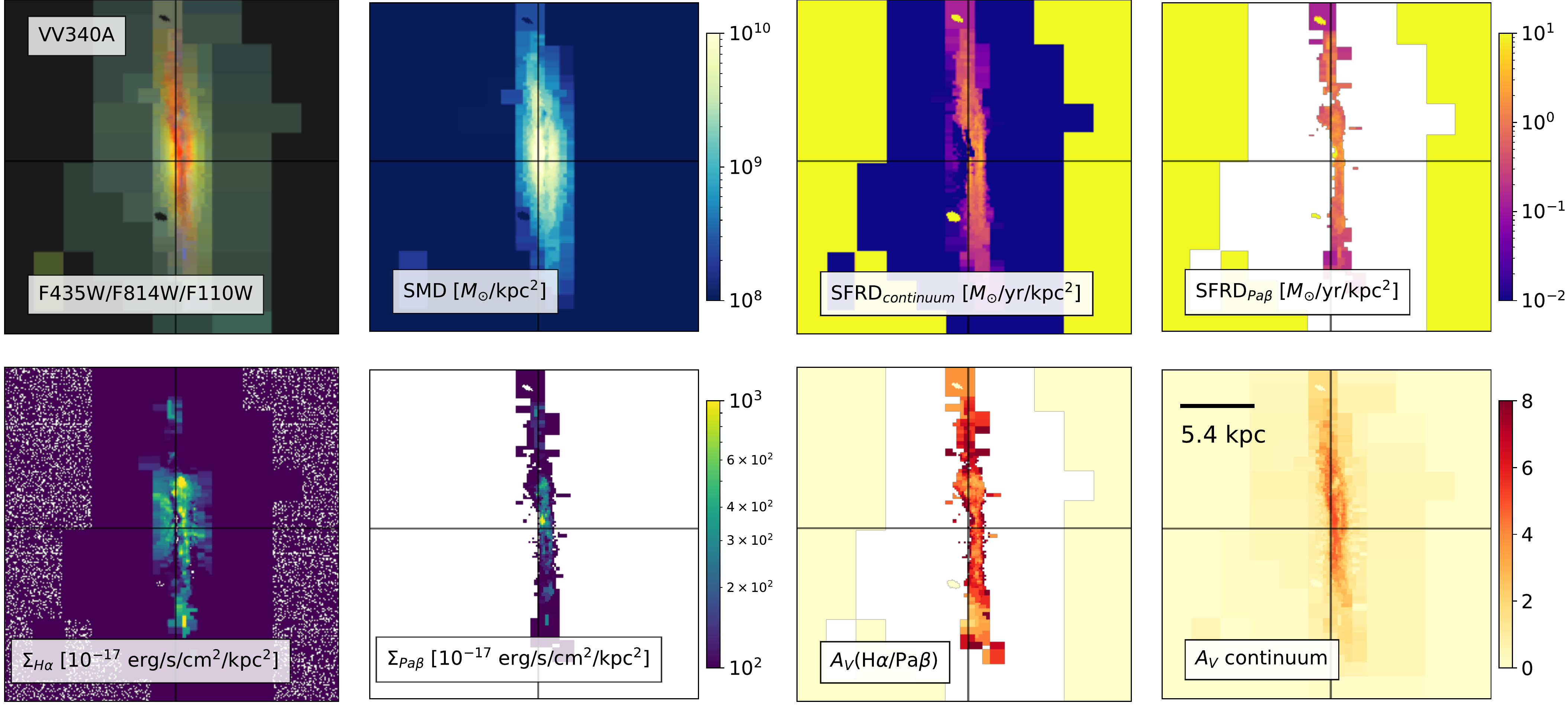}
\caption{Maps of the physical properties inferred with our SED-fitting code for VV340A. See Figure \ref{appendix_arp220} for more details.}
\label{appendix_vv340a}
\end{figure*}

\section{Additional Galaxies}

Due to the availability of $L_{IR}$ and 24$\mu$m flux measurements, we exclude from our analysis 29 additional galaxies. These targets also often have spatial extent larger than the Field of View (FoV) of the instrument. Nonetheless, we have processed the HST data of these additional targets following the same reduction process explained in \S\ref{sec2}, and the reduced mosaics are also publicly available\footnote{\url{http://cosmos.phy.tufts.edu/dustycosmos/}}. Table \ref{tab_sample2} summarises the filter coverage of the additional targets.

\begin{table*}[!th]
\centering
\begin{tabular}{l c c c c c c c}
\hline
\hline
\multirow{3}{*}{Target} & R.A. & Dec.& \multirow{3}{*}{$z$\textsuperscript{$\dagger$}} & \multirow{3}{*}{log$(\frac{L_{IR}}{L_{\odot}})$} & UV & Optical & IR \\ 
 & [deg] & [deg] & & & WFC3/ & WFC3/UVIS, ACS/WFC & WFC3/IR \\
 & & & & & UVIS & & \\
\hline
ESO338-IG004 & 291.99308 & -41.57523 & 0.0095 & - & & F550M, FR656N & f110w, f130n \\ 
MCG+00-29-023 & 170.30109 & -2.984167 & 0.0247 & 11.4\textsuperscript{*} & & f625w, f673n & f110w, f132n  \\ 
M51 & 202.46958 & 47.19526 & 0.0016 & 10.4\textsuperscript{*} & F225W, & F435W, F555W, F606W, & F110W, F128N \\ 
 & & & & & F275W, & F658N, F673N, F689M,& \\
 & & & & & F336W  & F814W & \\
M82 & 148.96846 & 69.67970 & 0.0007 & 10.8\textsuperscript{*} & F225W, & F435W, F487N, F502N, & f110w, f128n, F160W, \\ 
 & & & & & F280N, & F547M, F555W, F658N,& F164N \\
 & & & & & F336W,  & F660N, F673N, F814W & \\
 & & & & & F373N  &  & \\
M83 & 204.25383 & -29.86576 & 0.0017 & 10.1\textsuperscript{*} & F225W, & F435W, F438W, F487N, & F110W, F125W, F128N, \\ 
 & & & & & F336W, & F502N, F547M, F555W, &  F140W, F160W, F164N \\
 & & & & & F373N & F657N, F658N, F660N, & \\
 & & & & & & F673N, F814W & \\
NGC0253 & 11.88806 & -25.2888 & 0.0009 & 10.4\textsuperscript{*} & & F475W, F606W, F814W & F110W, F128N, F130N, \\ 
& & & & & & & F160W, F164N \\
NGC1140 & 43.63976 & -10.0285 & 0.0050 & -  & & F625W, F658N & F110W, F128N, \\ 
& & & & & & & F160W, F164N \\
NGC1396 & 54.52743 & -35.43992 & 0.0029 & - & & F475W, F606W, F625W, & f110w, f128n \\ 
& & & & & & F658N, F850LP & \\
NGC1482 & 58.6622 & -20.50245 & 0.0063 &  10.8\textsuperscript{*} & & F621M, F657N & F110W, F128N, \\ 
& & & & & & & F160W, F164N \\
NGC2551 & 126.20956 & 73.41200 & 0.0077 & - & & F625W, F658N & f110w, f128n \\ 
NGC2681 & 133.38641 & 51.31371 & 0.0023 & 9.5\textsuperscript{*} & & F658N, F814W & f128n, f139m \\ 
\hline
\multicolumn{8}{l}{\textbf{Table 3.} (Continues below) }\\
\end{tabular}
\end{table*}

\begin{table*}[!t]
\centering
\begin{tabular}{l c c c c c c c}
\hline
\hline
\multirow{3}{*}{Target} & R.A. & Dec.& \multirow{3}{*}{$z$\textsuperscript{$\dagger$}} & \multirow{3}{*}{log$(\frac{L_{IR}}{L_{\odot}})$} & UV & Optical & IR \\ 
 & [deg] & [deg] & & & WFC3/ & WFC3/UVIS, ACS/WFC & WFC3/IR \\
 & & & & & UVIS & & \\
\hline
NGC2841 & 140.51106 & 50.97648 & 0.0019 & 10.1\textsuperscript{**} & F225W, & F435W, F438W, F547M, & F110W, f128n, f139m, \\ 
& & & & & F336W & F657N, F658N, F814W & F160W \\
NGC2985 & 147.59264 & 72.27865 & 0.0043 & 10.2\textsuperscript{*} & & F658N, F814W & f110w, f128n  \\ 
NGC3358 & 160.88763 & -36.41071 & 0.0101 & -  & & f625w, f665n & f110w, f128n, f130n  \\
NGC3738 & 173.95085 & 54.52373 & 0.0007 & - & & F438W, F606W, F658N, & F110W, F128N, F160W  \\
& & & & & & F814W & \\
NGC4038 & 180.47084 & -18.86759 & 0.0056 & 10.8\textsuperscript{*} & F336W & F435W, F487N, F502N, & F110W, F128N, F160W, \\ 
& & & & & & F550M, F555W, F606W, & F164N \\
& & & & & & F625W, FR656N, F658N, & \\
& & & & & & F673N, F814W & \\
NGC4214 & 183.91323 & 36.32689 & 0.0010 & 8.9\textsuperscript{*} & F225W, & F438W, F487N, F502N, & F110W, F128N,  \\
& & & & &F336W, & F547M, F657N, F673N, & F160W, F164N \\
& & & & &F373N & F814W & \\
NGC5128 & 201.36506 & -43.01911 & 0.0018 & 10.1\textsuperscript{*} & F225W, & F438W, F487N, F502N, & F128N, F160W, F164N \\ 
& & & & & F336W & F547M, F657N, F673N, &  \\
& & & & & & F814W & \\
NGC6217 & 248.16340 & 78.19821 & 0.0046 & 10.3\textsuperscript{*} & & F435W, F625W, F658N, & f110w, f128n  \\
 &  &  & & & & F814W  &   \\
NGC6690 & 278.70935 & 70.52389 & 0.0016 & -  & & F625W, F658N & f110w, f128n  \\
NGC6946 & 308.718 & 60.1539 & 0.0001 & 10.9\textsuperscript{**} &  & F435W, F547M, F555W,  & F110W, F128N, F160W,  \\ 
& & & & & & F606W, F657N, F673N, & F164N \\
& & & & & & F814W & \\
NGC6951 & 309.30865 & 66.10564 & 0.0048 & 10.6\textsuperscript{*} & & F555W, F658N, F814W & f110w, f128n  \\
NGC7090 & 324.12027 & -54.55732 & 0.0028 & 9.2\textsuperscript{*} & & F606W, F625W, F658N, & f110w, f128n  \\ 
 &  &  & & & & F814W  &   \\
PGC4798 & 20.01106 & 14.36153 & 0.0312 & 11.6\textsuperscript{*} & & F435W, f673n, F814W & f110w, f132n  \\
SDSS-J110501.98 & 166.25825 & 59.68431 & 0.0338 & -  & F336W & F438W, F502N, F673N, & f110w, f132n  \\
+594103.5 &  &  & & & & F775W  &   \\
SDSS-J172823.84 & 262.09933 & 57.54539 & 0.0290 & -  & F336W & F438W, F502N, F673N, & f110w, f132n  \\
+573243.4  &  &  & & & & F775W  &   \\
SDSS220141.64 & 330.4235 & 11.85675 & 0.0296 & -  & & FQ508N, F621M, F673N, & f110w, f132n  \\
+115124.3  &  &  & & & & F763M  &   \\
UGC5626 & 156.11644 & 57.39251 & 0.0085 & - & & F625W, F658N & f110w, f130n  \\ 
\hline
\hline
\multicolumn{8}{l}{\textsuperscript{$\dagger$}\footnotesize{The redshifts were extracted from the NASA/IPAC Extragalactic Database (NED), though we note that some of the closest targets}}\\
\multicolumn{8}{l}{\footnotesize{are not necessarily in the Hubble flow.}}\\
\multicolumn{8}{l}{\textsuperscript{*}\footnotesize{Infrared luminosity from the IRAS Revised Bright Galaxy Sample \citep{2003AJ....126.1607S}.}}\\
\multicolumn{8}{l}{\textsuperscript{**}\footnotesize{Infrared luminosity from the KINGFISH Survey \citep{2011PASP..123.1347K}.}}\\
\end{tabular}
\caption{(Continued) Sources in the sample of nearby star-forming galaxies presented in this work that are not part of the GOALS Survey \citep{2009PASP..121..559A}. All targets have available reduced multi-band data, but are excluded in the analysis presented in this work. The filters added by the Cycle 23 SNAP program (HST-14095, \citealt{2015hst..prop14095B}) are indicated with small caps, whereas the rest are archival data.}
\label{tab_sample2}
\end{table*}

\clearpage

\bibliographystyle{aasjournal}

\end{document}